\documentclass[12pt,a4paper]{article} 
\usepackage{times}
\usepackage{epsfig}
\usepackage{pstricks}

\let\xepsfbox\epsfbox
\def\epsfbox#1{\scalebox{1.2}{\xepsfbox{#1}}}

\evensidemargin \oddsidemargin 
\makeatletter

\newcommand{\potmul}{3.9}

\DeclareSymbolFont{boldmath}{OML}{cmm}{b}{it}
\DeclareSymbolFontAlphabet{\mathb}{boldmath}
\DeclareMathAlphabet{\Bbb}{U}{msb}{m}{n}
\DeclareMathAlphabet{\euf}{U}{euf}{b}{n}

\DeclareMathSymbol{\balpha}{0}{boldmath}{"0B}
\DeclareMathSymbol{\bbeta}{0}{boldmath}{"0C}
\DeclareMathSymbol{\bgamma}{0}{boldmath}{"0D}
\DeclareMathSymbol{\bdelta}{0}{boldmath}{"0E}
\DeclareMathSymbol{\bepsilon}{0}{boldmath}{"0F}
\DeclareMathSymbol{\bzeta}{0}{boldmath}{"10}
\DeclareMathSymbol{\bfeta}{0}{boldmath}{"11}
\DeclareMathSymbol{\btheta}{0}{boldmath}{"12}
\DeclareMathSymbol{\biota}{0}{boldmath}{"13}
\DeclareMathSymbol{\bkappa}{0}{boldmath}{"14}
\DeclareMathSymbol{\blambda}{0}{boldmath}{"15}
\DeclareMathSymbol{\bmu}{0}{boldmath}{"16}
\DeclareMathSymbol{\bnu}{0}{boldmath}{"17}
\DeclareMathSymbol{\bxi}{0}{boldmath}{"18}
\DeclareMathSymbol{\bpi}{0}{boldmath}{"19}
\DeclareMathSymbol{\brho}{0}{boldmath}{"1A}
\DeclareMathSymbol{\bsigma}{0}{boldmath}{"1B}
\DeclareMathSymbol{\btau}{0}{boldmath}{"1C}
\DeclareMathSymbol{\bupsilon}{0}{boldmath}{"1D}
\DeclareMathSymbol{\bphi}{0}{boldmath}{"1E}
\DeclareMathSymbol{\bchi}{0}{boldmath}{"1F}
\DeclareMathSymbol{\bpsi}{0}{boldmath}{"20}
\DeclareMathSymbol{\bomega}{0}{boldmath}{"21}
\DeclareMathSymbol{\beps}{0}{boldmath}{"22}
\DeclareMathSymbol{\bthet}{0}{boldmath}{"23}
\DeclareMathSymbol{\bomeg}{0}{boldmath}{"24}
\DeclareMathSymbol{\bvphi}{0}{boldmath}{"27}
   
\newcommand{\nwl}{\nonumber\\} 

\newcommand{\txt}[1]{\quad\hbox{#1}\quad}

\@addtoreset{equation}{section}

\def\xcaption#1{\begin{quote}
               \def\normalsize{\small}\small
               \caption{#1}\global\edef\@currentlabel{\@currentlabel}
               \end{quote}\hrule}

\newcommand{\sref}[1]{section~\ref{#1}}

\newcommand{\fref}[1]{figure~\ref{#1}}

\newcommand{\eref}[1]{(\ref{#1})}

\newcommand{\ssty}{\scriptstyle}
\newcommand{\dsty}{\displaystyle}

\def\^#1{^{\hspace{.3em}#1}}

\newcommand{\follows}{\quad\Rightarrow\quad}
\newcommand{\equivalent}{\qquad\Leftrightarrow\qquad}

\newcommand{\sfrac}[2]{\frac{\ssty #1}{\ssty #2}}
\newcommand{\ft}[2]{{\textstyle{{#1}\over{#2}}}}

\newcommand{\expo}[1]{\mathrm{e}^{#1}}

\newcommand{\xsqrt}[1]{\sqrt{\dsty #1}}
\newcommand{\xfrac}[2]{\frac{\dsty #1}{\dsty #2}}

\newcommand{\comm}[2]{[#1,#2]}
\newcommand{\pois}[2]{\bigl\{#1,#2\bigr\}}

\newcommand{\dd}{\mathrm{d}}
\newcommand{\del}{\partial}
\newcommand{\deldel}[1]{\frac{\del}{\del #1}}

\newcommand{\ii}{\mathrm{i}}
\newcommand{\Tr}{\mathrm{Tr}}
\newcommand{\Trr}[1]{\Tr(#1)}

\newcommand{\grpSL}{\mathsf{SL}}
\newcommand{\algsl}{\euf{sl}}

\newcommand{\ZZ}{\Bbb{Z}}
\newcommand{\RR}{\Bbb{R}}

\newcommand{\pol}{\Delta}

\newcommand{\qsp}{\tilde\mathcal{P}}
\newcommand{\qpot}{\mathit{\tilde\Theta}}
\newcommand{\qsym}{\mathit{\tilde\Omega}}
\newcommand{\qham}{\tilde H}

\newcommand{\psp}{\mathcal{P}}
\newcommand{\ppot}{\mathit{\Theta}}

\newcommand{\pham}{H}

\newcommand{\mul}{\zeta}

\newcommand{\braket}[2]{\langle #1 | #2 \rangle}

\newcommand{\op}[1]{\widehat{#1}}

\newcommand{\con}{\mathcal{C}}
\newcommand{\tcon}{\mathcal{E}}
\newcommand{\rcon}{\mathcal{D}}

\newcommand{\gam}{\bgamma}
\newcommand{\one}{\mathbf{1}}

\newcommand{\hol}{\mathb{u}} \newcommand{\hols}{u}
\newcommand{\ehol}{\mathb{h}} 
\newcommand{\chol}{\mathb{g}} 
\newcommand{\mom}{\mathb{p}} \newcommand{\momv}{p}
\newcommand{\pos}{\mathb{x}} \newcommand{\posv}{x}
\newcommand{\dis}{\mathb{z}} 

\newcommand{\tmom}{\mathb{P}} 
\newcommand{\tang}{\mathb{J}} 

\newcommand{\vv}{\mathb{v}} 
\newcommand{\ww}{\mathb{w}} 
\newcommand{\uu}{\mathb{u}} 
\newcommand{\pp}{\mathb{p}} 

\newcommand{\cT}{\tilde T}
\newcommand{\cR}{\tilde R}
\newcommand{\cdir}{\tilde \phi}

\newcommand{\xdir}{\phi}
\newcommand{\pdir}{\beta}

\newcommand{\xinn}{\xdir_\mathrm{in}}
\newcommand{\xout}{\xdir_\mathrm{out}}

\newcommand{\vrh}{\varrho}

\newcommand{\snh}{\mathop{\operator@font snh}\nolimits}
\newcommand{\csh}{\mathop{\operator@font csh}\nolimits}
\newcommand{\tnh}{\mathop{\operator@font tnh}\nolimits}
\newcommand{\tn}{\mathop{\operator@font tn}\nolimits}
\newcommand{\atn}{\mathop{\operator@font atn}\nolimits}
\newcommand{\sn}{\mathop{\operator@font sn}\nolimits}
\newcommand{\cs}{\mathop{\operator@font cs}\nolimits}

\newcommand{\ph}{\varphi}
\newcommand{\stat}{\lambda}
\newcommand{\cut}{\lambda}
\newcommand{\ecut}{\eta}
\newcommand{\prt}{\pi}
\newcommand{\tm}{\tilde m}

\newcommand{\lam}{\Lambda}

\newcommand{\sgn}{\mathop{\operator@font sgn}\nolimits}

\newcommand{\eps}{\varepsilon}
\newcommand{\p}{\varphi}
\newcommand{\mm}{\omega}
\newcommand{\lap}{\triangle}

\newcommand{\Mpl}{M_\mathrm{Pl}}
\newcommand{\Mmin}{M_\mathrm{min}}
\newcommand{\Mmax}{M_\mathrm{max}}
\newcommand{\lpl}{\ell}

\newcommand{\Jmax}{J_{\mathrm{max}}}
\newcommand{\Rmin}{R_{\mathrm{min}}}

\makeatother

\begin{document}

\setcounter{page}{0}
\thispagestyle{empty}

\begin{flushright}
MZ-TH/00-45\\
gr-qc/0103085
\end{flushright}

\begin{center}
  \LARGE \textsc{The 2+1 Kepler Problem and Its Quantization}
\end{center}
 
\vspace*{8mm}

\begin{center}
  \textbf{Jorma Louko}\\[2ex]
  School of Mathematical Sciences, University of Nottingham,\\
  Nottingham NG7 2RD, United Kingdom\\
  jorma.louko@nottingham.ac.uk\\[2ex]
  and\\[2ex]
  \textbf{Hans-J\"urgen Matschull}\\[2ex]
  Institut f\"ur Physik, Johannes Gutenberg-Universit\"at\\
  55099 Mainz, Germany\\
  matschul@thep.physik.uni-mainz.de
\end{center}

\vspace*{10mm}

\begin{center}
  March 2001
\end{center}

\vspace*{10mm}

\begin{abstract}
  We study a system of two pointlike particles coupled to three
  dimensional Einstein gravity. The reduced phase space can be
  considered as a deformed version of the phase space of two
  special-relativistic point particles in the centre of mass frame.
  When the system is quantized, we find some possibly general effects
  of quantum gravity, such as a minimal distances and a foaminess of
  the spacetime at the order of the Planck length. We also obtain a
  quantization of geometry, which restricts the possible asymptotic
  geometries of the universe.
\end{abstract}

\newpage

\section*{Outline and summary}
The Kepler system is the simplest realistic example of a coupled two
body system, and belongs to the few systems that can be solved exactly
within the framework of Newtonian gravity. It consists of two
pointlike objects, characterized only by their masses, and interacting
with the gravitational field. Unfortunately, within the framework of
general relativity, the two body problem not only lacks of an exact
solution. It is not even well defined, because Einstein gravity in
four spacetime dimensions does not admit pointlike matter sources.
Clearly, this makes general relativity so interesting. But the obvious
drawback is the absence of a simple but still realistic toy model,
which is sometimes very useful. The Kepler system is, in a sense, the
hydrogen atom of gravity.

The situation is different in three spacetime dimensions, where
Einstein gravity is not only a much simpler field theory
\cite{witten,matrev,matcs}. It also admits pointlike matter sources
\cite{starus,djh,welrh,bcv,ms,cms,matmul}, and even a more or less
straightforward canonical quantization
\cite{witten,matrev,thooft,wel2p,matwel}. The vacuum Einstein
equations in three dimensions require the spacetime to be flat outside
the matter sources. There are neither gravitational waves, nor local
gravitational forces. However, the spacetime becomes curved if matter
is present. The simplest example of a non-trivial spacetime is the
gravitational field of a massive pointlike particle. The spacetime is
the direct product of a real line with a conical space. The particle
is sitting at the tip, and the deficit angle of the cone is $8\pi Gm$,
where $m$ is the rest mass of the particle, and $G$ is Newton's
constant.

In units where the velocity of light is one, $G$ has the dimension of
an inverse mass or energy. The inverse $\Mpl=1/G$ is the Planck mass.
It is a classical quantity in the sense that $\hbar$ is not involved
in the definition. There is thus a certain amount of curvature on the
world line of a point particle, which is proportional to its mass. But
this is just a very simple conical singularity in an otherwise flat
spacetime. There is a simple way to visualize a spacetime containing a
single point particle. One starts from a flat three dimensional
Minkowski space and cuts out a \emph{wedge}. A wedge is a subset which
is bounded by two timelike half planes, whose common boundary is a
timelike geodesic. This geodesic becomes the world line of the
particle.

\begin{figure}[t]
  \begin{center}
  \epsfbox{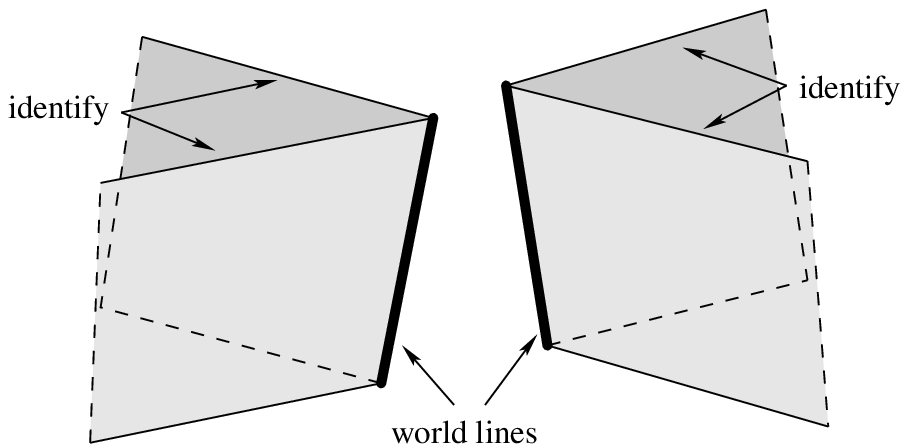}
    \xcaption{The Kepler spacetime can be constructed by cutting out two 
    wedges from a flat Minkowski space. The faces are identified, such
    that two conical singularities arise in an otherwise flat
    spacetime. In the rest frame of the each particle, the deficit
    angle of the conical space is proportional to the mass of the
    particle.}
  \end{center}
  \label{wdg}
\end{figure}

The two half planes are mapped onto each other by a certain isometry
of Minkowski space. It is a Lorentz rotation about the world line, and
the angle of rotation is $8\pi Gm$. The region inside the wedge is
taken away, and the points on the two half planes are identified,
according to the Lorentz rotation. The result is a locally flat
spacetime with a conical singularity on the world line. For a massless
particle, the same procedure can be applied to a pair of half planes,
whose common boundary is a lightlike geodesic. They are then mapped
onto each other by a null rotation. The Kepler spacetime contains two
particles, and therefore we have to apply this procedure twice. The
result is shown in \fref{wdg}.

Unless one of the particles has a deficit angle which is bigger than
$\pi$, hence a mass above $\Mpl/8$, every possible two particle
spacetime can be constructed in this way. The geometry of the
spacetime only depends on the relative motion of the particles, which
can be read off immediately from the relative orientation of the two
world lines in Minkowski space. It is therefore possible to get an
overview of all possible spacetime geometries very easily. However,
there are also some problems with this simple construction. First of
all, it does not work for masses bigger then $\Mpl/8$, although such
spacetimes do exist. In this case, the half planes defining the
boundary of the wedges must be replaced by curved surfaces, as
otherwise the wedges overlap. But this is actually not a serious
problem.

The more serious problem has to do with the asymptotic structure of
the spacetime at infinity. The region far away from the particles is
split into two segments in \fref{wdg}. Each segment is a subset of
Minkowski space. But on the wedges we have to apply non-trivial
transition functions, relating the Minkowski coordinates on one side
to those on the other side. To find out what the spacetime looks like
at infinity, it would be nicer to have a single coordinate chart
covering this region. There is in fact a particular reason why we are
interested in the asymptotic structure of the Kepler spacetime. In
order to quantize it in the end, we first have to set up a proper
classical Hamiltonian formulation. This requires a proper definition
of an action principle for the underlying field theory of Einstein
gravity. And this again requires some kind of asymptotical flatness
condition to be imposed on the metric at infinity \cite{nico}.

The asymptotic structure of the Kepler spacetime depends crucially on
the relative motion of the particles. If they are moving slowly, then
far away from the particles the spacetime is also conical. It looks
almost like the gravitational field of a single particle, whose mass
is equal to the sum of the two masses of the real particles. The rest
frame of this \emph{fictitious} particle can be identified with the
\emph{centre of mass frame} of the universe. If the particles are
moving faster, the apparent mass of the fictitious particle has to be
replaced by the total energy of the system. It also receives a spin,
which represents the total angular momentum. But still, the universe
looks like a cone at infinity, and this cone defines the centre of
mass frame.

Something strange happens when the relative motion of the particles
exceeds a certain threshold \cite{welwin}. The definition of a centre
of mass frame then breaks down, and the asymptotic structure of the
spacetime is no longer conical. Even more peculiar, the spacetime then
contains closed timelike curves \cite{gott,steif,holmat}. Clearly,
these are very interesting features of such a simple two particle
spacetime. But for our purpose we have to exclude them, again because
we want to set up a proper Hamiltonian framework. This requires a well
defined causal structure of the spacetime. Otherwise the Hamiltonian,
or ADM formulation cannot be applied to general relativity. It
requires the existence of a foliation of the spacetime by spacelike
slices \cite{mtw}

In order to get rid of these problems, we have to go over from the
simple construction of the Kepler spacetime in \fref{wdg}, to a
slightly more sophisticated one, which focuses more on the asymptotic
structure of spacetime at infinity. The basic idea is, first to fix
the asymptotic structure of the spacetime, and then insert the
particles. In \fref{wdg}, the sequence is the other way around. The
world lines are inserted first, then the actual spacetime is
constructed, and finally the asymptotic structure can be read off by
looking at the region at infinity. The transition from this picture to
an alternative description of the Kepler spacetime, based on its
asymptotic structure, is explicitly carried out in \cite{loumat}.
There we also give a comprehensive overview of all those spacetimes
that admit the definition of a centre of mass frame.

Somewhat schematically, the alternative construction of a two particle
spacetime is shown in \fref{tip}. One starts from a big cone, cuts off
the tip, and identifies the cut lines, which are two geodesics, such
that a conical surface with two tips arises. The three dimensional
version of this construction yields the Kepler spacetime, although the
technical details are a little bit more involved. The advantage of
this procedure is that the original cone immediately defines the
centre of mass frame of the universe, as seen by an observer at
infinity, and independent of the way the particles are inserted.
Moreover, we will be able also introduce position and momentum
coordinates of the particles, referring to the centre of mass frame
defined by the big cone.

In this way, the Kepler system can effectively be treated like a
simple two particle system in a fixed three dimensional background
spacetime, although the technical details are again slightly more
involved. The phase space becomes a finite dimensional manifold, and
the kinematical and dynamical properties of the Kepler system are
finally encoded in the usual way, in the symplectic structure, or the
Poisson bracket, and the Hamiltonian. The phase space structure is
very similar to a system of two special-relativistic point particles
in a flat Minkowski space, restricted to the centre of mass frame. And
in fact, it is possible to take the limit $G\to0$, where the
gravitational interaction is switched off, and then the Kepler system
reduces to this free particle system.

The article is therefore organized as follows. In \sref{free}, we
shall only study the free particle system. Of course, there is nothing
new about this to be learned, expect perhaps the relativistic
definition of a centre of mass frame, which is in this form not a very
common concept. It is actually a centre of energy frame, but we shall
stick to the more familiar notion of a centre of mass. After imposing
the appropriate restriction on the phase space, we shall go through
the usual canonical programme. Starting from the classical Hamiltonian
framework, we perform a phase space reduction, derive and solve the
classical equations of motion, and finally we quantize it, deriving
the energy eigenstates and the spectra of certain interesting
operators.  The idea behind this preparation, using a well known and
very simple toy model for our real toy model, is to keep the
conceptual aspects apart from the technical aspects.

The conceptual aspects are, for example, the definitions of the
various phase spaces, the way the mass shell constraints are imposed,
the principle idea of the phase space reduction, and finally also the
quantization methods. At the classical level, the free particle system
also provides a nice toy model for the Hamiltonian formulation of
general relativity within the ADM framework. This is indicated in
\fref{rpp}. At the quantum level, we are going to consider two
alternative quantization methods, the Schr\"odinger method applied to
the reduced classical phase space, where all gauge symmetries are
removed, and the Dirac method applied to an extended phase space,
where the dynamics of the system is defined by a generalized mass
shell constraint.

All these concepts can then be applied to the Kepler system in the
very same way. It is therefore useful first to explain them using a
much simpler model. The more technical aspects are then the
modifications that we have to make when the gravitational interaction
is switched on. They are divided into a classical part in
\sref{class}, and a quantum part in \sref{quant}. In the classical
section, we shall first look at the Kepler spacetime itself, in the
way explained above and described in \fref{tip}. We shall mainly focus
on a geometric point of view, without going into any technical details
and proofs. All these details can be found in the previously mentioned
and more comprehensive article \cite{loumat}. It includes, in
particular, the precise transition from \fref{wdg} to the alternative
description in \fref{tip}.

In a certain sense, the Kepler spacetime is a \emph{deformation} of
the free particle spacetime, with Newton's constant being the
deformation parameter. At any stage of the derivation we can always
get back to the free particle system by taking the limit $G\to0$. This
can be used as a cross check at various points, and sometimes it is
even possible to set up general rules, telling us how to deform the
free particle system to obtain the corresponding structures of the
coupled system. This also applies to the phase space of the Kepler
system, which is the subject of the second part of \sref{class}. It is
a \emph{deformed} version of the free particle phase space. The
interesting point is thereby that gravity affects the symplectic
structure of this phase space, and not only the Hamiltonian, which is
otherwise the typical feature of interactions between point particles.

The actual derivation of the deformed symplectic structure will not be
given in the article. We should however emphasize that this derivation
is a very important point. We do not want to make any special or
unmotivated assumptions. Instead, the only assumption that enters the
definition of the Kepler system is the following. The kinematical and
dynamical features of the gravitational field are completely defined
by the Einstein Hilbert action. All the relevant phase space
structures can then be derived from this action principle, by a
straightforward phase space reduction. However, apart from the
symplectic structure, all other features of the phase space can more
or less be inferred from geometric considerations. We shall therefore
restrict to these geometrical aspects here, and refer to \cite{matmul}
for the derivation of the symplectic structure for a more general
multi particle system.

In the last part of \sref{class}, we will go through the whole
canonical programme once again, deriving and solving the classical
equations of motion, and briefly describing the various kinds of
trajectories. At this point, we can actually forget about the general
relativistic nature of the system, and treat it as if it was a simple
two particle system living in a three dimensional background
spacetime. Or, if we do not want to give up the general relativistic
point of view completely, we may at least stick to the ADM picture,
and consider the Kepler system as a space which evolves in time. As
shown in \fref{emb}, the phase space variables define the geometry of
space at a moment of time, and this geometry changes with time.
Effectively, the particles are moving in a two dimensional space. We
can also think of a typical scattering process, and define quantities
like incoming and outgoing momenta, and scattering angles.

In the quantum section, we will first try to apply the same
quantization methods that we also applied to the free particle system.
We'll find that the straightforward Schr\"odinger quantization fails,
due to some peculiar features of the deformed classical phase space
and its symplectic structure. This already indicates that there are
some new effects to be expected, which are due to the gravitational
interaction, and which are fundamentally different from other
interactions. The Dirac method however works. It is possible to set up
a well defined operator representation, and to quantize and solve the
constraint equation, which is a generalized Klein Gordon equation. A
similar equation has also been found for a somewhat simpler single
particle system \cite{matwel}.

We can solve this constraint equation, and finally we are able to
express the energy eigenstates of the Kepler system explicitly as wave
functions on a suitably defined configuration space. Suitably thereby
means that the wave function has the usual physical interpretation as
a probability amplitude for the particles in space. Or, once again, if
we want to stick to the general relativistic point of view, it is a
probability amplitude for certain geometries of space. We have in this
sense a simple example for a truly quantized general relativistic
system. The quantum state gives us probabilities for geometries. But
nevertheless, it is somewhat more intuitive to consider the wave
function as a probability amplitude in an ordinary space. The space is
thereby the configuration space of the two particles in the centre of
mass frame. And it is also useful to have in mind the usual picture of
a scattering state in quantum mechanics.

The energy eigenstates are parameterized by two quantum numbers. The
inverse radial wavelength at infinity represents the eigenvalue of the
incoming and outgoing momenta of the particles. And the inverse
angular wavelength represents the eigenvalue of the angular momentum.
The latter is quantized in steps of $\hbar$, and the radial momentum
turns out to have a positive continuous spectrum. The energy, or the
frequency of the wave function, only depends on the radial quantum
number. So far, these are the typical features of scattering states in
quantum mechanics. However, the energy spectrum is bounded from below
and from above. This is actually not surprising, because the total
energy contained in a three dimensional universe is bounded from above
by $\Mpl/4$, which corresponds to the maximal deficit angle $2\pi$ of
a conical spacetime. The lower bound for the total energy is the sum
of the rest masses of the particles, which is also not surprising.

What is remarkable, however, is that the radial momentum of the
particles is nevertheless unbounded. For very large momenta of the
particles, the total energy of the Kepler system approaches the upper
bound $\Mpl/4$. This behaviour has also been found for a single
particle, where the upper bound is half as big, thus $\Mpl/8$
\cite{matwel}. The relation between the spatial momentum of the
particles in the centre of mass frame, and the total energy of the
system is shown in \fref{dsp}, where it is compared to the
corresponding free particle energy. The energy of the Kepler system is
always smaller than the free particle energy. The difference is a kind
of gravitational binding energy. At the quantum level, this has the
strange consequence that the wavelength of the wave function in space
can be arbitrarily small, but the frequency is bounded from above.

Finally, we shall then look at the wave functions themselves.
Unfortunately, it is hardly possible to read off any physically
interesting information directly from the analytic expressions. They
are somewhat complicated, involving hypergeometric functions. We shall
therefore look at the graphical representations of some typical wave
functions. They are shown in the figures~\ref{wav0}
through~\ref{wav3}, both for the free particles and the Kepler system.
We shall thereby find the following interesting features. At large
distances, the gravitational interaction has almost no effect. We have
the typical scattering wave function, which is a superposition of an
ingoing and an outgoing radial wave. The only difference between the
free and the coupled system is a phase shift, which indicates that
some interaction takes place when the particles are closer to each
other.

At small distances however, which are of the order of the \emph{Planck
length} $\lpl=G\hbar$, the wave function changes drastically. We
typically find that the particles avoid to be at the same point in
space. We even find that under certain very general circumstances, it
is impossible for the particles to get closer to each other than a
certain minimal distance. It is of the order of ten to hundred Planck
lengths, and depends on the rest masses of the particles and their
statistics. In three spacetime dimensions, there are not only bosons
and fermions, but also \emph{anyons}, and there is also a generalized
statistics if the particles are not identical \cite{anyons}. All this
can be taken into account very easily when the quantization is
performed.  The finite lower bound for the distance of the particles
in space arises whenever the particles are not two bosons, and it is
maximal for two identical fermions.

Referring again to the scattering picture, it is reasonable to say that
due to this minimal distance, it is impossible to probe the structure
of spacetime at small length scales, even if we increase the momentum
of the particles unboundedly. This is exactly the kind of limit that
quantum gravity is expected to impose on the ability to look at small
length scales in spacetime. Another feature of our toy model is
closely related to this, but a little bit more general. Even if the
particles are further apart than the minimal distance, it is still
impossible to localize them within a box that is smaller than a
certain size. More precisely, it is impossible to find a quantum state
where the relative position of the particles in space is arbitrarily
sharp at a given moment of time, even if the distance between the
particles is many orders of magnitude above the Planck scale.

Both features indicate that the quantized spacetime in which the
particles are living obtains a kind of foamy structure. Unfortunately,
it is not possible to derive a more explicit and intuitive
\emph{spacetime spectrum}, like the one for the single particle system
in \cite{matwel}. But in principle, we have a very similar situation,
and this is also expected to arise in a more realistic, or even in a
fully consistent theory of quantum gravity in higher dimensions.
Hence, although the Kepler system is only a very simple toy model,
some principle effects of quantum gravity can be seen. These are
really quantum gravity effects, because they disappear not only at the
classical level, hence in the limit $\hbar\to0$, whatever this
precisely means, but also in the well defined limit $G\to0$, where the
gravitational interaction is switched off.

Finally, a nice feature of this toy model is that, once the reduction
to a two particle system in three dimensions is accepted, everything
else can be derived exactly and without any further assumptions. It is
possible to keep the assumptions and simplifications clearly apart
from the mathematics and the physical conclusions. There are no hidden
points were additional, say, intuitive assumptions must made. The
\emph{only} assumption that enters the definition of the Kepler system
as a toy model is the Einstein Hilbert action with the appropriate
matter terms for the particles, which is assumed to define the
dynamics of the gravitational field. This is the concept on which the
derivation of a general multi particle phase space is based in
\cite{matmul}. The point where this assumption enters this article, is
the definition of the symplectic structure in \sref{class}.

\section{The free particle system}
\label{free}
Before we switch on the gravitational interaction, let us consider a
system of two uncoupled relativistic point particles $\prt_k$
($k=1,2$) in flat, three dimensional Minkowski space. As a vector
space, we identify this with the spinor representation $\algsl(2)$ of
the three dimensional Lorentz algebra. The twelve dimensional
\emph{kinematical} phase space is spanned by the positions
$\pos_k=\posv^a_k\gam_a$ and the momentum vectors
$\mom_k=\momv^a_k\gam_a$ of the particles, where $\gam_a$ ($a=0,1,2$)
is an orthonormal basis given by the usual gamma matrices
\eref{gamma}. Further conventions and some useful formulas regarding
the vector and matrix notation are given in the appendix.

We have the usual symplectic potential $\qpot$, from which we read off
the Poisson brackets, 
\begin{equation}
  \label{rpp-pot-pois}
  \qpot = \ft12 \sum_k \Trr{\mom_k \, \dd \pos_k } \follows
   \pois{\momv^a_k}{\posv^b_k} = \eta^{ab}.
\end{equation}
The Hamiltonian $\qham$ is a linear combination of the two mass shell
constraints $\con_k$, with Lagrange multipliers $\mul_k\in\RR$ as
coefficients,
\begin{equation}
  \label{rpp-ham}
   \qham = \sum_k \mul_k \, \con_k , \qquad
     \con_k = \ft14 \Trr{\mom_k\^2} + \ft12 m_k\^2.
\end{equation}
The \emph{physical} phase space is a subset of the kinematical phase
space, which is defined by the mass shell constraints and the positive
energy conditions,
\begin{equation}
  \label{rpp-mss}
  \ft12\Trr{\mom_k\^2} = - m_k\^2 , \qquad
  \momv^0_k = \ft12 \Trr{\mom_k \gam^0} > 0.
\end{equation}
The world lines are parameterized by a common, unphysical time
coordinate $t$, and the Hamiltonian generates the time evolution with
respect to this coordinate, 
\begin{equation}
  \label{rpp-evolve}
  \dot\mom_k = \pois{\qham}{\mom_k} = 0 , \qquad
  \dot\pos_k = \pois{\qham}{\pos_k} = \mul_k \, \mom_k.
\end{equation}
The freedom to choose the multipliers $\mul_k$ corresponds to the
gauge freedom to reparameterize the world lines. Finally, there are
some conserved charges which are of interest, namely the total
momentum vector and the total angular momentum vector,
\begin{equation}
  \label{rpp-tot-mom}
  \tmom = \sum_k \mom_k , \qquad
  \tang = \ft12 \sum_k \comm{\mom_k}{\pos_k} . 
\end{equation}
The associated rigid symmetries are the translations and Lorentz
rotations of the world lines with respect to the \emph{reference
frame}, which is defined by the coordinates of the embedding Minkowski
space.

\subsubsection*{The centre of mass frame}
So far, this is the standard Hamiltonian formulation of a
special-relativistic two particle system. Since we are only interested
in the relative motion of the particles, we shall now impose some
further restrictions on the phase space variables. Provided that the
total momentum vector is positive timelike, there always exists a
reference frame where
\begin{equation}
  \label{rpp-com}
  \tmom = M \, \gam_0, \qquad
  \tang = S \, \gam_0, \qquad M,S\in\RR.  
\end{equation}
We call this the \emph{centre of mass} frame. There is only one
special situation where a centre of mass frame does not exist. If both
particles are massless and if they move with the velocity of light
into the same direction, then the total momentum is lightlike. These
special states are excluded in the following.

\begin{figure}[t]
  \begin{center}
    \epsfbox{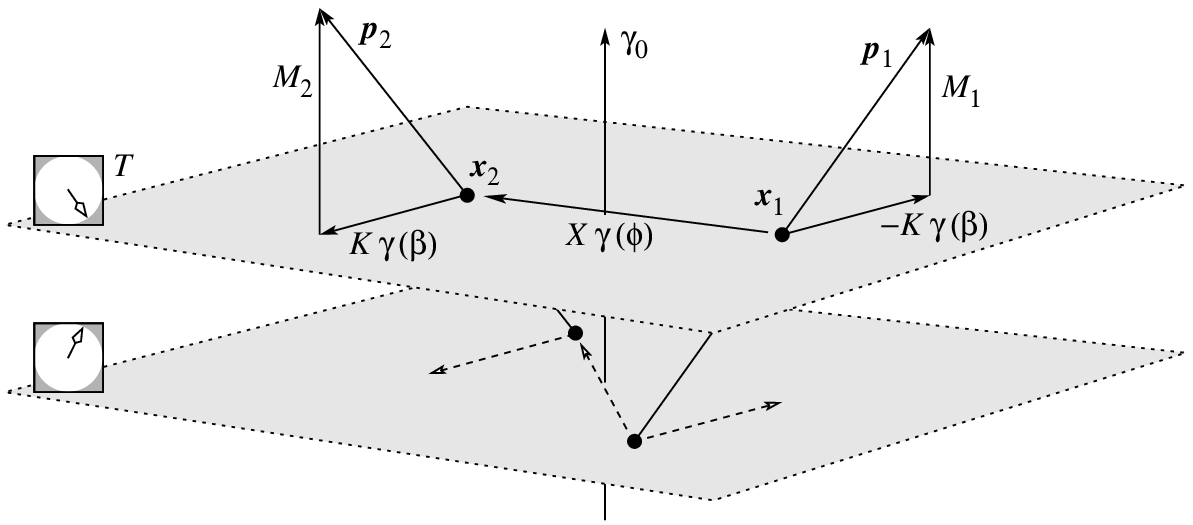}
    \xcaption{The free particle system in the centre of mass frame.
    The embedding Minkowski space is foliated by a family of equal
    time planes, labeled by an ADM time coordinate $t$. The relative
    position of the particles in space is defined by $X$ and $\xdir$,
    and the spatial momentum by $K$ and $\pdir$. The energies of the
    particles are $M_k$, and the clock $T$ represents the absolute
    time in the centre of mass frame, which can be regarded as a
    reference frame of some external observer.}
  \label{rpp}.
  \end{center}
\end{figure}

In the centre of mass frame, the rigid symmetries are reduced to a two
dimensional group of time translations and spatial rotations about the
$\gam_0$-axis. The associated charges are the total energy $M$ and the
spatial angular momentum $S$. It is also allowed to speak about an
absolute time, defined by the $\gam_0$-axis, and an absolute space
orthogonal to it, defined by the $\gam_{1,2}$-axes in Minkowski space.
Essentially, we have a non-relativistic system, although the dynamical
properties of the particles are still relativistic. We can use this to
impose a gauge condition, which restricts the way the world lines are
parameterized. It is reasonable to choose the parameterization such
that
\begin{equation}
  \label{rpp-eqt}
  \posv^0_1 = \posv^0_2.
\end{equation}
At each moment of time $t$, the particles are then located on the same
equal time plane, as indicated in \fref{rpp}. We can think of an ADM
like foliation of the embedding Minkowski space by equal time planes.
The planes are labeled by an ADM time coordinate $t$, and each plane
represents an instant of time in the centre of mass frame. For the
moment, we do not require the ADM time $t$ to be related in any
way to the absolute time in the centre of mass frame. Hence, there is
still one gauge degrees of freedom left, which is compactible with the
gauge condition \eref{rpp-eqt}. This is a simultaneous
reparameterization of both world lines.

From the phase space point of view, the various restrictions can be
regarded as additional constraints. All together, we have seven
constraints. The two mass shell constraints \eref{rpp-mss}, the gauge
condition \eref{rpp-eqt}, and four independent spatial components of
the definition \eref{rpp-com} of the center of mass frame. Note that
the time components of these equations are definitions of the phase
space functions $M$ and $S$. Since we have an odd number of
constraints, it is clear that at least one of them is a first class
constraint, which generates the simultaneous time evolution of the two
particles as a gauge symmetry. The other six are second class
constraints, as we are now going to show.

To easiest way to do this is to eliminate six of the seven
constraints, and to show that the reduced symplectic structure is
still non-degenerate. We do this by introducing new phase space
variables, which are similar to the usual non-relativistic centre of
mass coordinates and momenta of a two particle system. The general
solution to the equation $\tmom=M\gam_0$ can obviously be written as
\begin{equation}
  \label{rpp-mom-MVK}
  \mom_1 = M_1 \, \gam_0 - K \, \gam(\pdir), \qquad
  \mom_2 = M_2 \, \gam_0 + K \, \gam(\pdir).
\end{equation}
The new variables $M_k$ are the energies of the particles, and $K$ and
$\pdir$ are polar coordinates defining the spatial momentum and its
direction. The rotating unit vector $\gam(\pdir)$ is introduced in
\eref{gamma-rot}. It defines the angular direction $\pdir$ in
Minkowski space, thus in this case the direction of motion of the
particles in the centre of mass frame.

To express the mass shell constraints in terms of these variables, it
is convenient to replace the mass parameters $m_1$ and $m_2$ by a
total mass $\mu$ and a relative mass $\nu$,
\begin{equation}
  \label{rpp-mu-nu}
  \mu = m_2 + m_1 , \qquad \nu = m_2 - m_1 , \qquad
  0 \le \nu \le \mu. 
\end{equation}
Without loss of generality, we assume that $m_2\ge m_1$. The special
cases are $\nu=0$, where both particles have the same mass, and
$\nu=\mu$, where at least one particle is massless. The same
redefinition can be applied to the energy variables,
\begin{equation}
  \label{rpp-M-V}
  M = M_2 + M_1  , \qquad V = M_2 - M_1 , \qquad  |V| <  M ,
\end{equation}
where the inequality represents the positive energy condition for both
particles.  The mass shell constraints \eref{rpp-ham} are then given
by 
\begin{equation}
  \label{rpp-mss-K}
  \con_1 = \frac{K^2}{2} - \frac{(M-V)^2 - (\mu-\nu)^2}{8},
   \qquad 
  \con_2 = \frac{K^2}{2} - \frac{(M+V)^2 - (\mu+\nu)^2}{8}. 
\end{equation}
They can be simplified by taking the following linear combinations,
\begin{equation}
  \label{rpp-DE}
  \rcon = \con_2 - \con_1 = 
      \frac{\mu  \nu -  M  V}2 , \qquad
  \tcon  = \con_2 + \con_1 = 
        K^2 +\frac{M^2   + V^2 - \mu^2 - \nu^2}4. 
\end{equation}
Let us solve the constraint $\rcon=0$ for $V$, so that $V$ becomes a
simple function of $M$,
\begin{equation}
  \label{rpp-V-M}
  V = \frac{\mu\nu}{M}.
\end{equation}
What remains is a single mass shell constraint, which can be written
as 
\begin{equation}
  \label{rpp-EF}
  \tcon =   K^2 - F(M) \approx 0, \txt{where}
  F(M) = \frac{ (M^2 - \mu^2) \, (M^2 - \nu ^2) }{4 M^2}.
\end{equation}
We'll see later one that this is just a somewhat unusual way to write
the familiar relation between the momentum and the energy of two
relativistic point particles. The positive energy condition becomes a
non-trivial condition to be imposed on $M$, namely
\begin{equation}
  \label{rpp-erg-M}
  | V | < M \equivalent 
   \frac{\mu\nu} M < M \equivalent M > \sqrt{\mu\nu}.
\end{equation}
Note that this implies $M\neq0$, so that the various places where $M$
appears in the denominator are not problematic. All together, the
momentum vectors are parameterized by three independent variables $M$,
$K$, $\pdir$. They have an immediate physical interpretation as the
total energy, the spatial momentum, and the direction of motion of the
particles in the centre of mass frame.

The positions $\pos_k$ are subject to the constraints $\tang=S\gam_0$
and $\posv_1^0=\posv_2^0$. The general solution can be parameterized
by three other variables $T$, $X$, and $\xdir$, so that
\begin{equation}
  \label{rpp-T-dis}
  T = \posv^0_1 = \posv^0_2, \qquad
  \dis = \pos_2 - \pos_1 = X \, \gam(\xdir).
\end{equation}
The radial coordinate $X\ge0$ represents the relative position of the
particles in space, and $\xdir$ is the spatial orientation. It is also
useful to introduce the relative position vector $\dis$, which is
going to have a generalization for the interacting particles later on.
The \emph{clock} $T$ represents the absolute time in the centre of
mass frame. It is the $\gam_0$-coordinate of the equal time plane in
\fref{rpp}. So far, it is an arbitrary function of the ADM time $t$,
because we are still free to choose the parameterization of the world
lines, and thus the labeling of the equal time planes by the
unphysical coordinate $t$. 

Using all this, we can solve the equation $\tang=S\gam_0$, and express
the result in terms of the new phase space variables,
\begin{equation}
  \label{rpp-pos-XT}
  \pos_1 = T \, \gam_0 - \frac{M+V}{2M} \, X \, \gam(\xdir),
   \qquad 
  \pos_2 = T \, \gam_0 + \frac{M-V}{2M} \, X \, \gam(\xdir).
\end{equation}
The relativistic version of the centre of mass frame is thus actually
a centre of energy frame. At each moment of time, the centre of energy
is located on the $\gam_0$-axis in \fref{rpp}. But nevertheless, let
us stick to the notion centre of mass frame. The crucial point is that
the positions of the particles are, like the momenta, specified by
three independent variables. We have the clock $T$, the relative
position $X$, and the orientation $\xdir$.

To see that the six eliminated constraints were second class
constraints, we have to compute the reduced symplectic potential.
Inserting \eref{rpp-mom-MVK} and \eref{rpp-pos-XT} into
\eref{rpp-pot-pois} gives 
\begin{equation}
  \label{rpp-pot-K}
  \qpot =   K \, \cos(\pdir-\xdir) \, \dd X + 
           X \, K \, \sin(\pdir-\xdir)\, \dd \xdir - M \, \dd T .
\end{equation}
To simplify this, we replace the polar momentum coordinates $K$ and
$\pdir$ by Cartesian coordinates 
\begin{equation}
  \label{rpp-QS-K}
  Q = X \, K \, \cos(\pdir-\xdir), \qquad 
  S = X \, K \, \sin(\pdir-\xdir). 
\end{equation}
They define the \emph{radial} and \emph{angular} momentum. One can
easily verify that $S$ indeed satisfies $\tang=S\gam_0$, so it
coincides with the previous definition. Moreover, inserting this into
\eref{rpp-mom-MVK} gives
\begin{equation}
  \label{rpp-mom-MVQS}
  \mom_1 = \frac{M-V}{2} \, \gam_0 
           -\frac{ Q \, \gam(\xdir) + S \, \gam'(\xdir)}{X}, \qquad
  \mom_2 = \frac{M+V}{2} \, \gam_0 
           +\frac{ Q \, \gam(\xdir) + S \, \gam'(\xdir)}{X}.
\end{equation}
This tells us that $Q/X$ is the component of the momentum parallel to
the relative position, and $S/X$ is the component orthogonal to it.
This is the usual definition of a radial and angular momentum. If we
use $Q$ and $S$ as the basic phase space variables, then the
expression to be inserted into the mass shell constraint \eref{rpp-EF}
is
\begin{equation}
  \label{rpp-K-QS}
  K^2 = \frac{Q^2+S^2}{X^2}.
\end{equation}
And finally, the symplectic potential simplifies to 
\begin{equation}
  \label{rpp-pot-ext}
  \qpot =  X^{-1} Q \, \dd X + S \, \dd \xdir - M \, \dd T . 
\end{equation}
This defines a non-degenerate symplectic structure $\qsym=\dd\qpot$, and
we read off the following non-vanishing Poisson brackets,
\begin{equation}
  \label{rpp-pois-ext}
  \pois MT = -1, \qquad
  \pois QX = X, \qquad
  \pois S\xdir = 1.
\end{equation}
From this it is immediately obvious that $M$ is the charge associated
with time translations $T\mapsto T-\Delta T$, and $S$ is the charge
associated with spatial rotations $\xdir\mapsto\xdir+\Delta\xdir$. 

What remains from the Hamiltonian is a single mass shell constraint
$\tcon$, and a multiplier $\mul$, which is some not further
interesting linear combination of the original multipliers $\mul_k$,
\begin{equation}
  \label{rpp-ham-ext}
  \qham = \mul \, \tcon.
\end{equation}
It is still an unphysical Hamiltonian. Its value is zero for physical
states, and it generates the time evolution with respect to the
unphysical ADM time $t$, which is formally a gauge transformation.
Except for the reduced symmetry group, this looks very much like the
Hamiltonian description of a single relativistic point particle. We
have a six dimensional \emph{extended}, or kinematical phase space
\begin{equation}
  \label{rpp-qsp}
  \qsp = \big\{\quad ( M, Q, S ; T, X, \xdir ) \quad \big| \quad 
                X \ge 0 , \quad 
                \xdir \equiv \xdir + 2 \pi \quad \big\},
\end{equation}
and a single mass shell constraint $\tcon$, which defines the physical
subspace and an associated gauge symmetry. It is this feature that we
are going to exploit in the following, treating the two particle
system in the centre of mass frame as if it was a single particle
system. Of course, this is a well know way to solve a two particle
system in non-relativistic mechanics, and we'll see that it also works
for this almost trivial relativistic system.

\subsubsection*{Complete reduction}
We can also go over to an effectively non-relativistic formulation,
where no constraint and no gauge symmetry is left. We just have to
impose another gauge condition. The most natural one is to require the
ADM time $t$ to coincide with the absolute time $T$ in the centre of
mass frame. This provides another constraint, and together with the
mass shell constraint we get a new pair of second class constraints,
\begin{equation}
  \label{rpp-gauge-con}
  K^2 = F(M) , \qquad T = t.
\end{equation}
Let us assume that the function $F$ can be inverted. We can then
define a four dimensional \emph{reduced} phase space
\begin{equation}
  \label{rpp-psp}
  \psp = \big\{\quad ( Q, S ; X, \xdir ) \quad \big| \quad 
                X \ge 0 , \quad 
                \xdir \equiv \xdir + 2 \pi \quad \big\}.
\end{equation}
It is just the usual non-relativistic phase space of the relative
motion of two particles in a plane. To derive the symplectic structure
and the Hamiltonian on $\psp$, we have to take into account that the
constraints \eref{rpp-gauge-con} are explicitly time dependent. This
implies a mixing of the Hamiltonian and the symplectic structure when
we perform the reduction. We have to consider the extended symplectic
potential on $\qsp$,
\begin{equation}
  \label{rpp-epot-ext}
  \qpot - \qham \, \dd t = 
           X^{-1} Q \, \dd X + S \, \dd \xdir 
          - M \, \dd T - \mul \tcon \, \dd t .  
\end{equation}
To obtain the reduced structures on $\psp$, we insert the solutions to
the equations $K^2=F(M)$ and $T=t$, and write the result as a
combination of the reduced symplectic potential $\ppot$ and the
reduced Hamiltonian $\pham$,
\begin{equation}
  \label{rpp-epot-red}
  \ppot - \pham \, \dd t = 
           X^{-1} Q \, \dd X + S \, \dd \xdir 
          - F^{-1}(K^2) \, \dd t  . 
\end{equation}
Note that the last term in \eref{rpp-epot-ext} vanishes, because the
constraint $\tcon=0$ is now identically satisfied. What comes out is
\begin{eqnarray}
  \label{rpp-pot-red}
  \ppot = X^{-1} Q\, \dd X + S \, \dd \xdir
  \follows
  \pois QX = X, \qquad \pois S\xdir = 1.  
\end{eqnarray}
The Poisson brackets of the remaining variables are unchanged, and
they are the usual non-relativistic ones. 

So, we now have an \emph{unconstrained} Hamiltonian formulation of the
free particle system, where the Hamiltonian represents the physical
energy of the system, and generates the time evolution with respect to
the absolute time in the centre of mass frame. Explicitly, one finds
that
\begin{equation}
  \label{rpp-ham-red}
  \pham = F^{-1}(K^2) 
    = \xsqrt{K^2+m_1\^2} 
    + \xsqrt{K^2+m_2\^2}. 
\end{equation}
Not surprisingly, this is just the relativistic energy of two point
particles with the same spatial momentum $K$. The signs of the two
square roots are fixed by the positive energy condition. A minus sign
for one of the square roots would violate \eref{rpp-erg-M}, as one of
the particles would then be in an antiparticle state. To express
$\pham$ as a function of the reduced phase space variables, we have to
insert \eref{rpp-K-QS}.

It is also quite instructive to look at the range of $\pham$. The
right hand side of \eref{rpp-ham-red} is obviously minimal for $K=0$,
and it increases unboundedly with $K$. Thus, we have $\pham\ge\Mmin$,
where
\begin{equation}
  \label{rpp-Mmin}
      \Mmin = m_1 + m_2 . 
\end{equation}
For $K=0$ the particles are at rest with respect to each other, and
thus also with respect to the centre of mass. There is one exception,
however, where such a state cannot be realized. If one of the
particles is massless, then we have $\nu=\mu$, and the positive energy
condition \eref{rpp-erg-M} requires that $M>\Mmin$. In this case, the
states with $K=0$ are excluded, and we have the stronger condition
$\pham>\Mmin$. Clearly, this is because a massless particles cannot be
at rest, and consequently a state with vanishing momentum $K$ does not
exist.

Let us also consider the non-relativistic limit. For small momenta
$K$, the energy $\pham$ either starts off linearly or quadratically
with $K$, depending on whether a massless particle is present or not.
Let us consider the case where both particles are massive. If we then
expand the right hand side of \eref{rpp-ham-red} up to the second
order in $K$, we get
\begin{equation}
  \label{rpp-red-mss}
  \pham \approx \Mmin + \frac{1}{2m} \, K^2 , \txt{where} 
      m  = \frac{m_1\,m_2}{m_1+m_2}. 
\end{equation}
This is the usual non-relativistic relation between the spatial
momentum $K$ and the energy $\pham$ in the centre of mass frame. The
parameter $m$ is the reduced mass of the two particles system.  The
difference between the total energy $\pham$ and the total rest mass
$\Mmin$ is the non-relativistic kinetic energy. For large momenta $K$,
on the other hand, we find the usual relativistic behaviour
$\pham\approx2K$. The energy becomes a linear function of the spatial
momentum. The factor of two arises because we have two particles with
the same momentum. For some typical mass parameters, the relations
between $K$ and $\pham$ are shown as the broken lines in \fref{dsp}.

\subsubsection*{Trajectories}
There are now two alternative ways to describe the two particle
system. We can either use the \emph{constrained} Hamiltonian
formulation based on the \emph{extended} phase space $\qsp$. Or we may
use the \emph{unconstrained} formulation, based on the \emph{reduced}
phase space $\psp$. Consider first the constrained formulation. In
this case we have six independent phase space variables, and the
Hamiltonian is given by $\qham=\mul\tcon$, where $\tcon$ is the mass
shell constraint and $\mul$ is an arbitrarily chosen function of the
unphysical ADM time $t$. It is not difficult to derive the resulting
time evolution equations and to solve them.  

The energy $M$ and the angular momentum $S$ are of course preserved
charges. The same holds for the spatial momentum $K$, as it is only
this combination of the phase space variables $Q$, $S$, and $X$ that
enters the Hamiltonian. Thus,
\begin{equation}
  \label{rpp-t-MSK}
  \dot M = \pois\qham M = 0 , \qquad
  \dot S = \pois\qham S = 0 . \qquad
  \dot K = \pois\qham K = 0.
\end{equation}
For the relative position $X$ and the conjugate radial momentum $Q$,
we find
\begin{equation}
  \label{rpp-t-XQ}
  \dot X = \pois\qham X = 2 \, \mul \, \frac QX , \qquad
  \dot Q = \pois\qham Q = 2 \, \mul \, \frac{Q^2+S^2}{X^2} 
                      = 2 \, \mul \, K^2.  
\end{equation}
And finally, the brackets of $\qham$ with $\xdir$ and $T$ are given by
\begin{equation}
  \label{rpp-t-T}
  \dot \xdir = \pois\qham \xdir = 2 \, \mul \, \frac S{X^2}, \qquad
  \dot T = \pois\qham T = \mul \, F'(M) . 
\end{equation}
It is clear that the same equations of motion arise in the
unconstraint formulation, except that $M$ and $T$ are not independent
variables, $\qham$ is replaced by $\pham$, and instead of the
multiplier $\mul$ the function $1/F'(\pham)$ appears, which formally
implies that $\dot T=1$, which is consistent with the gauge condition
$T=t$. 

We can easily solve these differential equations step by step. Those
for $M$ and $S$ are trivial, stating that $M(t)=M_0$ and $S(t)=S_0$
for some constants $M_0$ and $S_0$. Moreover, $K(t)=K_0$ is also a
constant of motion. It is related to $M_0$ by the constraint
$K_0^2=F(M_0)$. Thus $K_0$ and $M_0$ are not independent, and they are
subject to the previously derived restrictions. For massive particles,
we have $M_0\ge\Mmin$ and $K_0\ge0$. If at least one massless particle
is present, then we have the stronger condition $M_0>\Mmin$ and
$K_0>0$.

Using this, it is easy to solve the equation of motion for $Q$. The
general solution is $Q(t)=Q_0+\epsilon(t)K_0^2$, where $Q_0$ is some
integration constant. The function $\epsilon(t)$ is determined up to
a constant by $\dot\epsilon(t)=2\mul(t)$. For $K_0>0$ we can obviously
choose this free constant so that $Q_0=0$. On the other hand,
$K_0=0$ implies, by definition \eref{rpp-K-QS}, that both $Q$ and $S$
must be zero. In this case we also have $Q_0=0$. Therefore, the
general solutions found so far are
\begin{equation}
  \label{rpp-MSKQ-t}
  M(t) = M_0, \qquad 
  S(t) = S_0, \qquad
  K(t) = K_0, \qquad
  Q(t) = \epsilon(t) \, K_0^2. 
\end{equation}
If we insert this into the time evolution equations for $X$ and
$\xdir$, then we can integrate them. The result is
\begin{equation}
  \label{rpp-X-t}
  X(t) = \xsqrt{ R_0^2 + \epsilon^2(t) \, K_0^2 }, \qquad
  \xdir(t) = \xdir_0 + 
             \arctan\left( \epsilon(t) \, \frac{K_0^2}{S_0} \right),
\end{equation}
where $R_0\ge0$ and $\xdir_0$ are two more integration constants. And
finally, we can also solve the equation of motion for $T$, which gives
\begin{equation}
  \label{rpp-T-t}
  T(t) = T_0 + \epsilon(t) \, \frac{  F'(M_0) }2 . 
\end{equation}
These are the most general solutions to the time evolution equations
on $\qsp$, provided by the Hamiltonian $\qham=\mul\tcon$. They are
parameterized by an arbitrary function $\epsilon(t)$, representing the
gauge freedom, and six integration constants $M_0$, $K_0$, $S_0$,
$T_0$, $R_0$, $\xdir_0$. Only four of them are independent.  The
constraint relates the momentum $K_0$ to the energy $M_0$, and
additionally there is a relation between $S_0$, $X_0$ and $K_0$, which
follows from the definition \eref{rpp-K-QS},
\begin{equation}
  \label{rpp-orb-con}
  K_0^2 = F(M_0) , \qquad
   K_0 \, R_0 = | S_0 |.
\end{equation}
There are two ways to look at these solutions. If we keep $t$ fixed
and vary $\epsilon(t)$, then we pass along a \emph{gauge orbit}
generated by the constraint $\tcon$. If we instead fix the function
$\epsilon(t)$ and vary $t$, then we see a \emph{trajectory} in the
phase space $\qsp$, which describes the time evolution generated by
the Hamiltonian $\qham$ in a particular gauge. 

The derivation of the trajectories on the reduced phase space $\psp$
yields the same result, with the same integration constants, however
with the gauge function replaced by 
\begin{eqnarray}
  \label{rpp-eps-gauge}
  \epsilon(t) = 2 \, \frac{t-T_0}{F'(M_0)},
\end{eqnarray}
which obviously implies $T(t)=t$. The inverse of $F'(M_0)$ is always
well defined because, within the range of $M_0$ allowed by the
positive energy condition, the function $F(M)$ is monotonically
increasing with $M$. So, the trajectories on $\psp$ are parameterized
by the same four independent integration constants, but we no longer
have any gauge freedom.

Let us now look a little bit closer at the various trajectories.
Consider first the case where $S_0\neq0$. The second relation in
\eref{rpp-orb-con} then implies that both $K_0>0$ and $R_0>0$, and
\eref{rpp-X-t} describes a straight line in polar coordinates. The
particles approach each other from a direction $\xinn$. At the time
$T_0$ they reach a minimal distance $R_0$ with spatial orientation
$\xdir_0$. And finally they separate again, moving into a spatial
direction $\xout$. The directions $\xinn$ and $\xout$ are always
antipodal, and both are orthogonal to $\xdir_0$,
\begin{equation}
  \label{rpp-in-out}
  \xinn = \xdir_0 - \frac{\pi}2 \, \sgn S_0 , \quad
  \xout = \xdir_0 + \frac{\pi}2 \, \sgn S_0  \follows
  \xout = \xinn + \pi \, \sgn S_0.  
\end{equation}
The sign of the angular momentum $S$ tells us on which side the
particles pass each other. This is a trivial scattering process,
because the directions $\xinn$ and $\xout$ always differ by $\pi$.
Nevertheless, the scattering picture is quite useful. There will by a
real scattering when the gravitational interaction is switched on. 

The constant of motion $K=K_0$ has a useful interpretation in this
picture. It represents the momentum which is canonically conjugate to
the distance between the particles, when this distance is very large.
More precisely, consider the pair of canonically conjugate variables
\begin{equation}
  \label{rpp-PR-QX}
   P = Q / X  , \qquad R =  X  \follows \pois PR = 1,
\end{equation}
replacing $Q$ and $X$. Thus, $R$ also represents the distance between
the particles in space, but $P$ is now the momentum which is
canonically conjugate to it, in contrast to $Q$ which is not exactly
conjugate. For the trajectories derived above, we have
\begin{equation}
  \label{rpp-RP-t}
  R(t) = \xsqrt{ R_0^2 + \epsilon^2(t) \, K_0^2 }, \qquad
  P(t) = \frac{\epsilon(t) \, K_0^2} 
              {\xsqrt{ R_0^2 + \epsilon^2(t) \, K_0^2 }} .
\end{equation}
In the limit $t\to\pm\infty$, or equivalently
$\epsilon(t)\to\pm\infty$, we have $R(t)\to\infty$ and $P(t)\to\pm
K_0$. The constant of motion $K=K_0$ is equal to the momentum $P$ when
the particles are far apart. This will be useful to know when the
system is quantized later on.  The eigenvalue of $K$ is then the
inverse radial wavelength of the wave function at spatial infinity. At
the classical level, $K_0$ is the momentum of the incoming and
outgoing particles in the scattering process.

A special situation arises when $S_0=0$. Then, according to
\eref{rpp-orb-con}, we have either $K_0=0$ or $R_0=0$. The first case,
$S_0=0$, $K_0=0$, and $R_0>0$, is the trivial one. In \eref{rpp-X-t},
we may replace $K_0^2/S_0$ by $S_0/R_0^2$. Then we get the simple
trajectory $X(t)=R_0$ and $\xdir(t)=\xdir_0$. The particles are at
rest, at a spatial distance $R_0$ and with angular orientation
$\xdir_0$. The integration constant $T_0$ is in this case redundant,
and the energy $M_0=\Mmin$ is equal to the sum of the rest masses. Of
course, such a trajectory only exists when both particles are massive,
as otherwise we must have $K_0>0$, because massless particles cannot
be at rest.

The more interesting case is $S_0=0$, $R_0=0$, and $K_0>0$. The
argument of the $\arctan$ is then ill defined. But we can still
consider the trajectory as a limit. In the limit $S_0\to0$ and
$R_0\to0$, with $K_0$ and all other integration constants fixed, we
get
\begin{equation}
  \label{rpp-X-t-coll}
  X(t) = |\epsilon(t)| \, K_0 , \qquad
  \xdir(t) = \xdir_0 + 
             \frac{\pi}{2} \, \sgn S_0 \, \sgn \epsilon(t) .
\end{equation}
The crucial point is now that the limit still depends on $\sgn S_0$,
thus on the direction of the limit. But if we change the definition of
the integration constants slightly, we may also write 
\begin{equation}
  \label{rpp-touch}
  \xdir(t) = \cases{ \xinn & for $T<T_0$, \cr
                     \xout & for $T>T_0$,} \qquad 
                    \xout = \xinn \pm\pi, 
\end{equation}
where the sign no longer matters. Clearly, this is just the definition
of a straight line through the origin of the polar coordinate system.
The particles approach each other from a direction $\xinn$, touch each
other at $t=T_0$ without interaction, and separate again into the
opposite direction $\xout$. The case $S_0=0$ is only special because
of a coordinate singularity of the phase space variables. We could
avoid this by introducing a globally well defined chart on $\psp$ and
$\qsp$, replacing the polar coordinates $X$ and $\xdir$ by Cartesian
coordinates, and the radial and angular momenta $Q$ and $S$ by the
appropriate conjugate variables. 

\subsubsection*{Schr\"odinger quantization}
Let us now, finally, perform the quantization of the free particle
system. The most straightforward way is to start from the reduced
classical phase space $\psp=\{(Q,S;X,\xdir)\}$ with symplectic
potential \eref{rpp-pot-red},
\begin{eqnarray}
  \label{rpp-pot-sch}
  \ppot = X^{-1} Q\, \dd X + S \, \dd \xdir.
\end{eqnarray}
This is an unconstrained formulation. We can apply the standard
Schr\"odinger quantization procedure. We use a position representation,
where the wave function is given by $\psi(x,\ph)$, with $x$ and $\ph$
being the eigenvalues of the operators $X$ and $\xdir$,
\begin{equation}
  \label{rpp-op-X-xdir}
  \op X \, \psi(x,\ph) = x \, \psi(x,\ph), \qquad
  \op \xdir \, \psi(x,\ph) = \ph \, \psi(x,\ph), 
\end{equation}
with $x\ge0$ and $\ph\equiv\ph+2\pi$. Actually, the second equation is
of course only well defined with $\xdir$ replaced by a periodic
function of $\xdir$, as only those functions are well defined on the
classical phase space. Concerning the coordinate singularity at $x=0$,
we shall follow the concept in \cite{bousor}. It can later also be
applied to the interacting system without any essential modification.
We impose a periodicity condition on the wave function,
\begin{equation}
  \label{rpp-stat}
  \psi(x,\ph+2\pi) = \expo{2\pi\ii\stat} \, \psi(x,\ph),
\end{equation}
and we require the wave function to be finite at $x=0$. The parameter
$\stat$ is some fixed real number. It represents a quantization
ambiguity that typically arises when the classical phase space is not
simply connected \cite{woodhs}. A special case occurs when the two
particles have the same mass. They can then be regarded as identical,
and replacing $\ph$ with $\ph+\pi$ already takes us back to the same
state.  In this case, we have the stronger relation
\begin{equation}
  \label{rpp-stat-id}
  \psi(x,\ph+\pi)  = \expo{\pi\ii\stat} \, \psi(x,\ph),
\end{equation}
and the parameter $\stat$ defines the statistics of the particles. If
$\stat$ is an even integer, then the particles are called
\emph{bosons}, and for odd integers they are called \emph{fermions}.
Note that this only refers to the statistics, not to the internal
structure of the particles. There is no spin statistics theorem for
this simple toy model.

For non-integer values of $\stat$, the particles are usually called
\emph{anyons} \cite{anyons}. Anyons can only arise in three spacetime
dimensions. In higher dimensions, the fundamental group of the phase
space of two identical particles is $\ZZ_2$ rather than $\ZZ$, which
implies that $\stat$ has to be an integer, which is even for bosons
and odd for fermions. Another special feature in three spacetime
dimensions is that the non-trivial phase factor in \eref{rpp-stat}
also shows up when the particles are not identical. Slightly abusing
the language, we shall then also refer to $\stat$ as the statistics
parameter. We should also note that for identical particles the real
number $\stat$ is defined modulo two, otherwise modulo one.

For $\psi(x,\ph)$ to represent the usual probability amplitude in
polar coordinates, we define the scalar product to be 
\begin{equation}
  \label{rpp-prod-sch}
  \braket{\psi_1}{\psi_2} = \int  x \, \dd x \, \dd \ph
               \, \bar\psi_1(x,\ph) 
               \, \psi_2(x,\ph). 
\end{equation}
The momentum operators are defined so that the commutators are
$-\ii\hbar$ times the Poisson brackets,
\begin{equation}
  \label{rpp-comm}
  \comm {\op Q}{\op X} = -\ii\hbar \, \op X, \qquad
  \comm {\op S}{\op \phi} = -\ii\hbar .
\end{equation}
They become self-adjoint operators if we set
\begin{equation}
  \label{rpp-sch-QS}
  \op Q  \, \psi(x,\ph)
           = - \ii\hbar \, \deldel x \, x  \,  \psi(x,\ph), 
  \qquad
  \op S  \, \psi(x,\ph) 
           = - \ii\hbar \, \deldel \ph  \,  \psi(x,\ph).
\end{equation}
Note the extra factor of $x$ in the operator representation of $Q$. It
has to appear in the given ordering, since otherwise the resulting
operator is not self-adjoint with respect to the inner product
\eref{rpp-prod-sch}. We should also note that the operators for $Q$
and $S$ are uniquely defined by the commutation relation, up to
redefinitions that can be compensated either by a phase transformation
of the wave function, or by redefining the periodicity condition
\eref{rpp-stat}. In other words, the statistics parameter $\stat$ is
the only relevant quantization ambiguity.

%%JL revision block begins
To derive the energy eigenstates let us first consider the
operator~$K^2$, of which the Hamiltonian $\pham=F^{-1}(K^2)$ is a
function. We choose the Hermitian ordering
\begin{equation}
  \label{rpp-op-K}
  \op K^2 = \op X^{-1} (\op Q^2 + \op S^2 ) \op X^{-1}. 
\end{equation}
In the representation~\eref{rpp-sch-QS}, $K^2$ is then proportional to
the usual Laplacian in the polar coordinates,
\begin{equation}
  \label{rpp-K-act}
  \op K^2 \, \psi(x,\p) = - \hbar^2 \lap \, \psi(x,\ph), \txt{where}
  \lap =  x^{-1} \deldel x \, x \, \deldel x 
              + x^{-2} \deldel \ph \, \deldel \ph.
\end{equation}
To make $K^2$ self-adjoint, a boundary condition at $x=0$ is needed.
Following \cite{bousor}, we choose this condition so that the wave
function remains finite at $x=0$. The normalized eigenfunctions are
then parametrised by two quantum numbers, $k$~and~$s$, and given by
the Bessel functions of the first kind,
\begin{equation}
  \label{rpp-chi-sch}
  \chi(k,s;x,\ph) = \frac1{\sqrt{2\pi}} \, 
                     \expo{\ii s\ph} \, J_{|s|}(kx).  
\end{equation}
They are normalized so that 
\begin{equation}
  \label{rpp-chi-norm-sch}
  \int x \, \dd x \, \dd\ph \, 
    \bar\chi(k_1,s_1;x,\ph) \, \chi(k_2,s_2;x,\ph) = 
          k^{-1} \, \delta(k_2-k_1) \, \delta_{s_2-s_1}. 
\end{equation}
%%JL revision block ends 
As expected, the quantum number $k$ is the inverse radial wavelength
of the wave function at infinity. We already expected the eigenvalues
of $K$ to have this property, because for large distances of the
particles the momentum $K$ is equal to the canonical radial momentum
$P=Q/X$. The quantum number $s$ is the inverse angular wavelength, and
of course it represents the eigenvalue of the angular momentum. The
possible values of $s$ can be inferred from the periodicity condition
\eref{rpp-stat}. It implies
\begin{equation}
  \label{rpp-S-eigen}
  \expo{\ii s(\ph + 2\pi )} = \expo{2\pi\ii\stat} \, \expo{\ii s\ph} 
   \follows s \in \stat + \ZZ.
\end{equation}
For identical particles, the stronger condition \eref{rpp-stat-id}
implies
\begin{equation}
  \label{rpp-S-eigen-ident}
  \expo{\ii s(\ph + \pi )} = \expo{\pi\ii\stat} \, \expo{\ii s\ph} 
   \follows s \in \stat + 2 \ZZ. 
\end{equation}
It follows that the eigenvalues of $K$ and $S$ are 
\begin{equation}
  \label{rpp-eigen-KS}
       K = \hbar k , \qquad   S = \hbar s .
\end{equation}
The spectrum of $K$ is continuous and positive, and $S$ is quantized
in steps of $\hbar$, or $2\hbar$ for identical particles. The actual
values of $S$ are determined by $\stat$. In the special case of two
identical bosons, for example, the total angular momentum is an even
multiple of $\hbar$, and for two identical fermions it is an odd
multiple of $\hbar$. This is what we usually find for bosons and
fermions in higher dimensions as well. The peculiar feature of anyons
is that the spectrum of the angular momentum is shifted by a
non-integer multiple $\stat\hbar$.

%%JL revision block begins 
The completeness relation of the eigenfunctions reads 
\begin{equation}
  \label{rpp-chi-ortho}
  \sum_s \int k \, \dd k \,
    \bar\chi(k,s;x_1,\ph_1) \, \chi(k,s;x_2,\ph_2) = 
   x^{-1} \, \delta(x_2-x_1) \, \delta_\stat(\ph_2-\ph_1), 
\end{equation}
where $s$ takes the values given above, and $\delta_\stat$ is a
periodic delta function. It fulfills the same periodicity condition as
the wave functions, hence \eref{rpp-stat} or \eref{rpp-stat-id},
depending on whether the particles are identical or not. 
%%JL revision block ends 

The momentum eigenstates $\chi(k,s;x,\ph)$ are also the energy
eigenstates. The Hamiltonian is a function $\pham=F^{-1}(K^2)$ of $K$,
so that the energy becomes a function of the quantum number $k$. It is
useful to introduce a \emph{dispersion relation}, which represents the
classical relation between the momentum $K$ and the energy $\pham$ as
a quantum relation between the inverse radial wavelength $k$ at
infinity and the frequency $\mm$ of the wave function,
\begin{equation}
  \label{rpp-disp}   
   \mm(k) = \hbar^{-1} \,  F^{-1}(\hbar^2 k^2) . 
   \follows 
    \op \pham \, \chi(k,s;x,\ph) = \hbar \mm(k) \, \chi(k,s;x,\ph) . 
\end{equation}
It follows that the spectrum of $\pham$ coincides with its classical
range $\Mmin\le\pham<\infty$. And finally, we can write down the
general solution to the time dependent Schr\"odinger equation, which is
a superposition of energy eigenstates with appropriate frequencies,
\begin{equation}
  \label{rpp-sch-wave}
   \psi(t;x,\ph) = \sum_s \int k \, \dd k \, \expo{-\ii\mm(k)t} 
       \, \psi(k,s) \, \chi(k,s;x,\ph). 
\end{equation}
The time-independent function $\psi(k,s)$ represents the probability
amplitude in momentum space. Clearly, this is a general superposition
of plane waves, written down in polar coordinates using the Bessel
functions. The Schr\"odinger quantization of the free particles system
is completely straightforward and gives the expected result. 

\subsubsection*{Dirac quantization}
There is not much more to be said about this very simple dynamical
system. However, in case of the Kepler system it turns out that a
simple Schr\"odinger quantization like this is not possible.  The
reasons are not immediately obvious at this point, so let us not
discuss them here. Instead, let us look for an alternative method that
leads to the same result. A possible alternative quantization is the
Dirac procedure, based on the six dimensional extended phase space
$\qsp=\{(M,Q,S;T,X,\xdir)\}$, with symplectic potential
\eref{rpp-pot-ext}, thus
\begin{equation}
  \label{rpp-qpot-3}
  \qpot =  X^{-1} Q \, \dd X + S \, \dd\xdir - M \, \dd T . 
\end{equation}
To quantize this phase space, we choose a wave function
$\psi(\tau,x,\ph)$, which additionally depends on the eigenvalue
$\tau$ of the clock $T$. In addition to the operators
\eref{rpp-op-X-xdir} and \eref{rpp-sch-QS} we have
\begin{equation}
  \label{rpp-op-dir}
  \op T \, \psi(\tau,x,\ph) = \tau \, \psi(\tau,x,\ph) ,\qquad
  \op M \, \psi(\tau,x,\ph) 
           = \ii\hbar \deldel \tau  \, \psi(\tau,x,\ph).
\end{equation}
All the basic operators are then self adjoint with respect to the
scalar product
\begin{equation}
  \label{rpp-prod-dir}
  \braket{\psi_1}{\psi_2} = \int  \dd\tau \, x \, \dd x \, \dd \ph
               \, \bar\psi_1(\tau,x,\ph) \, \psi_2(\tau,x,\ph). 
\end{equation}
On this extended Hilbert space, we have to impose the constraint
$\tcon=K^2-F(M)$, and the positive energy condition $M>\sqrt{\mu\nu}$.
The constraint is given as a function of $K$ and $M$, so it is useful
to diagonalize these operators first. Clearly, the eigenstates are
again given by the same Bessel function,
\begin{equation}
  \label{rpp-chi-dir}
  \chi(\mm,k,s;\tau,x,\ph) = \frac1{\sqrt{2\pi}} \, 
        \expo{-\ii\mm\tau} \, \expo{\ii s\ph} \, J_{|s|}(kx).  
\end{equation}
The normalization now reads 
\begin{eqnarray}
  \label{rpp-chi-norm-dir}
  \int \dd\tau \, x \, \dd x \, \dd\ph \, 
    \bar\chi(\mm_1,k_1,s_1;\tau,x,\ph) \, \chi(\mm_2,k_2,s_2;\tau,x,\ph) 
        = \qquad & &  \nwl 
   2\pi \, \delta(\mm_2-\mm_1) \, 
   k^{-1} \, \delta(k_2-k_1) \, \delta_{s_2-s_1}.&& 
\end{eqnarray}
The spectrum of $K$ and $S$ is the same as before, and that of $M$ is
real and continuous. The eigenvalues are 
\begin{equation}
  \label{rpp-eigen-MKS}
  M   = \hbar \mm  , \qquad
  K   = \hbar k  , \qquad
  S   = \hbar s . 
\end{equation}
It is then very easy to solve the constraint. The states which are
annihilated by the operator $\tcon=K^2-F(M)$ and satisfy the positive
energy condition $M>\sqrt{\mu\nu}$ are those where the quantum numbers
are related by the dispersion relation $\mm=\mm(k)$. A general
physical state is given by
\begin{equation}
  \label{rpp-dir-wave}
   \psi(\tau,x,\ph) = \sum_s \int k \, \dd k  \, 
             \psi(k,s) \, \chi(\mm(k),k,s;\tau,x,\ph), 
\end{equation}
where $\psi(k,s)$ is again the wave function in momentum space. It is
useful to split off a radial wave function and write the result as
\begin{equation}
  \label{rpp-wave-rad}
   \psi(\tau,x,\ph) = \sum_s \int k \, \dd k  \, 
          \expo{-\ii\mm(k)\tau} \, \expo{\ii s\ph} \, 
             \psi(k,s) \, \zeta(k,s;x), 
\end{equation}
where 
\begin{equation}
  \label{rpp-zeta}
  \zeta(k,s;x) =  \frac1{\sqrt{2\pi}} \, J_{|s|}(kx).
\end{equation}
We shall later compare this radial wave function to that of the Kepler
system, and from this we will be able to read off the basic effects of
quantum gravity in this toy model. 

Formally, the physical wave functions \eref{rpp-dir-wave} in the Dirac
approach are exactly the same as \eref{rpp-sch-wave} in the
Schr\"odinger approach. There is only the following conceptual
difference. The wave function in the Dirac approach is not a solution
to the time dependent Schr\"odinger equation, but a solution to a
generalized Klein Gordon equation, thus a constraint equation. In the
Schr\"odinger formulation, the wave function depends by definition on
the physical time $t$, whereas in the Dirac formulation, the time
dependence is encoded implicitly in the dependence of the wave
function on the eigenvalue $\tau$ of the clock, which is one of the
classical phase space variables.

In the Dirac approach, there is no time evolution of the state with
respect to the ADM time $t$, because this is an unphysical coordinate.
The Hamiltonian annihilates all physical states, and the Schr\"odinger
equation is void. To say something about the physical time evolution
in the Dirac quantization, we have to refer to an operator that
represents a clock. The free particle system is in this context a nice
toy model to explain how the problem of time arises in quantum
gravity, and how it can be solved by considering operators
representing physical clocks \cite{isham}.

\section{The classical Kepler system}
\label{class}
Now we are going to switch on the gravitational interaction. As
already mentioned in the introduction, we shall first consider an
alternative way to construct the Kepler spacetime, which includes the
definition of a centre of mass frame, based on the asymptotic
structure of the spacetime far away from the particles. We shall
thereby focus on the geometrical aspects, and skip all technical
details and proofs. They can be found in \cite{loumat}. We shall then
introduce a set of phase space variables. They provide position and
momentum coordinates, similar to those of the free particles. But at
the same time they also specify the geometry of space at a moment of
time, thus providing the of ADM variables of general relativity.

On the phase space spanned by these variables, we can then introduce a
Hamiltonian and a symplectic potential. The former can be inferred
from some straightforward geometric considerations. The latter can
only be derived from the full theory of Einstein gravity as a field
theory. As this is not part of this article, we can here only refer to
the general derivation of the phase space structures for a multi
particle model in \cite{matmul}. On the given phase space, we can then
derive and solve the equations of motion, and finally we shall briefly
discuss the various trajectories and compare them to the free particle
trajectories.

\subsubsection*{The static cone}
To explain the definition of the centre of mass frame, and to show how
the phase space variables of the Kepler system specify the geometry of
space, it is useful first to consider a simplified example. Let us
assume that the particles are at rest with respect to each other, so
that the spacetime is static. It is then possible to introduce a
globally defined absolute time coordinate $\cT$, and the spacetime can
be foliated by the surfaces of constant $\cT=T$. In the following, a
letter with a tilde always denotes a coordinate on the spacetime,
whereas all other symbols are configuration or phase space variables,
or functions thereof. 

In the static spacetime, the surface $\cT=T$ represents an instant of
time, like the equal time planes for the free particle system in
\fref{rpp}. All slices have the same geometry, and the spacetime is
just the direct product of a fixed space with a real line. The
Einstein equations require this space to be locally flat, and there
must be two conical singularities, representing the particles. In
other words, the space is a conical surface with two tips. We denote
them by $\prt_k$, with $k=1,2$. The deficit angles at the tips are
equal to $8\pi Gm_k$. What we would like to find is a suitable set of
configuration variables, describing the geometry of such a surface.
And we are also looking for an appropriate definition of a centre of
mass frame. 
\begin{figure}[t]
  \begin{center}
    \epsfbox{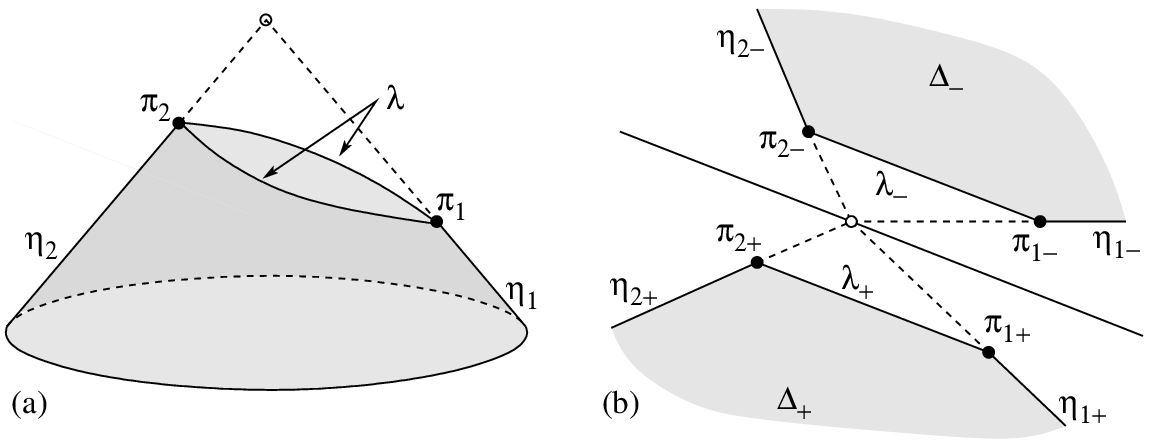}
    \xcaption{A double cone with two tips can be constructed by cutting
    off the tip from a single cone. The cut lines $\cut$, which are
    afterwards identified, are two geodesics connecting the points
    $\prt_1$ and $\prt_2$. If the spacetime is static, then it is the
    direct product of this space with a real line. If the conical
    surface is cut not only along the geodesic $\cut$, but also along
    two radial lines $\ecut_1$ and $\ecut_2$, then the two half spaces
    $\pol_\pm$ can be embedded symmetrically into a Euclidean plane.}
  \end{center}
  \label{tip}
\end{figure}

For this purpose, we embed the conical surface in a special way into a
Euclidean plane, as shown in \fref{tip}. The first step is to
introduce a geodesic $\cut$, which connects the two tips $\prt_1$ and
$\prt_2$ on the conical surface. There is always a unique such
geodesic, because apart from the two tips there is no curvature. When
the conical space is cut along the line $\cut$, it becomes a surface
with a hole in the middle. The points $\prt_k$ are located on a
circular boundary. As shown in \fref{tip}(a), the resulting surface is
a subset of an ordinary cone, with a certain region around the tip
taken away. The total deficit angle of the big cone is the sum of the
deficit angles at the tips $\prt_k$, thus $8\pi G(m_1+m_2)$.

Let us introduce on the big cone a polar coordinate system
$(\cR,\cdir)$, so that $\cR$ defines the radial distance from the
tip, and $\cdir$ is an angular coordinate with a period of $2\pi$.
Including the absolute time $\cT$ as a third coordinate, the spacetime
metric can be written as
\begin{equation}
  \label{static-cone}
  \dd s^2 = -\dd\cT^2 + \dd\cR^2 + 
        (1-4G(m_1+m_2))^2 \, \cR^2 \, \dd\cdir^2. 
\end{equation}
This defines a static cone with a total deficit angle of $8\pi
G(m_1+m_2)$. The conical surface in \fref{tip}(a) becomes a surface of
constant $\cT=T$ in this spacetime. Roughly speaking, we can say that
the static Kepler spacetime is a cone with a tip cut off, and with the
cut lines identified.

From the physical point of view, the situation is as follows. Far away
from the particles, say at spatial infinity, the spacetime looks like
the gravitational field of a single particle with mass $m_1+m_2$. It
is reasonable to identify the rest frame of this \emph{fictitious}
particle with the centre of mass frame of the universe. The centre of
mass frame is thus defined by a certain property of the spacetime
metric at spatial infinity, which does not directly refer to the
particles. It is a kind of asymptotical flatness, which is explicitly
defined in terms of the \emph{conical coordinates} $(\cT,\cR,\cdir)$.

The fictitious tip of the big cone in \fref{tip}(a) is the apparent
location of the centre of mass, as seen by an observer at infinity. In
analogy to the free particle system, let us identify the centre of
mass frame with the reference frame of such an external observer. With
respect to this reference frame, we define the \emph{absolute}
positions of the particles in spacetime as the conical coordinates
$(T_k,R_k,\xdir_k)$ of the points $\prt_k$ on the big cone. Again in
analogy to the free particles, these absolute coordinates are not
independent. They are specified by three independent \emph{relative}
coordinates.

At each moment of time, both points $\prt_k$ are located on the same
surface of constant $\cT=T$. Moreover, for the cut line $\cut$ to be
of the same length on both sides of the big cone, the points $\prt_k$
must be located on two opposite, or antipodal radial lines. Hence, we
must have
\begin{equation}
  \label{static-T-xdir}
  T_1 = T , \quad T_2 = T, \qquad 
  \xdir_1 = \xdir \pm\pi , \qquad \xdir_2 = \xdir.
\end{equation}
The variables $T$ and $\xdir$ are straightforward generalizations of
the corresponding free particle variables. The former defines a clock,
the latter represents the angular orientation of the particles with
respect to the reference frame. As a third independent coordinate, we
introduce the distance $R$ between the particles. It is the length of
the geodesic $\cut$. It is a simple exercise in conical geometry to
compute this. The result can be expressed as a function of the
conical radial coordinates $R_k$ and the total deficit angle,
\begin{equation}
  \label{static-len}
  R^2 = R_1^2 + R_2^2 + 
        2 \, R_1 \, R_2 \, \cos(4\pi G(m_1+m_2)).
\end{equation}
Finally, there is one more consistency condition. The deficit angles
at the points $\prt_k$ must be equal to $8\pi Gm_k$. So far we have
only fixed the total deficit angle of the big cone, which is the sum
of the two individual deficit angles. We still have to fix the way
this deficit angle is distributed over the two tips. This yields
another relation between the radial coordinates $R_k$. It is again a
simple exercise in conical geometry to show that
\begin{equation}
  \label{static-cR}
  R_1 = \xfrac{\sin(4\pi Gm_2)}{\sin(4\pi G(m_1+m_2))} \, R  \qquad
  R_2 = \xfrac{\sin(4\pi Gm_1)}{\sin(4\pi G(m_1+m_2))} \, R.
\end{equation}
So, what is the conclusion? There are obvious three independent
variables $T$, $R$, and $\xdir$. They define the geometry of the space
manifold at a moment of time, and the way this space manifold is
embedded into the spacetime manifold. In this sense, they provide some
kind of discretized ADM variables of general relativity. On the other
hand, from the particle point of view, the same variables define the
relative position of the particles with respect to each other, as
well as their absolute positions with respect to the centre of mass
frame. The definition of the relative coordinates is very similar to
\eref{rpp-T-dis} for the free particles.

We cannot say anything about the momentum variables at this point,
because we are only considering the static states. We still have to
generalize these concepts for moving particle. But let us stick to the
static case for a moment, and let us introduce a slightly different
representation of the same conical geometry of space. Let us not only
cut the space manifold along the geodesic $\cut$, but also along two
antipodal radial lines $\ecut_k$, extending from the points $\prt_k$
to infinity. Hence, $\ecut_k$ is a line of constant $\cT=T_k$ and
$\cdir=\xdir_k$ in the big cone, which lies inside the surface of
constant conical time $\cT=T$. The conical surface is then divided
symmetrically into two \emph{half spaces}, which we denote by $\pol_+$
and $\pol_-$.

Each half space $\pol_\pm$ is bounded by three edges, denoted by
$\ecut_{1\pm}$, $\cut_\pm$, and $\ecut_{2\pm}$, and it has two
corners, called $\prt_{1\pm}$ and $\prt_{2\pm}$. The two half spaces
are flat and simply connected. They can be embedded into a Euclidean
plane in the following unique way, which is shown in \fref{tip}(b).
The apparent position of the fictitious centre of mass, hence the tip
of the big cone, is mapped onto the origin of the plane. Moreover, the
edges $\cut_\pm$ are mapped onto two parallel straight lines, with
angular direction $\xdir$. This fixes the embedding of the half spaces
into the plane completely.

To describe the embedding more explicitly, it is convenient to think
of the Euclidean plane as an equal time plane in Minkowski space. We
choose it to be the plane with $\gam_0$-coordinate $T$. The geometry
of the two half spaces can then be specified as follows. The edges
$\cut_\pm$ are represented by two spacelike Minkowski vectors
$\dis_\pm$. They are given by
\begin{equation}
  \label{static-dis}
  \dis_\pm = R \, \gam(\xdir).
\end{equation}
The end points of the edges $\cut_\pm$ are the corners $\prt_{k\pm}$.
They are located at the following points in Minkowski space,
\begin{equation}
  \label{static-pos}
  \pos_{1\pm} = T \, \gam_0  
                - R_1 \, \gam(\xdir \mp 4\pi Gm_1) , \qquad
  \pos_{2\pm} = T \, \gam_0 
                + R_2 \, \gam(\xdir \pm 4\pi Gm_2),    
\end{equation}
where the coordinates $R_k$ given by \eref{static-cR}. One can easily
verify that this implies $\dis_\pm=\pos_{2\pm}-\pos_{1\pm}$. And
finally, the edges $\ecut_{k\pm}$ are radial lines, extending from the
corners $\prt_{k\pm}$ to infinity. A radial line in Minkowski space is
a geodesic which is orthogonal to the $\gam_0$-axis, and intersects
with it if it is extended beyond its end point.

The advantage of \fref{tip}(b) is that it uses less dimensions to
visualize the conical geometry of space. In contrast to \fref{tip}(a),
we do not need an auxiliary third Euclidean dimension to embed the big
cone. This will be useful when we now consider the moving particles,
because then the space is no longer flat. We can also read off the
deficit angles of the particles directly from \fref{tip}(b). The
deficit angle of the particle $\prt_1$ is the angle between the edges
$\ecut_{1+}$ and $\ecut_{1-}$ in the plane. According to
\eref{static-pos}, this angle is equal to $8\pi Gm_1$. The deficit
angle of the particle $\prt_2$ is the angle between the edges
$\ecut_{2-}$ and $\ecut_{2+}$ on the other side, which is equal to
$8\pi Gm_2$.

The disadvantage of \fref{tip}(b) is that the definition of the centre
of mass frame is less obvious. However, it is actually only the
relation between the geometry of space and the centre of mass frame,
which we need in the following. And this can still be read off from
\fref{tip}(b). First, we need to know which time slice is represented
by the given space. This is specified by the clock $T$, and according
to \eref{static-pos} this is also the $\gam_0$-coordinate of the plane
shown \fref{tip}(b). And secondly, we need to know the spatial
orientation of the particles with respect to the reference frame. This
is specified by the symmetry axis of \fref{tip}(b). It is the solid
line in the middle, which is a radial line with angular direction
$\gam(\xdir)$.

\subsubsection*{The spinning cone}
When the particles are moving, we have to generalize two aspects of
this construction. The geometry of space is no longer constant in
time. The previously introduced relative position coordinates become
time dependent, and we also have to introduce conjugate momentum
variables. The time evolution will finally be provided by some
Hamiltonian, which we still have to derive. Moreover, we also have to
replace the conical metric \eref{static-cone} of the big cone by a
more general conical geometry. Let us first consider this second
aspect.

If the particles are in motion with respect to each other, then the
fictitious centre of mass particle receives a variable mass $M$ and a
variable spin $S$, where $M$ is the total energy and $S$ is the total
angular momentum of the system. The gravitational field of such a
particle is a \emph{spinning cone} \cite{djh}. Using the same conical
coordinates as before, the metric of a spinning cone can be written as
\begin{equation}
  \label{spin-cone}
  \dd s^2 = - (\dd \cT + 4 G S \, \dd\cdir)^2 + \dd \cR^2
            + (1-4GM)^2 \, \cR^2 \, \dd\cdir^2.
\end{equation}
This is obviously a generalized version of \eref{static-cone}. The
previously considered static case is recovered if we set $M=m_1+m_2$
and $S=0$. But nevertheless, we can still use the conical coordinates
$(\cT,\cR,\cdir)$ to define the centre of mass frame of the universe,
and we may also identify this with the reference frame of some
external observer sitting at infinity.

To visualize the geometry of the spinning cone, let us consider a
surface of constant conical time $\cT=T$. We can still say that each
such surface represents an instant of time in the centre of mass
frame. Moreover, all such surfaces have the same geometry, as long as
we do not insert the moving particles. The spinning cone is no longer
static, but still stationary. A surface of constant conical time is
not flat, but it can locally be embedded into Minkowski space. The
following local isometry maps a segment of the spinning cone into
Minkowski space. We fix some angular direction $\xdir_0$ and define
\begin{equation}
  \label{iso}
  (\cT,\cR,\cdir) \quad \mapsto \quad
    ( \cT + 4 G S \, (\cdir-\xdir_0) ) \, \gam_0 + 
    \cR \, \gam( \cdir - 4GM \, (\cdir-\xdir_0) ). 
\end{equation}
One can easily check that this is indeed an isometry. It is only a
local isometry, because it does not respect the periodicity of
$\cdir$. A surface of constant $\cT=T$ of the spinning cone is mapped
onto a \emph{screw surface} in Minkowski space.

A screw surface in Minkowski space is defined by a family radial
lines, where the $\gam_0$-coordinate increases linearly with the
angular direction. The slope of this screw surface is determined by
the parameters $M$ and $S$. If we pass once around the spinning cone,
thus if we increase $\cdir$ by $2\pi$, then the total angle covered by
the screw surface in Minkowski space is $2\pi-8\pi GM$, and the amount
that we move forward in time is $8\pi GS$. The spinning cone has a
\emph{deficit angle} of $8\pi GM$ and a \emph{time offset} of $8\pi
GS$. In units where the velocity of light is one, the latter is a
length or time if $S$ has the dimension of an angular momentum.

The central axis of the spinning cone at $\cR=0$ is mapped onto the
$\gam_0$-axis in Minkowski space. This is the apparent world line of
the centre of mass, as seen by an observer at infinity. There is a
critical radius $R_0=4GS/(1-4GM)$. For $\cR<R_0$, the screw surface is
timelike, at $\cR=R_0$ it is lightlike, and for $\cR>R_0$ it is
spacelike. There are thus closed timelike curves in the spinning cone,
passing through the region $\cR<R_0$. When we insert the particles,
this critical region will be cut away, so that the actual Kepler
spacetime has no closed timelike curves. Outside the critical region
the causal structure of the spinning cone is well defined.

To insert the particles, we have to generalize the construction in
\fref{tip}(b). Remember that the plane in which this construction took
place was an equal time plane in Minkowski space, with
$\gam_0$-coordinate $T$. And remember also that the symmetry axis was
a radial line in this plane, with angular direction $\gam(\xdir)$. We
shall now \emph{lift} this picture into the third dimension. The half
spaces $\pol_\pm$ are tilted in a certain way, while the symmetry of
the figure is preserved. The deformation is so that finally the half
spaces become two segments of the previously considered screw surface.
Hence, they fit into the spinning cone as a surface of constant
conical time $\cT=T$.
\begin{figure}[t]
  \begin{center}
    \epsfbox{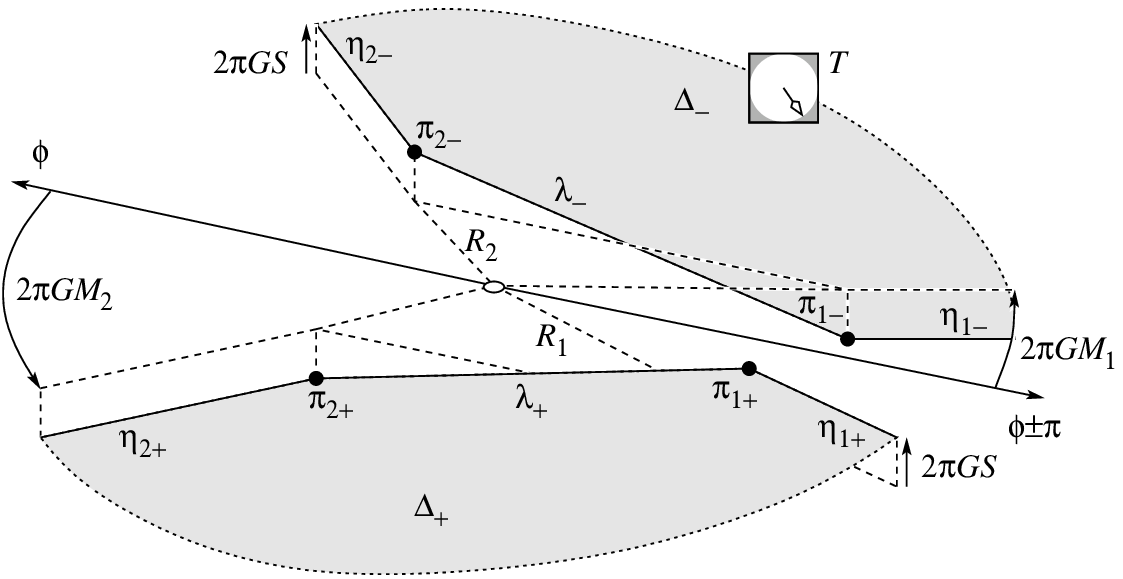}
    \xcaption{To define the geometry of space at a moment of time 
    if the particles are not at rest, the construction in
    \fref{tip}(b) has to be lifted into the third dimension in the
    embedding Minkowski space. The half spaces are then two segments
    of a screw surface. This surface represents an instant of absolute
    time in the centre of mass frame. The geometry of the two half
    spaces, the way they are embedded into the reference frame, and
    also the locations of the particles in space are specified by a
    set of six independent phase space variables.}
  \label{emb}
  \end{center}
\end{figure}

As indicated in \fref{emb}, the deformation is as follows. The corners
$\prt_{k\pm}$ are raised and lowered symmetrically by $2\pi GS$.
Moreover, the mass parameters $m_k$ appearing in \eref{static-pos},
which define the deficit angles of the particles, are replaced by two
independent \emph{energy} variables $M_k$. This is because for a
moving particle the deficit angle is actually not proportional to its
mass, but proportional to its energy \cite{matwel}. We shall look at
this more closely in a moment. As we want the symmetry axis of the
figure to be preserved, the locations of the corners $\prt_{k\pm}$ are
then given by
\begin{eqnarray}
  \label{pos-cone}
  \pos_{1\pm} &=& (T \pm 2\pi GS) \, \gam_0 
              - R_1 \, \gam( \xdir \mp 4\pi GM_1), \nwl
  \pos_{2\pm} &=& (T \mp 2\pi GS) \, \gam_0 
              + R_2 \, \gam( \xdir \pm 4\pi GM_2).
\end{eqnarray}
The radial coordinates $R_k$ are still to be determined. We recover
the static case \eref{static-pos} if we set $M_k=m_k$ and $S=0$. The
edges $\ecut_{k\pm}$ are still defined as the unique radial lines in
Minkowski space which pass through the corners $\prt_{k\pm}$. As they
are now located at different time levels in the embedding Minkowski
space, each half space becomes a segment of a screw surface. The time
offset of each half space is $4\pi GS$, and the angle in Minkowski
space covered by each half space is $\pi-4\pi GM$, where $M=M_1+M_2$.

Let us glue the two half spaces together along the edges
$\ecut_{k\pm}$. The result is a screw surface with a deficit angle of
$8\pi GM$ and a time offset of $8\pi GS$. We can still use
\fref{tip}(a) as a schematic representation of this surface. It is a
conical surface with a hole in the middle or, roughly speaking, a cone
with the tip cut off. It can be embedded into a spinning cone, as a
surface of constant conical time $\cT=T$. Thus it represents an
instant of time in the centre of mass frame, and the clock $T$
represents the absolute time. The angular orientation of this
embedding is fixed in the same way as before. The geodesics $\ecut_k$
are two antipodal radial lines with conical coordinates $\cT=T_k$ and
$\cdir=\xdir_k$, where
\begin{equation}
  \label{T-cone}
  T_1 = T , \quad T_2 = T, \qquad 
  \xdir_1 = \xdir \pm\pi , \qquad \xdir_2 = \xdir.
\end{equation}
The edges $\cut_\pm$ are the geodesics in Minkowski space, connecting
the corners $\prt_{k\pm}$. As these are in general not contained in
the screw surfaces, we actually have to deform the half spaces
slightly in the neighbourhood of these edges. But this does not affect
the embedding into the spinning cone. We can still assign two
Minkowski vectors $\dis_\pm$ to the edges $\cut_\pm$. They must be
spacelike for the half spaces to be spacelike surfaces. And they can
still be chosen so that both have the same direction $\xdir$,
parallel to the symmetry axis. Moreover, we already know that the
$\gam_0$-components of these vectors must be equal to the time offsets
$4\pi GS$ of the two half spaces. Hence, we have
\begin{equation}
  \label{dis-cone}
  \dis_\pm = \xsqrt{ R^2+(4\pi GS)^2} \, \gam(\xdir)
            \mp 4\pi GS\, \gam_0 ,
\end{equation}
where $R$ is again the distance between the particles. On the other
hand, also have $\dis_\pm=\pos_{2\pm}-\pos_{1\pm}$, which implies that
the radial coordinates $R_k$ are given by the following
generalization of \eref{static-cR},
\begin{eqnarray}
  \label{R-cone}
  R_1 &=& \xfrac{\sin(4\pi GM_2)}{\sin(4\pi G(M_1+M_2))} \, 
            \xsqrt{R^2+(4\pi GS)^2} , \nwl
  R_2 &=& \xfrac{\sin(4\pi GM_1)}{\sin(4\pi G(M_1+M_2))} \, 
            \xsqrt{R^2+(4\pi GS)^2} .
\end{eqnarray}
So, after this somewhat technical construction, what did we get? There
are now six independent configuration variables $M_1$, $M_2$, $S$,
$T$, $R$, and $\xdir$. From the point of view of general relativity,
they specify the geometry of space at a moment of time, providing a
kind of discretized ADM variables. From the particle point of view,
they define both the relative position of the particles with respect
to each other, as well as their absolute positions with respect to the
centre of mass frame. 

The relative position of the particles is specified by the geodesic
distance $R$ and the angular orientation $\xdir$. The absolute
positions are given by the conical coordinates $(T_k,R_k,\xdir_k)$ of
the points $\prt_k$ in the spinning cone. All this is analogous to the
previously considered static case, and also to the free particles,
where the relative coordinates are $X$ (or $R$) and $\xdir$, and the
absolute positions are given by \eref{rpp-pos-XT}. What makes the
definitions a little bit more complicated here is that the reference
frame is a conical frame, and not a Minkowski frame. What we do not
see so far is in which sense the energy and momentum variables are
conjugate to the position coordinates, how they evolve in time, and
what is also still missing is a variable that replaces the radial
momentum $Q$.

\subsubsection*{Transition functions}
Let us have a closer look at the way the edges in \fref{emb} are glued
together. The embedding Minkowski space can locally be identified with
the spacetime manifold, if the world lines are excluded. More
precisely, it provides an atlas of the spacetime manifold in the
neighbourhood of the embedded space manifold at the given moment of
time. The atlas consists of two charts, and each chart contains one
half space as a spacelike surface. There are three overlap regions
between the two charts, corresponding to the edges $\ecut_{1\pm}$,
$\cut_\pm$, and $\ecut_{2\pm}$. The transition functions in these
overlap regions are isometries of the embedding Minkowski space, hence
Poincar\'e transformations. 

Since in the following only the rotational components are of interest,
let us ignore the translations that are involved. We can then say that
the edge $\ecut_{1-}$ is mapped onto the edge $\ecut_{1+}$ by a
certain Lorentz rotation. It can be represented by an element
$\ehol_1\in\grpSL(2)$ of the spinor representation of the three
dimensional Lorentz group. Similarly, the edge $\ecut_{2-}$ is also
mapped onto the edge $\ecut_{2+}$ by some Lorentz rotation, which is
given by $\ehol_2\in\grpSL(2)$. And finally, $\cut_-$ is mapped onto
$\cut_+$ by yet another Lorentz rotation $\chol\in\grpSL(2)$. The
transition functions tell us how the half spaces are to be glued
together along the edges. The question is thus, are they already
determined by the variables introduced so far, or is there still some
ambiguity?

Consider first the edges $\ecut_{1-}$ and $\ecut_{1+}$, on the right
hand side of \fref{emb}. They are obviously mapped onto each other by
a rotation about the $\gam_0$-axis by $8\pi GM_1$ in clockwise
direction. Similarly, $\ecut_{2-}$ is mapped onto $\ecut_{2+}$ by a
rotation about the $\gam_0$-axis by $8\pi GM_2$ in counter clockwise
direction. It is shown in the appendix that the group elements
representing these rotations are
\begin{equation}
  \label{ehol-MV}
  \ehol_1 = \expo{4\pi GM_1\gam_0} , \qquad 
  \ehol_2 = \expo{-4\pi GM_2\gam_0} . 
\end{equation}
To be precise, there are more general Lorentz rotations mapping the
given edges onto each other. We may, for example, multiply $\ehol_1$
from the left with a boost whose axis is parallel to the edge
$\ecut_{1-}$, or from the right with boost whose axis is parallel to
the edge $\ecut_{1+}$. However, there is yet another consistency
condition. The two half spaces have to fit together as two halves of a
single screw surface, without a kink, and this is only the case if the
transition functions are pure rotations.

Regarding the transition function $\chol$, which specifies how the
edges $\cut_+$ and $\cut_-$ are glued together, there is no such
additional restriction. The only consistency condition is that the
Lorentz rotation $\chol$ maps the Minkowski vector $\dis_-$ onto the
vector $\dis_+$, hence
\begin{equation}
  \label{dis-chol}
  \dis_+ = \chol^{-1} \dis_- \, \chol \equivalent
  \dis_- = \chol \, \dis_+ \, \chol^{-1}. 
\end{equation}
There is a one parameter family of such group elements, and therefore
there is one additional degree of freedom that has to be fixed. As an
ansatz, we expand the matrix $\chol$ in terms of the unit and gamma
matrices, using $\gam_0$, $\gam(\xdir)$ and $\gam'(\xdir)$ as an
orthonormal basis,
\begin{equation}
  \chol = u \one + w \, \gam_0 
          + q \, \gam(\xdir) + s \, \gam'(\xdir).
\end{equation}
This basis is adapted to the definition \eref{dis-cone} of the vectors
$\dis_\pm$, so that the equations \eref{dis-chol} for the coefficients
$u,w,q,s$ become as simple as possible. Additionally, we have the
equation $\det(\chol)=1$. The resulting system of quadratic equations
can be easily solved. There is a one-parameter family of solutions,
which is given by $w=0$ and
\begin{equation}
  \label{uqs}
  u = \frac{\xsqrt{R^2+(4\pi GS)^2}}{\xsqrt{R^2-(4\pi GQ)^2}}, \quad
  q = \frac{4\pi G Q }{\xsqrt{R^2-(4\pi GQ)^2}}, \quad 
  s = \frac{4\pi G S }{\xsqrt{R^2-(4\pi GQ)^2}}. 
\end{equation}
The reason for choosing the parameter $Q$ in this particular way
will become clear in a moment. The motivation is to have a certain
symmetry between the spatial components $q$ and $s$ of the matrix
$\chol$ on one side, and the variables $Q$ and $S$ on the other side.

Due to the square roots in the denominator, there is obviously a
restricted range $-R<4\pi GQ<R$ for the new parameter. To avoid such a
non-trivial restriction on the range of the phase space variables
later on, it is useful to replace the variable $R$ by a new variable
$X$, so that
\begin{equation}
  \label{R-QX}
  R^2 = X^2 +(4\pi GQ)^2.
\end{equation}
The expression for the transition function $\chol$ then simplifies
slightly and becomes
\begin{equation}
  \label{chol-QS}
  \chol = U \, \one 
      + 4 \pi G \, \frac{Q \, \gam(\xdir) + \, S \, \gam'(\xdir)}X,
\end{equation}
where $U\ge0$ is defined so that the $\chol$ has a unit determinant,
\begin{equation}
  \label{U-QS}
  U^2 = 1 + (4\pi G)^2 \, \frac{Q^2+S^2}{X^2}. 
\end{equation}
Here we should note the similarity to the formula \eref{rpp-mom-MVQS},
defining the momentum vectors of the free particles. If the group
element $\chol$ is considered as the spatial momentum of the
particles, the $Q$ and $S$ are generalized versions of the radial and
angular momentum.

To see what kind of group element $\chol$ it is, let us rewrite it in
the following alternative way. We replace the Cartesian coordinates
$Q$ and $S$ by polar coordinates $K$ and $\pdir$, so that
\begin{equation}
  \label{QS-K}
  4\pi G Q = X \sinh(4\pi GK) \cos(\pdir-\xdir), \quad
  4\pi G S = X \sinh(4\pi GK) \sin(\pdir-\xdir). 
\end{equation}
This is obviously a generalization of \eref{rpp-QS-K}, to which it
reduces in the limit $G\to0$. It follows that $U=\cosh(4\pi GK)$, and
the transition function becomes 
\begin{equation}
  \label{chol-K}
  \chol = \expo{4\pi G K\gam(\pdir)} = 
          \cosh(4\pi GK) \, \one + \sinh(4\pi GK) \, \gam(\pdir).
\end{equation}
It represents a boost with rapidity $4\pi GK$ and angular direction
$\pdir$. There is thus a certain relation between the transition
functions $\ehol_k$ and the energies $M_k$ of the particles on one
side, and the transition function $\chol$ and the spatial momentum $K$
of the particles on the other side. Of course, we thereby assume that
$K$ still defines the spatial momentum, in the same way as previously
for the free particles. 

\subsubsection*{Mass shell constraints}
To see that $M_k$ are in fact the energies, and $K$ is the spatial
momentum of the particles, let us now have a closer look at the
conical singularities, which arise at the corners of the half spaces
in \fref{emb} when they are glued together. We somehow have to ensure
that the spacetime metric in the neighbourhood of a world line
represents the gravitational field of a point particle with a fixed
mass $m_k$. It has to be a cone with a deficit angle of $8\pi Gm_k$,
when this angle is measured in the rest frame of the particle. In the
static case, we can read off this deficit angle directly from the
geometry of space in \fref{emb}, or \fref{tip}(b). In general,
however, we first have to transform to the rest frame of the particle,
as otherwise the deficit angle is proportional to the energy and not
to the rest mass.

An alternative way to derive the deficit angle is to consider the
\emph{holonomy} of the particle. This is, by definition, the Lorentz
rotation that acts on a spacetime vector which is transported once
around the particle, say, in clockwise direction. If the deficit angle
of the particle $\prt_k$ is $8\pi Gm_k$, then the holonomy is a
rotation by $8\pi Gm_k$ about some timelike axis. The axis is parallel
to the world line, and thus the holonomy also defines the direction of
motion of the particle. Massless particles can also be included. The
holonomy is then a null rotation, and the world line is lightlike. The
holonomy is in this sense a generalized, group valued momentum of the
particle \cite{matwel}.

If we think of the embedding Minkowski space in \fref{emb} as a
spacetime atlas consisting of two charts, then a spacetime vector is
simply represented by a Minkowski vector, which is attached to some
point on one of the two half spaces. As long as we stick to this half
space, the parallel transport is trivial. But whenever we pass across
one of the edges $\ecut_k$ or $\cut$ from $\pol_-$ to $\pol_+$, we
have to act on the vector with the appropriate transition function,
hence $\ehol_k$ or $\chol$. And when we pass from $\pol_+$ to
$\pol_-$, we have to act on the vector with the inverse transition
function. There are then all together four holonomies that can be
defined. We can choose the particle $\prt_k$ to be surrounded, and we
can choose the half space $\pol_\pm$ in which the path begins and
ends.

Let us define $\hol_{k\pm}\in\grpSL(2)$ to be the Lorentz rotation
acting on a vector which is first defined in the half space
$\pol_\pm$, and then transported in clockwise direction around the
particle $\prt_k$. Sorting out the factor ordering and the signs, we
find that
\begin{equation}
  \label{hol-1-2-pm}
  \hol_{1+} = \chol^{-1} \, \ehol_1, \quad 
  \hol_{1-} = \ehol_1 \, \chol^{-1} , \qquad 
  \hol_{2+} = \ehol_2\^{-1} \, \chol, \quad
  \hol_{2-} = \chol \, \ehol_2\^{-1}.
\end{equation}
For $\hol_{k\pm}$ to be a rotation by $8\pi Gm_k$ in clockwise
direction, it has to be an element of the conjugacy class of
$\expo{4\pi Gm_k\gam_0}$ in $\grpSL(2)$. This special element
represents a clockwise rotation by $8\pi Gm_k$ about the
$\gam_0$-axis. All others are obtained by acting on this with a proper
Lorentz rotation, hence a conjugation with some other element of the
Lorentz group. The conjugacy class is defined by the following
\emph{mass shell} and \emph{positive energy} condition
\cite{matwel,matmul}
\begin{equation}
  \label{hol-con}
  \hols_k = \ft12\Trr{\hol_{k\pm}} = \cos(4\pi Gm_k), 
   \qquad \ft12\Trr{\hol_{k\pm}\gam^0} > 0. 
\end{equation}
Since the trace of the holonomy is independent of the factor ordering,
this is one equation to be satisfied for each particle. Actually, each
particle has only one holonomy $\hol_k$. The two different values
$\hol_{k+}$ and $\hol_{k-}$ arise because the same physical object is
represented in two coordinate charts. 

If the holonomy is interpreted as a generalized momentum, then the
mass shell constraints are obviously generalizations of the free
particle constraints \eref{rpp-mss}. There is now, however, a
restriction on the mass parameters $m_k$, which has no counterpart for
the free particles. The deficit angle of a cone must be smaller then
$2\pi$, and consequently the rest mass of a particle is bounded from
above by $1/4G=\Mpl/4$, where $\Mpl=1/G$ is the Planck mass. At this
upper bound, the cosine in \eref{hol-con} takes its minimum $-1$, and
the holonomy becomes a full rotation by $2\pi$. For a massless
particle, on the other hand, the cosine is equal to one, which means
that the holonomy is a null rotation.

The same restriction applies to the total energy $M$, which defines
the total deficit angle of universe. We'll see later on that the
energy $M$ is in fact also bounded from below by $\Mmin=m_1+m_2$,
which is the total energy for a static state. Hence, the allowed range
for the mass parameters is
\begin{equation}
  \label{mss-range}
  m_k \ge 0 , \qquad m_1+m_2 < \Mpl/4 . 
\end{equation}
To see that the free particle mass shell constraints are recovered in
the limit $G\to0$, let us write the holonomies as functions of the
configuration variables. It is thereby useful to introduce the
following notation. We define a modified set of trigonometric
functions with rescaled arguments,
\begin{equation}
  \label{cs-sn-tn}
  \cs\vrh = \cos(2\pi G\vrh), \qquad
  \sn\vrh = \frac{\sin(2\pi G\vrh)}{2\pi G}, \qquad
  \tn\vrh = \frac{\tan(2\pi G\vrh)}{2\pi G}.
\end{equation}
Some useful properties of these functions are that the relation
$\tn\vrh={\sn\vrh}/{\cs\vrh}$ remains valid, and in the limit $G\to0$
we have $\cs\vrh\to1$, $\sn\vrh\to\vrh$, $\tn\vrh\to\vrh$. We also
introduce an analogous set of hyperbolic functions, but this time it
is useful to rescale the argument by a different factor,
\begin{equation}
  \label{csh-snh-tnh}
  \csh\vrh = \cosh(4\pi G\vrh), \qquad
  \snh\vrh = \frac{\sinh(4\pi G\vrh)}{4\pi G}, \qquad
  \tnh\vrh = \frac{\tanh(4\pi G\vrh)}{4\pi G}.
\end{equation}
Again, we have $\tnh\vrh={\snh\vrh}/{\csh\vrh}$, and in the limit
$G\to0$ we get $\csh\vrh\to1$, $\snh\vrh\to\vrh$, $\tnh\vrh\to\vrh$.
As an example for the application of these rescaled function, consider
the definition \eref{QS-K}, relating $Q$ and $S$ to $K$ and $\pdir$.
They simplify to
\begin{equation}
  \label{QS-K-x}
  Q = X \snh K \, \cos(\pdir-\xdir), \qquad
  S = X \snh K \, \sin(\pdir-\xdir),
\end{equation}
and $K$ can be written as a function of $Q$, $S$, and $X$, 
\begin{equation}
  \label{K-QS}
  \snh^2 K = \frac{Q^2+S^2}{X^2}.   
\end{equation}
According to the general rule $\snh\vrh\to\vrh$, this reduces to
\eref{rpp-K-QS} in the limit $G\to0$. Similar rules apply to all other
formulas below. They always reduce to the free particle counterparts
in the limit where the gravitational interaction is switched off.

To express the holonomies and finally the mass shell constraints in
terms of the energy and momentum variables, it is useful to make the
same redefinition that we previously also made for the free particles.
We replace the mass parameters $m_k$ by a total mass $\mu$ and a
relative mass $\nu$, assuming without loss of generality that $m_2\ge
m_1$,
\begin{equation}
  \label{mu-nu}
  \mu = m_2 + m_1 , \qquad 
  \nu = m_2 - m_1 , \qquad 
  0 \le \nu \le \mu < \Mpl/4 .
\end{equation}
Like for the free particles, the special cases are $\nu=0$, where both
particles have the same mass, and $\nu=\mu$, where at least one of the
particles is massless. The only new feature is the upper bound for
$\mu$, which goes to infinity in the limit $G\to0$. Similarly, we also
replace the energy variables $M_k$ by the total energy $M$ and a
relative energy $V$, so that
\begin{equation}
  \label{erg-MV}
  M = M_2 + M_1, \qquad V = M_2 - M_1 , \qquad 
   | V | < M < \Mpl/4 .
\end{equation}
The upper bound on $M$ follows from the fact that the total deficit
angle of the spinning cone must be smaller than $2\pi$. The
restriction on $V$ is the positive energy condition, which implies
that $M_k>0$ for both particles. 

Using all this, we can finally express the transition functions
\eref{ehol-MV} and \eref{chol-K} in terms of the new energy and
momentum variables. What we get is
\begin{eqnarray}
  \label{hol-MVK}
  \ehol_1 &=& \cs(M-V) \, \one + 2\pi G \, \sn(M-V) \, \gam_0, \nwl 
  \ehol_2 &=& \cs(M+V) \, \one - 2\pi G \, \sn(M+V) \, \gam_0, \nwl  
  \chol &=& \csh K \, \one + 4\pi G \, \snh K \, \gam(\pdir).
\end{eqnarray}
It is then not difficult to evaluate the traces of the holonomies, and
to express the mass shell constraints in terms of the same variables.
They are given by
\begin{equation}
  \label{hols-MVK}
  \hols_1 = \csh K \cs(M-V) = \cs(\mu-\nu), \qquad
  \hols_2 = \csh K \cs(M+V) = \cs(\mu+\nu). 
\end{equation}
This is the same as (2.36) in \cite{matwel}, if $K$ is interpreted as
the common spatial momentum of both particles in the centre of mass
frame, and $(M\pm V)/2$ are the energies of the particles, which are
in general different. Finally, we have to rescale the actual mass
shell constraints by an appropriate power of $G$, in order to get the
correct limit $G\to0$, 
\begin{equation}
  \label{con-X}
  \con_1 = \frac{ \csh K \cs(M-V) - \cs(\mu - \nu )}
                    {(4\pi G)^2}, \qquad
  \con_2 = \frac{ \csh K \cs(M+V) - \cs(\mu + \nu )}
                    {(4\pi G)^2}.
\end{equation}
Expanding this up to the second order in $G$, we recover the free
particle constraints \eref{rpp-mss-K} in the limit $G\to0$. And the
same applies to the positive energy condition \eref{erg-MV}, where the
upper bound on $M$ goes to infinity. 

We can then proceed in the same way as before. We use one of the mass
shell constraints to eliminate the relative energy $V$. There is then
a single mass shell constraint left, which is going to be the
generator of the time evolution. And we also get an even dimensional
phase space which is spanned by six independent variables. The first
step is again to define the linear combinations
\begin{eqnarray}
  \label{DE}
  \rcon' &=&  \con_2 -  \con_1 = 
       \frac{\csh K  \, \sn M \, \sn V -  \sn\mu \, \sn\nu}2 , \nwl  
  \tcon' &=&  \con_2 +  \con_1 = 
       \frac{ \csh K  \, \cs M \, \cs V -  \cs\mu \, \cs\nu }
            {2 (2\pi G)^2}.
\end{eqnarray}
Here we used some trigonometric identities to simplify the results. In
the limit $G\to0$, these constraints are equal to \eref{rpp-DE}.
However, it is now a little bit more complicated to decouple the
variables $K$ and $V$. We have to take another linear combination,
namely
\begin{equation}
  \label{D}
  \rcon =  \frac{ \rcon' + (2\pi G)^2 \, \tn M \, \tn V \, \tcon' }
                {\cs\mu\,\cs\nu} 
        = \frac{\tn\mu \, \tn\nu - \tn M \tn V}2 . 
\end{equation}
Note that the functions $1/{\cs\mu}$ and $1/{\cs\nu}$, and similarly
$\tn M$ and $\tn V$ are well defined within the range \eref{mu-nu} and
\eref{erg-MV}. The first poles of these functions are just outside the
allowed range of the arguments. Now, the positive energy condition
implies $\tn M>0$, and the function $\tn$ is invertible within the
allowed range of $V$. We can solve the equation $\rcon=0$ for $V$,
setting
\begin{equation}
  \label{V-M}
   V = \atn \left( \frac{\tn \mu \, \tn \nu}{\tn M} \right) . 
\end{equation}
The function $\atn$ is the inverse of $\tn$,
\begin{equation}
  \label{atn}
    \atn\vrh  = \frac{\arctan(2\pi G \vrh)}{2\pi G} \follows
    -\Mpl/4 < \atn\vrh < \Mpl/4,
\end{equation}
and in the limit $G\to0$ we have $\atn\vrh\to\vrh$. Using this and the
fact that $\tn$ and $\atn$ are both monotonically increasing, the
positive energy condition becomes a non-trivial condition to be
imposed on $M$, namely
\begin{equation}
  \label{erg-M}
    |\tn V| = \frac{\tn \mu \, \tn \nu}{\tn M} < \tn M  \equivalent
      \atn \left(\sqrt{\tn \mu \tn \nu}\right) <  M  < \Mpl/4 .    
\end{equation}
This is obviously a generalized version on \eref{rpp-erg-M}, with all
mass parameters and energy variables replaced by their rescaled
tangent. This is in fact a rather general rule, which tells us how to
obtain the various structures of the Kepler system from those of the
free particle system. Consider for example the mass shell constraint
that remains after eliminating $V$ from \eref{DE}. To derive this, we
have to take yet another linear combination, namely
\begin{equation}
  \label{E-def}
   \tcon = \frac{\csh K  \, \cs M \, \cs V +  \cs\mu \, \cs\nu}
                {2\cs^2 M} \, \tcon' 
         + \frac{\csh K  \, \sn M \, \sn V +  \sn\mu \, \sn\nu}
                {2\sn^2 M} \, \rcon'. 
\end{equation}
Again, the coefficients are well defined, because the range
\eref{erg-MV} of $M$ is exactly the interval where both $\sn M$ and
$\cs M$ are different from zero. Evaluating this linear combination,
$V$ drops out, and after some trigonometric simplifications we get
\begin{equation}
  \label{E}
  \tcon = \snh^2 K - \lam^2 \, F(M) \approx 0 . 
\end{equation}
Here, $0<\lam\le1$ is a constant, which only depends on the mass
parameters $\mu$ and $\nu$, and $F$ is the same function that also
appears in \eref{rpp-EF}, but once again with the masses $\mu$, $\nu$,
and the energy $M$ are replaced by $\tn\mu$, $\tn\nu$, and $\tn M$,
respectively,
\begin{equation}
  \label{F-def}
  \lam = \cs\mu \, \cs\nu,\qquad 
   F( M ) =   
    \frac{(\tn^2M-\tn^2\mu) \, (\tn^2M-\tn^2\nu)} 
          {4 \, \tn^2M}.
\end{equation}
Roughly speaking, the rule is to replace all mass and energy
quantities $\vrh$ by $\tn\vrh$, and to replace the momentum $K$ by
$\snh K$. Then we obtain the mass shell constraints of the Kepler
system from those of the free particle system, and also, for example,
the relation \eref{K-QS} between the radial momentum $Q$, the angular
momentum $S$, and the total spatial momentum $K$ of the particles. As
all these modifications become identities in the limit $G\to0$, we can
say that the Kepler system is a \emph{deformation} of the free
particle system, and Newton's constant is the deformation parameter.
We'll now see that this also applies to the various phase space
structures, such as the symplectic potential, the Poisson brackets,
and the Hamiltonian.

\subsubsection*{The phase space}
So far, we have seen that all physically relevant quantities,
including the geometry of space, the positions of the particles with
respect to the centre of mass frame, as well as the holonomies and the
mass shell constraints, can be expressed as a function of six
independent variables, the energy and momentum variables $M$, $Q$,
$S$, and the time and position variables $T$, $X$, $\xdir$. In analogy
to the free particles, we define the \emph{extended} phase space 
\begin{equation}
  \label{qsp}
  \qsp = \big\{\quad ( M, Q, S ; T, X, \xdir ) \quad \big| \quad 
                X \ge 0 , \quad 
                \xdir \equiv \xdir + 2 \pi \quad \big\}.
\end{equation}
Note that we do not impose any restriction on the energy $M$ at this
point. The positive energy condition \eref{erg-M} is imposed together
with the constraint \eref{E}, and this defines the physical subspace.
The Hamiltonian on $\qsp$ is again proportional to the mass shell
constraint. Thus if we introduce a multiplier $\mul$, we have in
analogy to \eref{rpp-ham-ext}
\begin{equation}
  \label{ham-ext}
  \qham = \mul\, \tcon.
\end{equation}
To derive the equations of motion, we also have to know the symplectic
structure on $\qsp$. But so far, there is no natural way to define it,
and there is also no reason why it should be the same as that of the
free particle system. Since we do not want to make any additional
assumptions, we can only \emph{derive} the symplectic structure from
the underlying field theory of Einstein gravity. Thus, we have to
apply a straightforward phase space reduction to the Einstein Hilbert
action.

The actual derivation is rather involved and technical, but on the
other hand it can be carried out without further complications for a
general multi particle model. It is therefore given in a separate
article \cite{matmul}. We are here not going to say anything in detail
about this derivation. We just take the general result, and adapt it
to the special case of a two particle system. The general expression
(\potmul) in \cite{matmul}, adapted to the special situation given in
\fref{emb}, is
\begin{equation}
  \label{pot-chol+}
  \qpot = - \sum_k M_k \, ( \dd T_k + 4GS \, \dd\xdir_k )
     - \frac1{8\pi G} \, \Trr{\chol^{-1} \dd\chol \, \dis_+ }.
\end{equation}
Here, $M_k$ are the energies of the particles $\prt_k$, $T_k$ and
$\xdir_k$ are their absolute positions with respect to the conical
reference frame, $\dis_+$ is the relative position vector of the
particles in the half space $\pol_+$, and $\chol$ is the transition
function mapping the edge $\cut_-$ onto $\cut_+$.

This expression does not look symmetric, but in fact it is, which can
be seen by replacing $\dis_+$ and $\chol$ with $\dis_-$ and
$\chol^{-1}$. To express the symplectic potential in terms of our six
independent phase space variables, we have to insert the expressions
\eref{T-cone} for $T_k$ and $\xdir_k$, \eref{dis-cone} for $\dis_\pm$,
and \eref{chol-QS} for $\chol$. The result is very simple and reads,
up to a total derivative that can be neglected,
\begin{equation}
  \label{pot-ext}
  \qpot = X^{-1} Q \, \dd X  + (1-4GM) \, S \, \dd\xdir - M \, \dd T.
\end{equation}
This is almost the same as the free particle expression
\eref{rpp-pot-ext}. There is only one modification. The angular
momentum $S$, which is conjugate to the orientation $\xdir$ of the
particles, is rescaled by a factor $1-4GM$. This factor takes values
between zero and one, and obviously it has to do with the conical
geometry of the spacetime at spatial infinity. As a consequence, we
have the following almost canonical Poisson brackets
\begin{equation}
  \label{pois-X}
  \pois MT = -1 , \qquad
  \pois QX = X ,
\end{equation}
but the following non-canonical brackets involving the angular
momentum $S$, 
\begin{equation}
  \label{pois-S}
  \pois S\xdir = \frac1{1-4GM}, \qquad
  \pois ST = - \frac{4GS}{1-4GM}. 
\end{equation}
The last one results from an off-diagonal term in the symplectic
two-form $\qsym=\dd\qpot$, involving a product of $\dd M$ and
$\dd\xdir$. The brackets can most easily be derived if we notice that
the canonically conjugate angular momentum is actually
\begin{equation}
  \label{J-S}
  J = (1-4GM) \, S \follows \pois J\xdir = 1.
\end{equation}
All other brackets with $J$ are zero. The brackets \eref{pois-S} are
then easily found by expressing $S$ as a function of $J$ and $M$. But
let us nevertheless stick to $S$ as one of the basic phase space
variables, because it has an immediate geometric interpretation,
defining the time offset of the spinning cone. It also shows up in the
transition function $\chol$ defined in \eref{chol-QS}, and in the mass
shell constraint \eref{E}, implicitly through the definition
\eref{K-QS} of $K$.

Now we have all the structures at hand which we need to describe the
kinematical and dynamical features of the Kepler system. In
particular, we can now derive the equations of motion, study the
classical trajectories, and finally we can quantize the Kepler system.
At each step, the result will be a deformed version of the
corresponding free particle result. The most interesting feature is
thereby the deformation of the symplectic structure by the factor in
front of the angular momentum. Apparently, this is only a marginal
modification. However, it turns out that it is responsible for some
unexpected effects at the quantum level. For example, the particles
can no longer be localized in space, and they cannot come closer to
each other than a specific distance.

But for the moment we shall stick to the classical phase space. Some
basic features of the Kepler system can be inferred immediately from
the given phase space structures. For example, we have a two
dimensional rigid symmetry group of time translations $T\mapsto
T-\Delta T$ and spatial rotations $\xdir\mapsto\xdir+\Delta\xdir$. The
Hamiltonian and the symplectic potential are both invariant. At the
spacetime level, these are the Killing symmetries of the spinning
cone. The rigid symmetries are the possible translations and rotations
of the universe with respect to the reference frame \cite{nico}. The
conserved charges associated with these symmetries are obviously $M$
and $J$.

Actually, we expected $S$ to be the angular momentum. It is a function
of $M$ and $J$, thus also a conserved charge, but the associated
symmetry is a combination of a time translation and a spatial
rotation. It is a \emph{screw rotation} of the spinning cone. The
Killing vector of this symmetry is a linear combination of the Killing
vectors of time translations and spatial rotations. It is orthogonal
to the Killing vector of time translations, whereas the rotational
Killing vector associated with $J$ has closed orbits of affine length
$2\pi$. Unless the spinning cone is static, this is not the same. The
definition of an angular momentum is therefore somewhat ambiguous.

\subsubsection*{Complete reduction}
We can also define a \emph{reduced} phase space $\psp$, in the very
same way as before. Instead of a mass shell constraint, the time
evolution is then provided by a Hamiltonian that represents the
physical energy. To go over from the \emph{constrained} to the
\emph{unconstrained} formulation, we have to solve the mass shell
constraint $\tcon=0$, and additionally we have to impose a suitable
gauge condition. As all this is completely analogous to the free
particles. We can choose the same natural gauge condition, requiring
the ADM time $t$ to be equal to the absolute time $T$ in the
centre of mass frame. What we get is a pair of second class
constraints,
\begin{equation}
  \label{gauge-con}
  \snh^2 K = \lam^2 \, F(M), \qquad T = t. 
\end{equation}
Since $F$ is more or less the same function as before, just with some
modified mass parameters, it can still be inverted, defining $M$ as a
function of $K$. The explicit solution is given in \eref{ham-red}
below. It is thus possible to remove the canonical pair $(M,T)$, and
to go over to a four dimensional, completely reduced phase space
\begin{equation}
  \label{psp}
  \psp = \big\{\quad ( Q, S ; X, \xdir ) \quad \big| \quad 
                X \ge 0 , \quad 
                \xdir \equiv \xdir + 2 \pi \quad \big\}.
\end{equation}
To derive the symplectic potential $\ppot$ and the Hamiltonian $\pham$
on $\psp$, we have to start from the extended symplectic potential on
$\qsp$, again because the gauge condition above is explicitly time
dependent. According to \eref{ham-ext} and \eref{pot-ext}, it is given
by
\begin{equation}
  \label{epot-ext}
  \qpot - \qham \, \dd t  
    = X^{-1} Q \, \dd X  + (1-4GM)\, S \, \dd\xdir - M \, \dd T
      - \mul \tcon \, \dd t. 
\end{equation}
On the reduced phase space, this becomes 
\begin{equation}
  \label{epot-red}
  \ppot - \pham \, \dd t 
        = X^{-1} Q \, \dd X  + (1-4G \pham) \, S \, \dd\xdir - 
        \pham \, \dd t.
\end{equation}
To derive the reduced Hamiltonian $\pham$, we have to solve the
equation \eref{gauge-con} for $M$, so that the positive energy
condition \eref{erg-M} is satisfied. The unique solution is 
\begin{equation}
  \label{ham-red}
  \pham =  F^{-1}\left( \dsty \lam^{-2} \, \snh^2 K \right) = 
    \atn\left(\xsqrt{ \lam^{-2} \snh^2 K  +  \tm_1\^2 }
            + \xsqrt{ \lam^{-2} \snh^2 K  +  \tm_2\^2 } \,\,\right) ,
\end{equation}
where $K$ is given by \eref{K-QS} as a function of the $Q$, $S$, and
$X$. The deformed mass parameters $\tm_k$ are just some useful
abbreviations. They are given by
\begin{equation}
  \label{tm-mu-nu}
  \tm_1 = \frac{\tn\mu - \tn\nu}{2}, \qquad
  \tm_2 = \frac{\tn\mu + \tn\nu}{2}.
\end{equation}
So, we find a somewhat deformed Hamiltonian, as compared to the free
particle definition \eref{rpp-ham-red}. It still depends only on the
momentum $K$, and not, for example, on the relative position $X$. This
we might have expected as a kind of gravitational potential. However,
in three dimensional Einstein gravity, there are no local
gravitational forces, and therefore we do not have a gravitational
potential. But nevertheless, there is something like a gravitational
binding energy. To see this, let us briefly discuss the range of
$\pham$, and the way it depends on the momentum $K$. First of all, we
easily see that $\pham$ is minimal for $K=0$, where it takes the value
\begin{equation}
  \label{M-min}
  \Mmin = m_1 + m_2.
\end{equation}
As for the free particles, the static states with $K=0$ are excluded
if at least one of the particles is massless. In this case, the
positive energy condition \eref{erg-M} requires $M>\Mmin$. Moreover,
for small momenta $K$, the energy also increases linearly if a
massless particle is present, and quadratically if both particles are
massive. For massive particles, we can also consider a
non-relativistic limit, where $K$ is small and the particles are
moving slowly compared to the speed of light. Expanding \eref{ham-red}
up to the second order in $K$ gives
\begin{equation}
  \label{red-mss}
  M \approx \Mmin + \frac1{2\tm} \, K^2, \txt{with}
  \tm = \frac{\cs^2\nu \, \tn^2\mu - \sn^2\nu}{4\tn\mu} .
\end{equation}
Apart from the deformed expression for the reduced mass $\tm$, we have
the usual non-relativistic behaviour of the kinetic energy. It can be
shown that $\tm>m$, thus the reduced mass of the coupled system is
always larger than the reduced mass of the free particles in
\eref{rpp-red-mss}. This is the only effect of the gravitational
coupling for low momenta. For a given value of $K$, the energy of the
coupled system is slightly smaller than the free energy. There is
still no gravitational potential. In particular, we do not recover
Newtonian gravity. In three spacetime dimensions, this is not the
non-relativistic limit of Einstein gravity. The relation between the
two theories only exists in higher dimensions.

\begin{figure}[t]
  \begin{center}
    \epsfbox{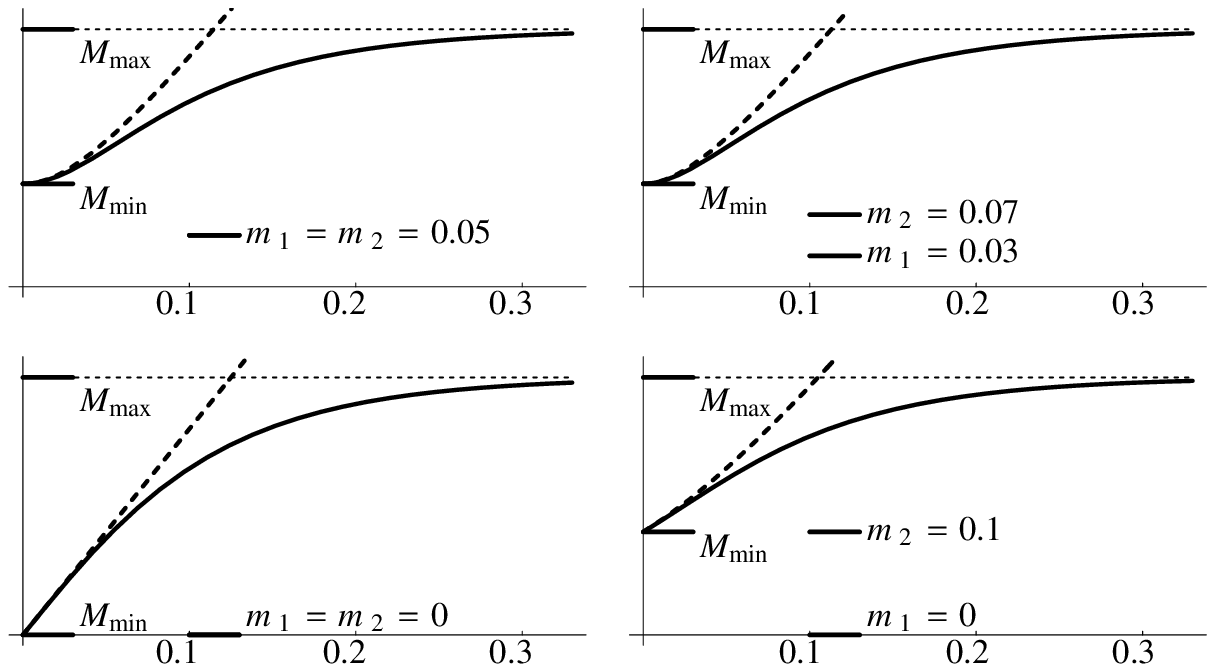}
    \xcaption{The energy momentum relation for the Kepler system (solid lines), 
    and the free particles (broken lines). For low momenta, the
    behaviour is almost the same. For large momenta, the energy of the
    free particle system increases linearly with the momentum, whereas
    that of the coupled system approaches the Planck energy from
    below. All quantities are given in Planck units $\Mpl=1/G$, and
    the upper bound for the total energy is $\Mmax=\Mpl/4$.}
  \label{dsp}
  \end{center}
\end{figure}

The behaviour of $\pham$ for large momenta $K$ is changed more
drastically, as compared to the free particle system. Asymptotically,
the energy is no longer linear in $K$. The right hand side in
\eref{ham-red} is bounded from above by $\Mpl/4$. The energy
asymptotically approaches the upper bound for large momenta $K$. This
is also a non-trivial result, because we might have expected that the
upper bound $\Mpl/4$ for the total energy $M$ imposes some upper bound
on the spatial momentum $K$ as well. But it seems that the
gravitational interaction knows about this upper bound, and takes it
into account in the relation between momentum and energy. To compare
the energy momentum relation for the free and the coupled particles,
we have plotted them for some typical mass parameters in \fref{dsp}.

For low momenta, the behaviour is very similar, we just have a
slightly modified reduced mass $\tm>m$ for massive particles. This
implies that the energy of the coupled system is a little bit smaller
than the free particle energy. For large momenta, however, we see that
the energy of the coupled system approaches $\Mmax=\Mpl/4$, whereas
the energy of the free particles increases unboundedly. We can think
of the difference between the free particle energy and the total
energy of the coupled system as the negative energy of the
gravitational field. This obviously compensates for the increasing
kinetic energy of the particles, so that the total energy remains
below the upper bound, which a quarter of the Planck energy.

If the momentum $K$ is of the order of one tenth of the Planck scale,
then the negative binding energy is already of the same order of
magnitude as the free particle kinetic energy. For momenta which are
larger than the Planck scale, the energy has more or less saturated at
the upper bound. The same behaviour of the energy can be found for a
single particle system \cite{matwel}, where the upper bound is
$\Mpl/8$, thus half of the maximal energy here. This is a generic
feature of three dimensional general relativity. It has to do with its
topological nature. The gravitational field does not fall off at
spatial infinity, which means that the spacetime is not asymptotically
flat. Instead, it is asymptotically conical, and therefore it is not
possible for the universe to contain more energy than $\Mmax=\Mpl/4$.

But now, let us turn to the more interesting object on the reduced
phase space $\psp$, the symplectic structure. It follows from
\eref{epot-red} that the reduced symplectic potential is given by
\begin{equation}
  \label{pot-red}
  \ppot = X^{-1} Q \, \dd X + (1-4G\pham) \, S \, \dd \xdir. 
\end{equation}
The somewhat curious feature of this expression is the Hamiltonian
$\pham$, which appears at the place where originally the energy $M$
has shown up. Since $\pham$ depends on $Q$, $S$ and $X$, the resulting
Poisson brackets become very complicated, because the two form
$\qsym=\dd\qpot$ is far from being diagonal in the given variables.

%%JL revision block begins 
It is therefore more appropriate first to look for a set of
canonically conjugate variables and then to define the Poisson
brackets. Clearly, what we have to do is to replace the variable $S$
by the canonical angular momentum $J=(1-4G\pham)S$. Doing so, the
symplectic potential and the resulting Poisson brackets become
\begin{equation}
  \label{pot-red-J}
  \ppot = X^{-1} Q \, \dd X + J \, \dd \xdir \follows
   \pois QX = X, \qquad \pois J\xdir = 1.  
\end{equation}
All other brackets of $Q$, $J$, $X$, and $\xdir$ are zero. We can
further replace $X$ and $Q$ by the geodesic distance $R$ of the
particles, given by \eref{R-QX}, and its canonically conjugate
momentum~$P$. The transformation reads
\begin{equation}
  \label{RP-def}
  R^2 = X^2 + (4\pi GQ)^2 , \qquad 
  \snh P = \frac QX ,
\end{equation}
and the inverse transformation is
\begin{equation}
  \label{RP-inv}
  X = \frac{R}{\csh P}, \qquad
  Q = R \, \tnh P .
\end{equation}
Inserting this into the symplectic potential gives
\begin{equation}
  \label{pot-red-P}
  \ppot = P \, \dd R + J \, \dd \xdir \follows
   \pois PR = 1, \qquad \pois J\xdir = 1.  
\end{equation}
The chart $(P,J;R,\xdir)$ is thus canonical on~$\psp$, and the
coordinates have a straightforward physical interpretation. This is
also the canonical chart introduced in \cite{loumat}.

In the chart $(Q,J;X,\xdir)$ the Hamiltonian $\pham$ is determined
implicitly as the solution to
\begin{equation}
\label{ham-J}
   \lam^2 \, F(\pham) =  \snh^2 K = \frac{Q^2+S^2}{X^2} 
      = \frac{Q^2+(1-4G\pham)^{-2}J^2}{X^2}, 
\end{equation}
and the corresponding equation in the chart $(R,\xdir, P, J)$ follows
by the substitution~\eref{RP-inv}.  As $F(\pham)$ involves the
rescaled tangent of~$\pham$, the equation is transcendental in either
chart.  The price for simplifying the symplectic potential
\eref{pot-red} into \eref{pot-red-J} or into the fully canonical form
\eref{pot-red-P} therefore is that $\pham$ cannot be expressed as an
elementary function of the variables.

\subsubsection*{Trajectories}
To find the classical trajectories, we return to the constrained
formulation based on the six dimensional phase space $\qsp$, with the
additional independent variables $M$ and~$T$, the symplectic
structure~\eref{pot-ext}, and the Hamiltonian
constraint~\eref{ham-ext}. As $\qsp$ is the gravitating version of the
special-relativistic phase space of \sref{free}, we can follow the
analysis of \sref{free} with the appropriate deformations.

The starting point is the Hamiltonian, which generates the time
evolution with respect to the unphysical ADM time $t$. It is given by
the mass shell constraint \eref{E}, and a multiplier $\mul$, which is
some arbitrary function of $t$,
\begin{equation}
  \label{E-QS}
  \qham= \mul \, \tcon , \qquad 
    \tcon = \frac{Q^2+S^2}{X^2} - \lam^2 \, F(M). 
\end{equation}
This is almost identical to \eref{rpp-EF}, and we can now use the
brackets \eref{pois-X} to derive the time evolution equations. The
total energy $M$ and the angular momentum $S$ are of course conserved
charges, and the same holds for the total spatial momentum $K$,
\begin{equation}
  \label{t-MS}
  \dot M = \pois \qham M = 0, \qquad 
  \dot S = \pois \qham S = 0  \qquad
  \dot K = \pois \qham K = 0.
\end{equation}
The time evolution equations for the radial coordinates are almost the
same as \eref{rpp-t-XQ},
\begin{equation}
  \label{t-XQ}
  \dot X = \pois\qham X = 2 \, \mul \, \frac QX , \qquad
  \dot Q = \pois\qham Q = 2 \, \mul \, \frac{Q^2+S^2}{X^2} 
                       = 2 \, \mul \, \snh^2 K.  
\end{equation}
The only essential modifications arise when we derive the evolution
equations for $\xdir$ and $T$. They differ from \eref{rpp-t-T},
because of the non-canonical brackets \eref{pois-S} of $S$ with
$\xdir$ and $T$. For the angular orientation, we find
\begin{equation}
  \label{t-xdir}
  \dot \xdir = \pois\qham \xdir = 2 \, \mul \,
                    \frac S{(1-4GM) X^2}, \qquad
\end{equation}
and for the clock $T$ we get
\begin{equation}
  \label{t-T}
  \dot T = \pois\qham T = 
        \mul \,  \left( \lam^2 \, F'(M) - 
                  \frac{8 G}{1-4GM} \, \frac{S^2}{X^2} \right) .  
\end{equation}
At first sight, these equations of motion are somewhat complicated,
but it is still possible to solve them explicitly, in the same way as
in \sref{free}. First, we observe that $M$, $S$ and $K$ are constants
of motion. Then, we introduce a function $\epsilon(t)$ so that
$\dot\epsilon(t)=2\mul(t)$. Using the same arguments that led to
\eref{rpp-MSKQ-t}, we conclude that
\begin{equation}
  \label{MSQ-t}
  M(t) = M_0, \qquad
  S(t) = S_0, \qquad
  K(t) = K_0, \qquad
  Q(t) = \epsilon(t) \, \snh^2 K_0, 
\end{equation}
where $M_0$, $S_0$ and $K_0$ are integration constants. It is then
straightforward to solve the equation of motion for $X$. The solution
is
\begin{equation}
  \label{X-t}
  X(t) = \xsqrt{ R_0\^2 + \epsilon^2(t) \snh^2 K_0 }, 
\end{equation}
where $R_0$ is some integration constant. And yet another constant
$\xdir_0$ arises when we then solve the equation of motion for
$\xdir$,
\begin{equation}
  \label{dir-t}
  \xdir(t) = \xdir_0 + \frac{1}{1-4GM_0}
             \arctan\left( \epsilon(t) \, \frac{\snh^2K_0}{S_0} \right).
\end{equation}
And finally, we can also solve the equation of motion for $T$. The
solution is
\begin{equation}
  \label{T-t}
  T(t) = T_0 + \epsilon(t) \, \frac{ \lam^2 \, F'(M_0) }{2} 
           - \frac{4 G S_0}{1-4GM_0} 
         \arctan\left(\epsilon(t) \, \frac{\snh^2K_0}{S_0} \right)  ,  
\end{equation}
where $T_0$ is yet another integration constant. All together, the
resulting trajectory is parameterized by the gauge function
$\epsilon(t)$, and six integration constants $M_0$, $K_0$, $S_0$,
$T_0$, $R_0$, $\xdir_0$. Again, only four of them are independent. It
follows from the constraint $\tcon=0$, and the definition of $K$ as a
function of $Q$, $S$, and $X$ that
\begin{equation}
  \label{orb-par}
   \snh^2 K_0 = \lam^2 \, F(M_0), \qquad 
   R_0 \, \snh K_0  = |S_0|.
\end{equation}
These relations and all the formulas above reduce to their free
particle counterparts in the limit $G\to0$. What applies to the phase
space structure, also applies to the classical trajectories of the
Kepler system. They reduce to those of the free particles when the
gravitational interaction is switched off. This is also a non-trivial
statement. It implies, for example, that there are no bound states,
again as a consequence of the absence of local gravitational forces.

%%JL revision block begins 
As seen in \sref{free}, the trajectories on the reduced phase space
$\psp$ are obtained from those on $\qsp$ by imposing the gauge
condition $T(t)=t$. We shall now verify that this gauge is always
accessible and unique. By construction, the trajectories on $\psp$
then solve Hamilton's equations with the Hamiltonian defined implictly
by \eref{ham-J}. We have to show is that the right hand side of
\eref{T-t} is a monotonically increasing function of $\epsilon(t)$,
and that its range is the whole real line. If this is the case, then
the equation $T(t)=t$ can always be solved for $\epsilon(t)$. 

That the range is the whole real line can be seen quite easily. In the
range of $M_0$ which is allowed by the positive energy condition,
$F'(M_0)$ is positive. Hence, if $\epsilon(t)$ is large, then the
$\arctan$ term can be neglected as compared to the linear term, which
is unbounded. So, what remains to be shown is that the derivative of
the right hand side of \eref{T-t} with respect to $\epsilon(t)$ is
positive. We know already that, up to a factor of
$2\mul=\dot\epsilon$, this derivative is equal to the right hand side
of \eref{t-T}. This is how it was derived. Hence, we have to show that
for all physical states
\begin{equation}
       \lam^2 \, F'(M) > \frac{8 G}{1-4GM} \, \frac{S^2}{X^2} .  
\end{equation}
The constraint \eref{E-QS} implies that  
\begin{equation}
  \frac{S^2}{X^2} \le \frac{S^2+Q^2}{X^2} = \lam^2 \, F(M). 
\end{equation}
It is therefore sufficient to show that
\begin{equation}
       F'(M) > \frac{8 G}{1-4GM} \, F(M) .  
\end{equation}
This can be simplified further, using the following estimation.  For
$0<\alpha<\pi$ we have $\pi-\alpha>\sin\alpha$. With $\alpha=4\pi GM$,
this implies
\begin{equation}
     1 - 4 G M > \frac{\sin(4\pi G M)}\pi = 4 G \sn M \, \cs M
   \follows 
     \frac{ 8 G }{1-4GM} < \frac{2}{\sn M \cs M}. 
\end{equation}
Hence, it is also sufficient to show
\begin{equation}
       F'(M) > \frac{2F(M)}{\sn M \cs M}  
\end{equation}
Now, all we have to do is to insert the definition \eref{F-def} for
$F$ and compute its derivative. It is then very easy to verify that
this inequality is indeed satisfied, provided that the positive energy
condition holds, that is $\tn M>\sqrt{\tn\mu\tn\mu}$. Thus, we
conclude that a gauge function $\epsilon(t)$ exists, so that
$T(t)=t$, although we do not know it explicitly.

\subsubsection*{Scattering}
Let us now describe the physical properties of the trajectories, in
particular the way they deviate from the free particle trajectories.
Since the relative position $X$ is still an auxiliary variable, we
should first replace it by the actual geodesic distance $R$ between
the particles. And it is also useful to consider the previously
defined canonically conjugate momentum $P$. Using \eref{RP-def}, we
find that on the given trajectories we have
\begin{equation}
  \label{RP-t}
  R(t) = \xsqrt{ R_0\^2 + \ft14 \epsilon^2(t) \snh^2(2 K_0) }, 
     \qquad
  \snh P(t) = \frac{\epsilon(t)\snh^2 K_0}
              {\xsqrt{ R_0\^2 + \epsilon^2(t) \snh^2 K_0 }}.
\end{equation}
Here, we can first of all see that the assumption was correct that $K$
represents the ingoing and outgoing momentum of the particles. If
$R(t)$ is large, then the canonically conjugate momentum $P(t)$
approaches $\pm K_0$. As for the free particles, we expect the
eigenvalues of $K$ to specify the inverse radial wavelength at
infinity, when the system is quantized. 

But now, let us consider the actual trajectories. First we assume that
$S_0\neq0$. According to \eref{orb-par}, this implies that $R_0>0$ and
$K_0>0$. In this case, the particles approach each other from infinity
from some direction $\xinn$, they reach a minimal distance $R_0$ at
the absolute time $T_0$ and with angular orientation $\xdir_0$, and
afterwards the separate again, moving to infinity into a direction
$\xout$. The interpretation of the integration constants $M_0$ and
$S_0$ is also obvious. These are the total energy and the angular
momentum, which are constant along the trajectory. So far, everything
is the same as before. 

However, there is now a crucial difference, because now we have a real
scattering. The angular directions $\xinn$ and $\xout$ are given by
\begin{equation}
  \label{inn-out}
  \xinn = \xdir_0 - \frac{\pi}2 \, \frac{\sgn S_0}{1-4GM_0}, \qquad
  \xout = \xdir_0 + \frac{\pi}2 \, \frac{\sgn S_0}{1-4GM_0}.
\end{equation}
It is no longer so that $\xout=\xinn\pm\pi$. Instead, the scattering
angle is 
\begin{equation}
  \label{scatter}
  \xout = \xinn + \pi \, \frac{\sgn S_0}{1-4GM_0}.
\end{equation}
The scattering angle depends on the deficit angle of the spinning
cone, thus on the energy of the particles, and the direction of the
scattering depends on the sign of the angular momentum, thus on which
side the two particles pass each other. This is not a surprise,
because it is the typical behaviour of a pair geodesics on a cone, and
of course also on a spinning cone.

In fact, the world lines of the particles can effectively be
considered as a pair of geodesics on antipodal sides of the spinning
cone \cite{loumat}. However, the crucial point is that the geometry of
this spinning cone depends on the energy and the angular momentum of
the particles. The particles are not just moving on a fixed background
cone. It is also interesting to note that for large momenta $K_0$,
thus for $M_0\approx\Mmax$, the factor in the denominator in
\eref{scatter} becomes arbitrarily small. In this case, the particles
wind around each other many times before they separate again. We also
see that the scattering angle is always larger than $\pi$. This means,
in a sense, that gravity is attractive, even though there are no local
forces.

The case $S_0=0$ is again somewhat special. According to
\eref{orb-par}, we either have $K_0=0$ or $R_0=0$. For $K_0=0$, we
have a static state. The particles are at rest, with a fixed distance
$R_0>0$, and a fixed angular orientation $\xdir_0$. The geometry of
space is that of \fref{tip}, and it is constant in time. The total
energy is $M_0=\Mmin$, and the integration constant $T_0$ is
redundant. The more interesting case is $S_0=0$ and $R_0=0$. Let us
once again consider the limit $S_0\to0$ and $R_0\to0$, with $K_0>0$
and all other integration constant fixed. What we find is similar to
\eref{rpp-X-t-coll}, 
\begin{equation}
  \label{X-t-coll}
  R(t) =  |\epsilon(t)| \, \frac{\snh(2K_0)}{2} , \qquad
  \xdir(t) = \xdir_0 + 
             \frac{\pi}{2} \, 
             \frac{ \sgn S_0 }{ 1-4 G M_0 } \, \sgn \epsilon(t).
\end{equation}
Again, the particles approach each other on a radial line with
direction $\xinn$, then they hit in a head on collision at the
absolute time $T_0$, and after that they continue on a radial line
with direction $\xout$. However, there is now a crucial difference to
the free particles. The relation between $\xinn$ and $\xout$ is no
longer independent of the sign of $S_0$, and thus the limit is
actually not well defined. We have 
\begin{equation}
  \label{touch}
  \xdir(t) = \cases{ \xinn & for $T<T_0$, \cr
                     \xout & for $T>T_0$,} \qquad 
                    \xout = \xinn \pm \frac{\pi}{1+4GM_0}, 
\end{equation}
which depends on the sign, unless $1/(1-4GM_0)\in\ZZ$. These are very
special cases where the deficit angle of the universe takes a value
in a certain discrete set. 

Clearly, this is the typical behaviour of a geodesic on a cone. If the
geodesic hits the tip, then it does not have a unique continuation. If
it is considered as a limit of a family of geodesics missing the tip,
then its continuation depends on the way we take the limit. To get a
unique time evolution, we have to say explicitly what the particles
have to do when they hit each other. Quite remarkably, this ambiguity
disappears at the quantum level. The following semi-classical argument
explains why. The problem only arises when the angular momentum $S$
vanishes. But then, the angular orientation $\xdir$ is smeared out
uniformly over all directions. Hence, for $S=0$ there is anyway no
well defined scattering angle, because both $\xinn$ and $\xout$ are
maximally uncertain.

Apart from this ambiguity, the classical dynamics of the Kepler system
is well defined on the extended phase space $\qsp$, and the
trajectories are easy to derive. Implicitly, this also defines the
trajectories on the reduced phase space $\psp$. We know that they
exist, and of course they have the same physical properties, but we
cannot write them down explicitly. To go back to the spacetime picture,
we have to perform the construction in \fref{emb} at each moment of
ADM time $t$. What we get is a space that evolves in time, and this
provides a foliation of the spacetime manifold by slices of constant
absolute time in the centre of mass frame. We can either take this
general relativistic point of view, but we may also stick to a
simplified point of view where we just think of two particles moving
in a two dimensional space. 

\section{The quantum Kepler system}
\label{quant}
With all the preparations made in the previous sections, we can now,
finally, quantize the Kepler system. As an ansatz, we shall first try
the Schr\"odinger method, based on the reduced phase space $\psp$. This
turns out to fail, due to a somewhat peculiar feature of the classical
phase space. Its global structure is not as simple as it appears,
which implies that a straightforward choice of a set of commuting
configuration variables does not provide a proper basis of the quantum
Hilbert space. We already noticed that the definition of the Poisson
brackets and the Hamiltonian was technically somewhat involved, and we
were not able, for example, to define the Hamiltonian explicitly as a
functions of a set of canonically conjugate variables. 

This technical problem turns into a principle one at the quantum
level. The way out is to use the Dirac quantization method, based on
the extended phase space $\qsp$. On this phase space, the mass shell
constraint can easily be expressed as a function of a set of
canonically conjugate variables. Instead of solving the Schr\"odinger
equation, we have to impose a constraint. It becomes a generalized
Klein Gordon equation and defines the physical Hilbert space. Just
like the classical phase space, the quantum Hilbert space turns out to
be a deformed version of the free particle Hilbert space. Finally, we
are going to write down the energy eigenstates explicitly as wave
functions in space and time, and from them we read off some
interesting effects of quantum gravity.

\subsubsection*{Schr\"odinger approach}
The most straightforward way to quantize the Kepler system seems to be
a Schr\"odinger quantization, based on the completely reduced phase
space $\psp$ defined in \eref{psp}, and spanned by the variables $Q$,
$S$, $X$, and $\xdir$. Apparently, the only technical problem is to
define the Hamiltonian $\pham$ explicitly as a quantum operator, which
is given by \eref{ham-red}. However, a first obstacle already arises
when we look at the symplectic potential \eref{pot-red}, which
explicitly depends on this function,
\begin{equation}
  \label{s-pot}
  \ppot = X^{-1} Q \, \dd X + (1-4G\pham) \, S \, \dd \xdir.
\end{equation}
This we have to take into account when we set up the operator
representation. We saw, however, that we may equally well replace the
variable $S$ by 
\begin{equation}
  \label{J-def}
  J = (1-4G\pham) \,  S \follows
  \ppot = X^{-1} Q \, \dd X + J \, \dd \xdir, 
\end{equation}
and consequently we have the canonical brackets
\begin{equation}
  \label{s-pois}
  \pois QX = X , \qquad 
  \pois J\xdir = 1.
\end{equation}
Now the quantization is straightforward. We choose a basis where $X$
and $\xdir$ are diagonal. The wave functions is $\psi(x,\ph)$, with
$x\ge0$, and $\ph$ has a period of $2\pi$ if the particles are
distinguishable, or the period is $\pi$ if the particles are
identical. We introduce a statistics parameter $\stat$, in the same
way as previously for the free particles. We may also define the same
scalar product,
\begin{equation}
  \label{s-prod}
  \braket{\psi_1}{\psi_2} = 
    \int x \, \dd x \, \dd \ph \, \bar\psi_1(x,\ph) \, \psi_2(x,\ph),
\end{equation}
which implies that $\psi(x,\ph)$ is the usual probability amplitude in
polar coordinates. And it is also straightforward to define the
operator representation, which is given by
\begin{equation}
  \label{s-basis}
  \op{X}     \, \psi(x,\ph) = x   \, \psi(x,\ph), \qquad 
  \op{\xdir} \, \psi(x,\ph) = \ph \, \psi(x,\ph),
\end{equation}
for the position operators, and the momentum operators are
\begin{equation}
  \label{s-QJ}
  \op Q \, \, \psi(x,\ph) 
     = - \ii\hbar \, \deldel x \, x \, \psi(x,\ph) , \qquad 
  \op J \, \, \psi(x,\ph) 
     = - \ii\hbar \, \deldel\ph \, \psi(x,\ph). 
\end{equation}
So far, everything is the same as before. It seems that the only
technical problem is to express the Hamiltonian $\pham$ in terms of
$X$, $Q$, and $J$. It is now implicitly defined by the equation
\eref{ham-J}, thus
\begin{equation}
  \label{s-ham}
  \frac{Q^2 + (1-4G\pham)^{-2} J^2}{X^2} = \lam^2 F(\pham).
\end{equation}
Since we could not even solve this equation explicitly at the classical
level, it is of course also impossible to derive an explicit operator
representation for $\pham$ at the quantum level. Nevertheless, since
we know that the Hamiltonian is well defined on the classical phase
space, this only seems to be a technical problem. But if we look at
all this more closely, we find that the problem is actually not a
technical, but a principle one. In fact, we made a wrong assumption in
the above derivation.

To set up a quantum representation like the one given above, it is not
sufficient to pick out a complete set of commuting phase space
variables like $X$ and $\xdir$, and to define the wave function
$\psi(x,\ph)$ to be a function on this \emph{configuration space}.
There is a second consistency condition, which is sometimes overlooked
because it is usually immediately obvious. The classical phase space
has to be the \emph{cotangent bundle} of the configuration space,
equipped with its canonical symplectic structure. At each point
$(X,\xdir)$ of the configuration space, the canonically conjugate
momenta, which are $(Q/X,J)$ in our case, have to span a vector space.
Splitting the phase space into a configuration and a momentum space
like this is called a \emph{polarization} \cite{woodhs}.

Without going into any details, let us roughly show what goes wrong
when the second criterion is not satisfied. Given a Hilbert space such
that to each point $x$ in the configuration space there corresponds a
basis state $\psi_x$, then for every bounded function $F$ on the
configuration space there exists a bounded self adjoint operator,
whose eigenstates are $\psi_x$ with eigenvalues $F(x)$. This is just
usual quantum mechanics, for example in the position representation.
Consequently, there is also a family of unitary operators
$\expo{\ii\tau F/\hbar}$, representing the flow of $F$. On the
classical phase space, this flow is a translation in the conjugate
momentum space. Therefore, the whole construction is consistent only
if the momentum space at each point $x$ of the configuration space is
at least an affine space. It can be made a vector space by choosing
some origin.

Now, coming back to the Kepler system, let us check whether the second
criterion is satisfied or not. We have to find out whether for every
fixed point in the configuration space $(X,\xdir)$, the canonically
conjugate momenta $(Q/X,J)$ span a vector space. Unfortunately, this
is not the case. To see this, consider the following estimation,
\begin{equation}
  \frac{J^2}{X^2} = (1-4G\pham)^2 \, \frac{S^2}{X^2}
                  \le (1-4G\pham)^2 \, \frac{S^2+Q^2}{X^2}
                   = (1-4G\pham)^2 \, \lam^2 \, F(\pham). 
\end{equation}
The first factor can be estimated further by
\begin{equation}
   1 - 4 G \pham < \frac1{\pi^2 G \tn \pham},
\end{equation}
which follows from $\cot\alpha>(\pi/2-\alpha)$ for $0<\alpha<\pi/2$,
and with $\alpha=2\pi G\pham$.  Using this, we find that
\begin{equation}
  \frac{J^2}{X^2} < \frac{\lam^2 \, F(\pham)}
                    {\pi^4 G^2 \, \tn^2\pham} 
                  < \frac{\lam^2}{4\pi^4G^2}.
\end{equation}
The last inequality follows from the following property of the
function $F$ defined in \eref{F-def}. For
$\tn\pham\ge\tn\mu\ge\tn\nu$, which follows from the positive energy
condition, we have
\begin{equation}
  \frac{F(\pham)}{\tn^2\pham} = 
      \frac{(\tn^2\pham-\tn^2\mu) \, (\tn^2\pham-\tn^2\nu)}
           {4\tn^4\pham}
       < \frac14.
\end{equation}
Hence, all together it follows that
\begin{equation}
  \label{J-bound}
   |J| < \Jmax, \txt{where} \Jmax = \frac{\lam}{2\pi^2G} \, X.
\end{equation}
The canonical angular momentum $J$ is restricted to some interval
around zero, whose size is determined by the configuration variable
$X$. Thus, at a fixed point $(X,\xdir)$ of the configuration space,
the canonical momenta $(Q/X,J)$ do \emph{not} span a vector space. We
have to conclude that the above definition of a position
representation is not the correct quantization of the given
classical phase space.

Of course, we could now argue that we simply picked out the wrong
configuration space. We just have to find an appropriate configuration
space to perform the correct quantization. But then we would lose the
simple physical interpretation of the configuration variables, and
consequently the direct interpretation of the wave function as a
probability amplitude in the usual sense. Therefore, to avoid this
unnecessary complication, we shall not try to quantize the Kepler
system bases on the reduced phase space $\psp$

\subsubsection*{Dirac approach}
The whole problem can be circumvented when we use the Dirac method
instead. The extended phase space $\qsp$ defined in \eref{qsp} has the
additional independent variables $M$ and $T$. It is a proper cotangent
bundle with its canonical symplectic potential \eref{pot-ext},
\begin{equation}
  \label{d-pot}
  \qpot =  X^{-1} Q \, \dd X  + (1-4GM) \, S \, \dd\xdir - M \, \dd T . 
\end{equation}
The configuration space is spanned by $(T,X,\xdir)$, and the
canonically conjugate momenta $(-M,Q/X,(1-4GM)S)$ have an unrestricted
range for fixed $T$, $X$, and $\xdir$. The quantization is
straightforward. We have a wave function $\psi(\tau,x,\ph)$, with
$x\ge0$, and regarding the periodicity in $\ph$ we have the same
relations as before. If the particles are distinguishable, we have
\begin{equation}
  \label{stat}
  \psi(\tau,x,\ph+2\pi) = \expo{2\pi\ii\stat} \, \psi(\tau,x,\ph), 
\end{equation}
and for identical particles the stronger relation is 
\begin{equation}
  \label{stat-id}
  \psi(\tau,x,\ph+\pi) = \expo{\pi\ii\stat} \, \psi(\tau,x,\ph), 
\end{equation}
where the statistics parameter $\stat$ is a fixed real number. The
position operators are given by
\begin{equation}
  \label{op-X}
  \op X \, \psi(\tau,x,\ph) = x \, \psi(\tau,x,\ph) , \qquad
  \op\xdir \, \psi(\tau,x,\ph) = \ph \, \psi(\tau,x,\ph),
\end{equation}
where the last equation is again to be understood with $\xdir$
replaced by a periodic function thereof. For the clock and the
canonically conjugate energy we have 
\begin{equation}
  \label{op-TM}
  \op T \, \psi(\tau,x,\ph) = \tau \, \psi(\tau,x,\ph) , \qquad 
  \op M \, \psi(\tau,x,\ph) = 
    \ii \hbar \, \deldel \tau \, \psi(\tau,x,\ph).
\end{equation}
And finally, we need the momentum operators $Q$ and $S$. As $S$ can
be expressed as a simple function of $M$ and $J$, let us first define
\begin{equation}
  \label{op-QJ}
  \op Q \, \psi(\tau,x,\ph) = 
    - \ii \hbar \, \deldel x \, x \, \psi(\tau,x,\ph), \qquad
  \op J \, \psi(\tau,x,\ph) =
    - \ii \hbar \, \deldel \ph \, \psi(\tau,x,\ph) .  
\end{equation}
The operator for $S$ can then be expressed as
\begin{equation}
  \label{op-S}
  \op S \, \psi(\tau,x,\ph) = 
    - \ii \hbar \, \left(1 - 4 \ii \lpl \, \deldel \tau \right)^{-1} 
       \, \deldel \ph \, \psi(\tau,x,\ph).
\end{equation}
The constant $\lpl=G\hbar$ is the \emph{Planck length}, which is here
actually interpreted as a \emph{Planck time}. So, we have a somewhat
unusual operator representation for the angular momentum $S$, which is
non-local in $\tau$. But this is not a serious problem. Finally, we
have the following scalar product, with respect to which all these
operators are self adjoint, and which is formally the same as
\eref{rpp-prod-dir},
\begin{equation}
  \label{prod-dir}
  \braket{\psi_1}{\psi_2} = \int  \dd\tau \, x \, \dd x \, \dd \ph
               \, \bar\psi_1(\tau,x,\ph) \, \psi_2(\tau,x,\ph). 
\end{equation}
This completes the definition of the extended Hilbert space. Now we
have to impose the mass shell constraint to define the physical
Hilbert space.  The constraint is given by \eref{E-QS}, so we should
first try to diagonalize the operator
\begin{equation}
  \label{op-K}
  \snh^2 \op K 
      =  \op X^{-1} \left( \op Q^2 + \op S^2 \right) \op X^{-1}.
\end{equation}
It acts on a wave function as 
\begin{equation}
  \label{op-K-basis}
  \snh^2 \op K \, \psi(\tau,x,\ph) = 
   - \hbar^2 \lap \, \psi(\tau,x,\ph), 
\end{equation}
where $\lap$ is a deformed Laplacian,
\begin{equation}
  \label{lap}
    \lap = x^{-1} \deldel x \, x \, \deldel x 
         + x^{-2} \, \left( 1 - 4 \ii \lpl \, \deldel \tau \right)^{-2}
                  \, \deldel\ph \, \deldel\ph.  
\end{equation}
Now, suppose that the operator $1-4GM$ in parenthesis is also
diagonal. Then $\lap$ is the Laplacian on a cone with a deficit angle
of $8\pi GM$. This is not surprising. We found that the classical
particles are effectively moving on such a cone. The eigenfunctions
are still Bessel function, however with slightly modified indices and
arguments. The normalized eigenstates are parameterized by three
quantum numbers $\mm$, $k$, and $s$, and they are explicitly given by
\begin{equation}
  \label{chi-K-X}
  \chi(\mm,k,s;\tau,x,\ph) =
     \xsqrt{\frac{\snh(2\hbar k)}{4\pi \, \hbar \, k}} 
     \, \expo{-\ii\mm\tau} \, \expo{\ii s\ph} \,  
     J_{\sfrac{|s|}{1-4\lpl\mm}}
            \left(\frac{\snh(\hbar k)}\hbar \, x \right).  
\end{equation}
This reduces to \eref{rpp-chi-dir} in the limit $G\to0$, which implies
$\lpl\to0$ and $\snh\vrh\to\vrh$. We are here imposing the same
regularity condition at $x=0$ that we also imposed in \sref{free}.
Following the arguments in \cite{bousor}, we require the wave function
to stay finite. This makes \eref{op-K-basis} a well defined self
adjoint operator. The eigenstates are normalized in the same way as
\eref{rpp-chi-norm-dir}, thus
\begin{eqnarray}
  \label{chi-norm}
  \int \dd\tau \, x \, \dd x \, \dd\ph \, 
    \bar\chi(\mm_1,k_1,s_1;\tau,x,\ph) 
     \, \chi(\mm_2,k_2,s_2;\tau,x,\ph)  = \qquad & &  \nwl 
   2\pi \, \delta(\mm_2-\mm_1) \, 
   k^{-1} \, \delta(k_2-k_1) \, \delta_{s_2-s_1}.&& 
\end{eqnarray}
The quantum numbers $k>0$ and $\mm$ are continuous, and $s$ is
discrete. It takes the values $s\in\ZZ+\stat$ or $s\in2\ZZ+\stat$,
depending on whether the particles are distinguishable or not. This
follows again from the periodicity condition imposed on the wave
function. Taking this into account when summing over $s$, we have the
following completeness relation for the eigenstates,
\begin{eqnarray}
  \label{chi-complete}
  \sum_s \int \dd\mm \,  k \, \dd k \, 
     \bar\chi(\mm,k,s;\tau_1,x_1,\ph_1) 
       \, \chi(\mm,k,s;\tau_2,x_2,\ph_2)
     = \qquad && \nwl 2\pi \, \delta(\tau_2-\tau_1) \, 
       x^{-1} \delta(x_2-x_1) \, \delta_\stat(\ph_2-\ph_1), &&                 
\end{eqnarray}
where $\delta_\stat$ is again the periodic delta function satisfying
the \eref{stat} or \eref{stat-id}. So, we now have a basis of the
extended Hilbert space, where the energy and momentum operators are
diagonal, with eigenvalues
\begin{equation}
  \label{eigen-KMJS}
   K   = \hbar k    , \qquad
   M   = \hbar \mm  , \qquad
   J   = \hbar s    , \qquad
   S   = \frac{\hbar s}{1-4\lpl\mm}.
\end{equation}
We see that the spectrum of $J$ is the one which is quantized in steps
of $\hbar$, like the angular momentum of the free particles in
\eref{rpp-eigen-MKS}. This is reasonable. On the classical phase space
it is the charge $J$ which generates a rotation with closed orbits,
and a period of $2\pi$. The spectrum of $S$ is also discrete, but the
eigenvalues additionally depend on the eigenvalues of $M$. The closer
the energy gets to the maximal physical energy $\Mmax=\Mpl/4=1/4G$,
the larger the steps are between the eigenvalues of $S$. We shall
discuss this spectrum and its physical implications at the very end of
this section.

It is now straightforward to impose the mass shell constraint. We can
simply repeat all the steps from \sref{free}. The constraint acts on
the energy momentum eigenstates as
\begin{equation}
  \label{tcon-op}
  \op \tcon \, \chi(\mm,k,s;\tau,x,\ph) = 
  \left( \snh^2(\hbar k) - \lam^2 F(\hbar\mm) \right) \, 
   \chi(\mm,k,s;\tau,x,\ph) .
\end{equation}
It is again useful to introduce a \emph{dispersion relation}
\begin{equation}
  \label{om-k}
  \mm(k) = \hbar^{-1} \, 
         F^{-1}\left( \frac{\snh^2(\hbar k)}{\lam^2} \right).
\end{equation}
It is the quantum version of the classical relation between the
momentum $K$ and the energy $M$, which is shown in \fref{dsp}. We
conclude that the physical states are those energy momentum
eigenstates, where the quantum numbers satisfy $\mm=\mm(k)$. These
states are annihilated by the constraint \eref{tcon-op}. The most
general physical state is a superposition
\begin{equation}
  \label{dir-wave}
   \psi(\tau,x,\ph) = \sum_s \int k \, \dd k  \, 
             \psi(k,s) \, \chi(\mm(k),k,s;\tau,x,\ph), 
\end{equation}
where $\psi(k,s)$ is again the wave function in momentum space. Let us
once again split off the dependence on the radial coordinate, writing 
\begin{equation}
  \label{wave-rad}
   \psi(\tau,x,\ph) = \sum_s \int k \, \dd k  \, 
          \expo{-\ii\mm(k)\tau} \, \expo{\ii s\ph} \, 
             \psi(k,s) \, \zeta(k,s;x).
\end{equation}
The radial wave function is then given by  
\begin{equation}
  \label{zeta-x}
  \zeta(k,s;x) = \frac1{4\pi} \,  
     \xsqrt{\frac{\sinh(8\pi\lpl k)}{\lpl k}} \, \, \, 
     J_{\sfrac{|s|}{1-4\lpl\mm(k)}}
       \left( \frac{\sinh(4\pi\lpl k)}{4\pi\lpl} \, x \right).  
\end{equation}
Here we replaced the rescaled hyperbolic functions by the usual ones,
to see that only the Planck length $\lpl$ appears as a dimensionful
constant. In the limit $\lpl\to0$, we still recover the free particle
expression \eref{rpp-zeta}.

\subsubsection*{Position representation}
Before we read off any physical information from the wave functions,
we have to transform to a representation where the geodesic distance
$R$ between the particles is diagonal, and not the auxiliary variable
$X$. So, the task is to find the eigenstates of $R$ and then to
transform the wave functions into this representation. According to
\eref{R-QX}, we have $R^2=X^2+(4\pi GQ)^2$, thus 
\begin{equation}
  \label{op-R}
  \op R^2 \, \psi(\tau,x,\ph) = 
      \left( x^2 - (4\pi \lpl)^2 \deldel x \, x \, \deldel x \, x \right) 
         \, \psi(\tau,x,\ph). 
\end{equation}
The eigenfunctions of this operator are modified Bessel functions of
the second kind, 
\begin{equation}
  \label{chi-R-X}
  \eta(r;x) = \frac{\xsqrt{\sinh(r/4\lpl)}}{2\sqrt2 \pi^2 \lpl\, x} \, 
          K_{\sfrac{\ii r}{4\pi\lpl}}\left(\frac{x}{4\pi\lpl}\right)
   \follows \op R \, \eta(r;x) = r \, \eta(r;x).            
\end{equation}
The eigenvalues $r$ are positive, as it should be, and continuous. The
eigenstates are orthonormal,
\begin{equation}
  \int x \, \dd x \, \eta(r_1;x) \, \eta(r_2;x) =
                            r^{-1} \delta(r_2-r_1),
\end{equation}
and complete
\begin{equation}
  \int r \, \dd r \, \eta(r;x_1) \, \eta(r;x_2) =
                            x^{-1} \delta(x_2-x_1).
\end{equation}
The transformation of a wave function from the $X$-representation to
the $R$-representation is given by
\begin{equation}
  \label{X-to-R}
  \psi(\tau,r,\ph)  = 
   \int x \, \dd x \, \eta(r;x) \, \psi(\tau,x,\ph). 
\end{equation}
For simplicity, we use the same symbol for the wave function in both
representations. This transformation is known as the Kontorovich
Lebedev transform \cite{tables}. The scalar product in the
$R$-representation becomes
\begin{equation}
  \label{prod-R}
  \braket{\psi_1}{\psi_2} = \int  \dd\tau \, r \, \dd r \, \dd \ph
               \, \bar\psi_1(\tau,r,\ph) \, \psi_2(\tau,r,\ph), 
\end{equation}
so that the transformed wave function $\psi(\tau,r,\ph)$ represents
the correct probability amplitude in the relative position space of
the particles. We can finally write down the most general physical
state as a wave function in the position representation. All we have
to do is to transform the radial wave functions \eref{zeta-x} from the
$X$-representation to the $R$-representation, thus
\begin{equation}
  \label{zeta-trans}
   \zeta(k,s;r) = \int x \, \dd x \, 
                  \eta(r;x) \, \zeta(k,s;x). 
\end{equation}
This convolution can be carried out explicitly, and the result can be
expressed in terms of a hypergeometric function \cite{math},
\begin{eqnarray}
  \label{zeta-r}
  \zeta(k,s;r) &=&  
      \frac{\Gamma(A) \, \Gamma(\bar A)}{\Gamma(A+\bar A)} \,
      \frac{ \xsqrt{ \sinh(r/4\lpl) \sinh(8\pi\lpl k)} }
           { (2\pi)^2 \, \sqrt{ 2\lpl\,k } } \, \times \nwl 
     {} && \qquad  \phantom\int {} \times
       \sinh^{A+\bar A-1}(4\pi\lpl k) \,\, 
       {}_2F_1\left(A ,\bar A;A+\bar A; - \sinh^2(4\pi\lpl k) \right), 
\end{eqnarray}
where 
\begin{equation}
  A = \frac12 \left( 1 + \frac{|s|}{1-4\lpl\mm(k)}  
                + \frac{\ii r}{4\pi\lpl} \right), \quad
  \bar A = \frac12 \left( 1 + \frac{|s|}{1-4\lpl\mm(k)}  
                - \frac{\ii r}{4\pi\lpl} \right). 
\end{equation}
So, we finally have a rather complicated but explicit representation
of the physical energy eigenstates as wave functions in the relative
position space of the particles. The operators for $R$ and $\xdir$,
representing the distance and the orientation of the particles, are
diagonal in this representation. It is therefore possible to read off
the physical properties of the states directly from the wave
functions.

\subsubsection*{Wave functions}
We shall in the following study the radial wave functions
$\zeta(k,s;r)$, representing the physical states with eigenvalues
\begin{equation}
  \label{wave-eigen}
  M = \hbar \mm(k), \qquad
  K = \hbar k , \qquad
  J = \hbar s , \qquad
  S = \frac{\hbar s}{1-4\lpl\mm(k)}.
\end{equation}
We shall compare them to the corresponding radial wave functions
$\zeta(k,s;r)$ of the free particle system, as given by
\eref{rpp-zeta}. There we may simply identify the radial coordinate
$x$ with $r$, as in the limit $G\to0$ the transformation from the
$X$-representation to the $R$-representation is trivial. In both
cases, the physical Hilbert space is spanned by a two parameter family
of energy momentum eigenstates, and the quantum numbers $k$ and $s$
have the same interpretation. The former is the inverse radial
wavelength at infinity, representing the momentum of the incoming and
outgoing particles, and the latter is the inverse angular wavelength,
representing the angular momentum.

Actually, we haven't check this yet for the Kepler system. For the
free particles, it is a well know feature of the Bessel function
\eref{rpp-zeta}. For large $r$ it falls off with $\sqrt{r}$,
oscillating with a wavelength of $2\pi/k$. It is not at all obvious
that the function \eref{zeta-r} has the same property. But we can use
the following semi-classical argument. We have seen in \eref{RP-t}
that for the classical trajectories, the canonically conjugate
momentum $P$ of the distance $R$ is equal to $\pm K$ for large $R$.
Therefore, we expect the eigenfunction of $K$ with eigenvalue $\hbar
k$ to be a superposition of an ingoing and an outgoing radial wave,
with wavelength $2\pi/k$ for large $r$. This argument holds for both
the free particles, where the corresponding classical relation is
\eref{rpp-RP-t}, and for the coupled particles.

An alternative way to see this is as follows. This argument can be
made rigorous by taking into account the correct operator ordering.
But for simplicity let us stick to a semi-classical level. Consider
the eigenvalue equation for $K$ in the $R$-representation. As a phase
space function, we have
\begin{equation}
  \label{K-RP}
  \snh^2 K = \frac{Q^2+S^2}{X^2}  
           = \frac{R^2 \, \tnh^2 P + S^2}{R^2/{\csh^2P}}
           = \snh^2 P + \frac{S^2}{R^2} \, \csh^2 P. 
\end{equation}
Hence, for large $R$ we have $P\approx\pm K$. Since $P$ is canonically
conjugate to $R$, and thus represented by the operator
$-\ii\hbar\del/\del r$, it follows that an eigenfunction of $K$ with
eigenvalue $\hbar k$ has a radial wavelength of $2\pi/k$ for large
$r$. At this point, we see what the advantage of the auxiliary
variable $X$ was. Of course, we could have used $R$ and $P$ as phase
space variables from the very beginning. Then, however, we had to
solve the eigenvalue equation for \eref{K-RP}, with the appropriate
operator ordering. Obviously, this is much more difficult than solving
the eigenvalue equation for \eref{op-K}, because the hyperbolic
functions become a non-local operators in the complex $r$-plane.

But now, let us come to the actual physical questions. We wanted to
find out what the basic differences are between the radial wave
functions $\zeta(k,s;r)$ for the free particles and those for the
coupled particles. So far, we know that both represent a scattering
state, where the incoming and outgoing particles have a radial
wavelength of $2\pi/k$, and a fixed angular momentum. In other words,
the wave functions are almost equal for large $r$, where large means
some orders of magnitude above the Planck length $\lpl$. The
interesting question is what happens at small distances. To see this,
we have plotted some typical wave functions, for different values of
$k$ and $s$.

In each of the following figures, the value of $s$ is fixed, and that
of $k$ varies. The upper part of the figure always shows the free
particle wave functions, the lower one those for the Kepler system,
both as a function of the distance $r$, which is measured in units of
the Planck length $\lpl$. To distinguish the wave functions for
different momenta $k$, we use the following rule. The larger $k$ is,
the smaller are the gaps in the curves. The broken curve with the
largest gaps represents the smallest momentum, and a solid curve
corresponds to the largest momentum in the figure. The quantum numbers
$k$ are given in units of $1/\lpl$, so that $k=1$ corresponds to a
radial wavelength of $2\pi\lpl$ at infinity, and to a momentum in
physical units of $1/G=\Mpl$.

The mass parameters $m_k$ are given in units of the Plank mass $\Mpl$.
They only have a marginal influence on the qualitative behaviour of
the wave functions. We therefore keep them fixed in the following, and
we choose two particles with the same mass $m_1=m_2=0.02\Mpl$, thus
still far below the maximally allowed rest mass. For simplicity, we
shall also restrict to states with integer $s$, thus particles with
conventional statistics. There are no principally different effects
occurring for anyons, only the numerical values of some quantities
considered below are slightly different. We shall briefly comment on
this at the appropriate place.
\begin{figure}[t]
  \begin{center}
    \epsfbox{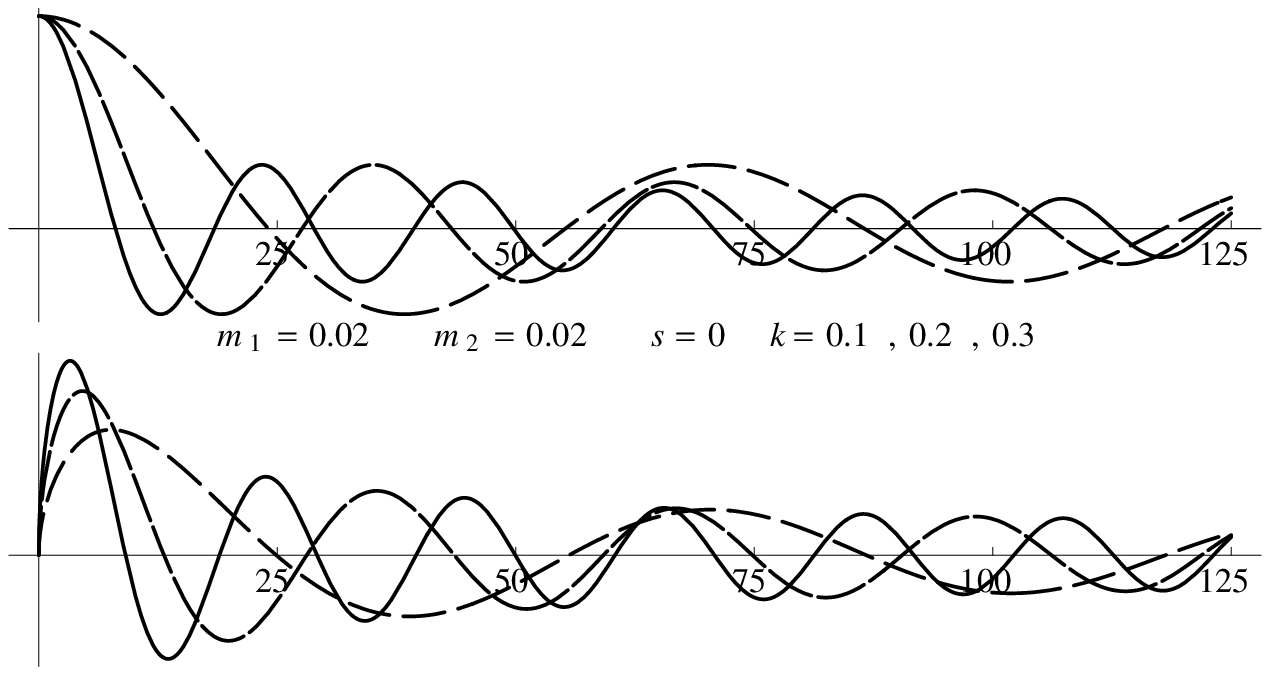}
    \xcaption{The wave functions for $s=0$. The coupled particles
    avoid being at the same point in space, but otherwise the wave
    functions are almost equal.}
    \label{wav0}
  \end{center}
\end{figure}

As a first example, consider the states with $s=0$ in \fref{wav0}. In
the upper part we see the Bessel functions $J_0(kr)$. They take a
maximum at $r=0$, and fall off with the square root of $r$ for large
$r$, oscillating with a wavelength of $2\pi/k$. Except for a small
region near the origin, which is of the order of ten Planck lengths,
the wave functions of the Kepler system are almost the same. For a
vanishing angular momentum, the coupled particles behave almost like
the free particles. There is not even a phase shift between the
ingoing and outgoing particles, so that far away from the origin the
interaction of the particles cannot even be detected. However, the
wave function of the coupled particles goes to zero at $r=0$, so that
the probability to find the two particles at the same point in space
vanishes.
\begin{figure}[t]
  \begin{center}
    \epsfbox{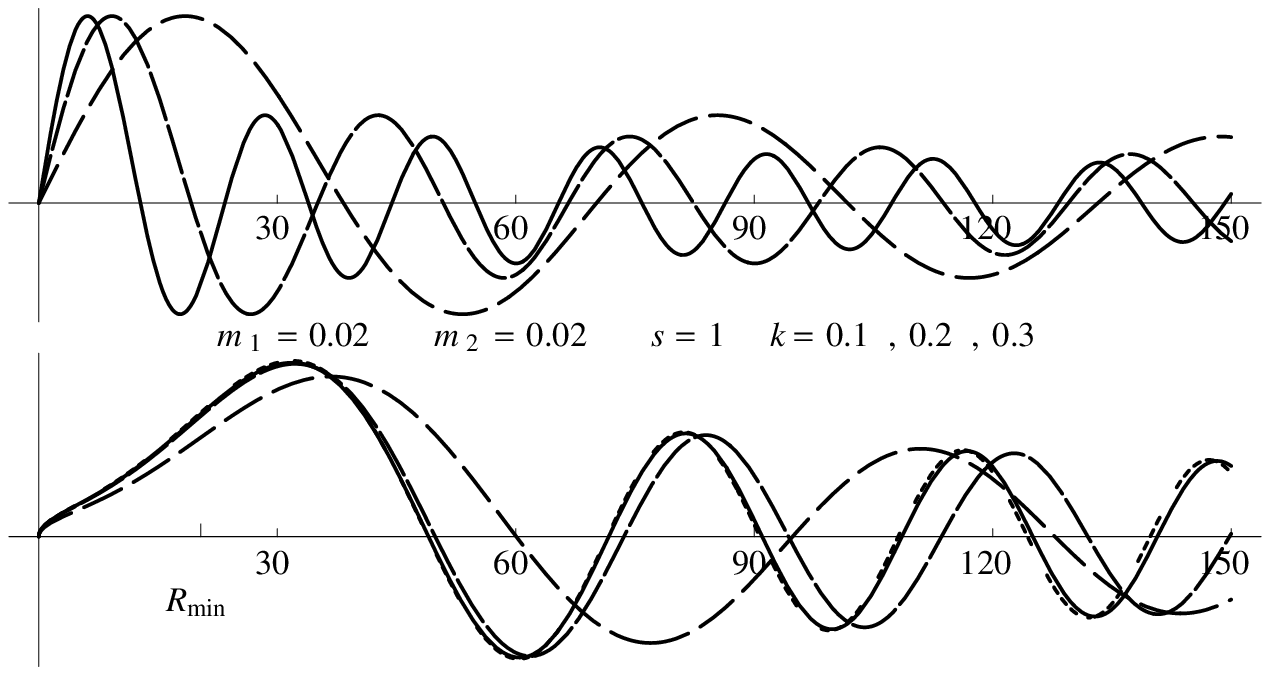}
    \xcaption{The wave functions for $s=1$. The free particles can still
    get arbitrarily close to the origin, whereas the coupled particles
    alway stay beyond the minimal distance $\Rmin$.}
  \label{wav1}
  \end{center}
\end{figure}

\begin{figure}[t]
  \begin{center}
    \epsfbox{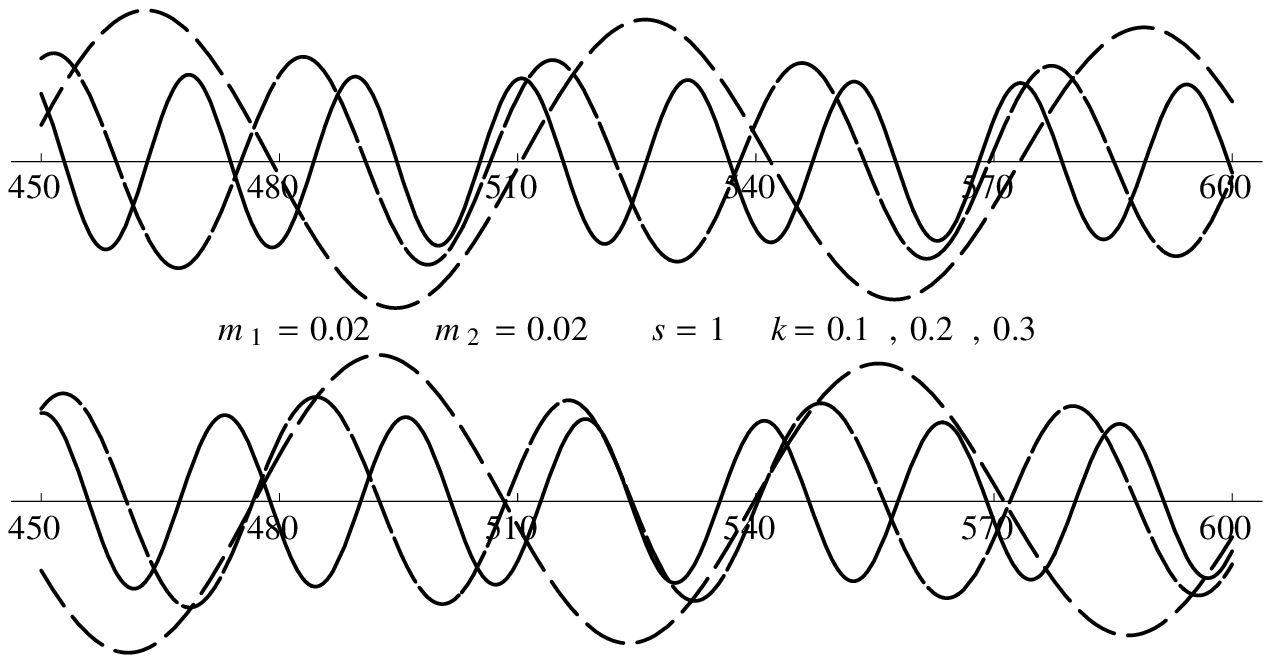}
    \xcaption{For large distances from the origin, the wave functions
    of the free and the coupled system are always the same, except for
    a phase shift between the ingoing and outgoing wave.}
  \label{wav2}
  \end{center}
\end{figure}

Obviously, the coupled particles try to avoid being both in a region
of space which is of the order of the Planck scale. This becomes even
more obvious when we look at the states with $s=1$ in \fref{wav1}.
For the free particles, we now have the Bessel functions $J_1(kr)$.
They are zero at $r=0$, and the first maximum is approximately at
$r\approx1/k$. For large $r$, they still fall off with the square root
of $r$, oscillating with a wavelength of $2\pi/k$. Apart from a
different phase of the oscillation, this is also the large $r$
behaviour of the wave functions of the Kepler system. To see this more
clearly, we have plotted the same wave functions for large distances
in \fref{wav2}. The phase shift indicates that an interaction between
the ingoing and outgoing particles has taken place. 

The more interesting feature is again the behaviour of the wave
function for small $r$. For the free particles, the maximum of the
wave function is getting closer to the origin with increasing momentum
$k$. For the coupled particles however, the maximum never gets beyond
a certain minimal distance $\Rmin$. In fact, we already found such a
minimal distance at the classical level. Remember that the quantum
number $s$ specifies the eigenvalue $\hbar s$ of the canonical angular
momentum $J$. On the other hand, we have seen that the classical phase
space function $J$ is restricted by \eref{J-bound}. We used this to
show that a Schr\"odinger quantization, in a representation where $X$
and $\xdir$ are diagonal, is not possible.

Turning the argument around, it follows that for a fixed $J$, there is
a minimal $X$, and thus also a minimal $R$, since $R^2=X^2+(4\pi
GQ)^2\ge X^2$. Consequently, we have the semi-classical relation
\begin{equation}
  \label{R-min}
  R > \Rmin(s) = \frac{2\pi^2 \lpl}{\lam} \, |s|. 
\end{equation}
It follows directly from \eref{J-bound} with $J=\hbar s$. For a fixed
angular momentum, there is thus a minimal distance between the
particles. It is of the order of the Planck length $\lpl$ if the
angular momentum is of the order of $\hbar$, and it increases linearly
with the angular momentum. The numerical factor in \eref{R-min}
depends on the masses of the particles, defining the number
$0<\lam\le1$ in \eref{F-def}. We see in \fref{wav1} that beyond the
point $\Rmin$ the wave function is no longer oscillating, but falls
off exponentially. In the classically forbidden region, the wave
function shows the typical tunneling behaviour.

Another peculiar observation is the following. For large momenta $k$,
the wave function of the coupled particles \emph{freezes} at small
distances $r$. More precisely, for small $r$ and large $k$ the radial
wave function $\zeta(k,s;r)$ no longer depends on $k$. The limiting
wave function is shown as the dotted curve in \fref{wav1}. This is
quite strange, because it means that the short distance behaviour of
the wave function will not change anymore with increasing momentum,
whereas for large distances we still have an oscillation with a
wavelength of $2\pi/k$. One can show that the limiting wave function
is $\eta(r;\Rmin)$, as defined in \eref{chi-R-X}. It is the
$R$-representation of the eigenstate of $X$ with eigenvalue $x=\Rmin$.
Hence, for small $r$ and large $k$ we have
\begin{equation}
  \label{freeze}
  \zeta_{\mm(k)}(k,s;r,\ph) \approx \expo{\ii s\ph} \, \chi(r;\Rmin),  
\end{equation}
with $\Rmin$ given by \eref{R-min}. Let us only briefly sketch how
this can be proven. If we return from the $R$-representation to the
$X$-representation, then the radial wave functions are given by the
Bessel functions \eref{zeta-x}. In the limit $k\to\infty$, one can
show that these functions provide an approximation of the delta
function $\delta(x-\Rmin)$. Transforming again to the
$R$-Representation, this implies \eref{freeze}. But let us not go into
any details of the proof here. Numerically it can be verified easily
that \eref{freeze} is in fact a good approximation.

What does it mean physically that the wave function freezes at small
$r$, and that the first maximum never goes beyond the point $r=\Rmin$?
Consider again the scattering process described by these wave
functions. In case of the free particles, we can make the following
statement. The larger we make the momentum of the incoming particles,
the closer the particles get to each other. In other words, to probe
small distances we need large momenta. This is a well known rule in
elementary particle physics or actually quantum physics in general. It
is one way to express Heisenberg's uncertainty relation. However, this
statement is no longer true when gravity is switched on. We can make
the momentum as large as we like, we never get beyond the minimal
distance $\Rmin$.  Apparently, there is some repulsion that keeps the
particles apart.

To see that this is a real \emph{quantum gravity} effect, observe that
$\Rmin$ is of the order of the Planck length, which involves both $G$
and $\hbar$. Hence, the effect disappears both if we switch off
gravity, as then we have the wave functions of the free particles, and
it also disappears in the classical limit. There is no restriction on
the integration constant $R_0$ for the classical trajectories in
\sref{class}, thus the particles can reach any arbitrarily small
distance. But what does it mean that the minimal distance $\Rmin$
depends on the quantum number $s$? First of all, one could argue that
in order to bring the particles closer together we just have to
consider states with smaller $s$. And in fact, we have seen in
\fref{wav0}, that no such minimal distance exists for $s=0$, in
agreement with \eref{R-min}.

But now assume that the particles are two identical fermions with
masses $m_1=m_2=m$. Then we have $s\in1+2\ZZ$, and consequently the
minimal value of $|s|$ is $1$. In this case, we have
\begin{equation}
  \label{R-min-abs}
  R > \Rmin = \frac{2\pi^2 \lpl}{\lam}
            = \frac{2\pi^2 \lpl}{\cos(4\pi Gm)}, 
\end{equation}
and this is an \emph{absolute} minimum for the distance, which can
never be reached by any wave function. This absolute minimum only
depends on the mass $m$ of the particles. For $m=0$ we have
$\Rmin=2\pi^2\lpl\approx20\lpl$. With increasing mass it increases
unboundedly. For $m\to\Mpl/8$, which is the maximal mass of two
identical particles, the cosine goes to zero. For anyons the absolute
minimum is smaller, because the minimal $|s|$ lies between zero and
one, and for bosons this curious effect disappears.

\begin{figure}[t]
  \begin{center}
    \epsfbox{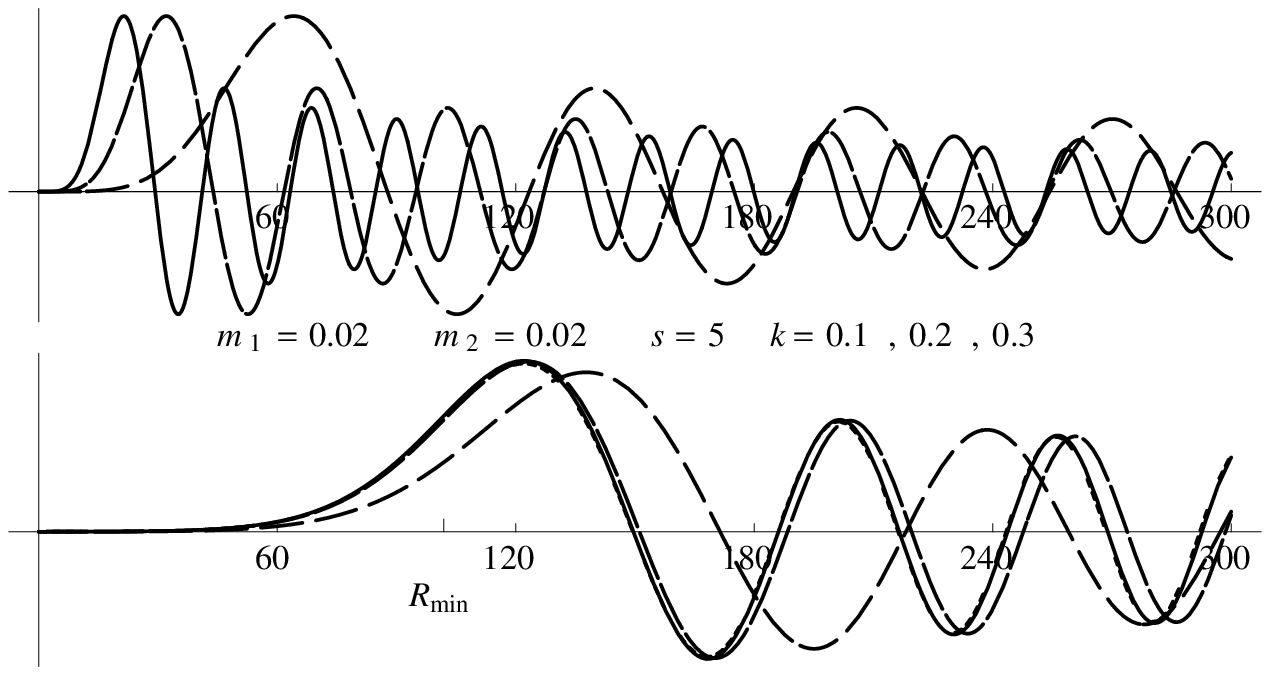}
    \xcaption{The wave functions for $s=5$. The minimal distance $\Rmin$
    increases linearly with $s$. As a consequence, it is not possible
    to localize the particles relative to each other, by superposing
    states with different $k$ and $s$}
  \label{wav3}
  \end{center}
\end{figure}

So, we find the following remarkable physical effect of quantum
gravity. If the statistics parameter $\stat$ is different from zero,
then the particles are no longer able to approach each other closer
than some minimal distance, which is of the order of the Planck
length. But this is not the only feature of the wave functions, from
which we can learn something about the short scale structure of
spacetime. A second effect also results from the relation
\eref{R-min}, and this is even independent of the statistics. Consider
the wave functions for larger angular momenta, for example $s=5$ in
\fref{wav3}. For large $r$, outside the figure, the free particle wave
functions and those of the coupled particles are still of the same
form, just with a phase shift, as in \fref{wav2}.

The minimal distance $\Rmin$ is now however five times as large as in
\fref{wav1}. Now, suppose that we want to localize the particles
relative to each other, not at the origin, but at some finite distance
$R=r_0$, and with some angular orientation $\xdir=\ph_0$. Clearly, for
this purpose we need a superposition of states with different quantum
numbers $k$ and $s$. We need large quantum numbers $s$ to get a sharp
peak in the angular orientation $\xdir$. And we need large quantum
numbers $k$ to get a sharp peak in the distance $R$. However, we have
just seen that for large angular momenta $s$, there is also a large
minimal distance $\Rmin$. And moreover, for large momenta $k$ the wave
function freezes when the distance is of the order of this minimal
distance. 

Roughly speaking, we can say that once we have a sharp angular
direction $\xdir$ of the particles, we do no longer have enough radial
states to superpose, in order to get a sharp distance $R$. Of course,
this is a somewhat heuristic argument, so let us make it more precise.
Consider first the following analogy. Suppose that we have a box
potential in flat space, which is zero outside and positive inside the
box. There are then two types of energy eigenstates, those tunneling
through the box, and those with sufficiently high energy passing the
box without tunneling. Here we have the situation that actually only
the tunneling states exist, because the region $r<\Rmin$ is
classically forbidden for all physical states. The physical Hilbert
space only consists of the tunneling states. 

But this subspace does not provide a complete basis of states in the
position space of the system. More precisely, it is not possible to
superpose the states, so that an arbitrarily sharp peak in position
space comes out. To make this a little bit more explicit, let us try
to define a physical state, which describes a localized pair of
particles. In case of the free particles, there is no problem. In the
Schr\"odinger quantization, we simply have to consider the special wave
function
\begin{equation}
  \label{rpp-loc}
  \psi(r,\ph) = r^{-1} \delta(r-r_0) \, \delta_\stat(\ph-\ph_0).
\end{equation}
It can of course be written a superposition of the energy momentum
eigenstates. We can even say explicitly what the corresponding wave
function $\psi(k,s)$ in momentum space is, which has to be inserted
into \eref{rpp-sch-wave}. It is
\begin{equation}
  \label{mom-loc}
  \psi(k,s) = \expo{-\ii s \ph_0} \, \zeta(k,s;r_0),
\end{equation}
because we have the completeness relation \eref{rpp-chi-ortho}. The
corresponding wave function in the Dirac quantization is obtained by
inserting the same wave function $\psi(k,s)$ in momentum space into
the formula \eref{rpp-wave-rad}. The resulting wave function
$\psi(\tau,r,\ph)$ is equal to \eref{rpp-loc} for $\tau=0$, thus
describing a localized pair of particle at a given moment of time. As
this is not an energy eigenstate, the particles will of course be
localized only at one moment of time.

So far, this is just ordinary quantum mechanics of free particles. But
now, let us ask whether such a localized state also exists for the
coupled particles. Are there physical states describing a localized
pair of particles, with a wave function of the form \eref{rpp-loc}? We
could try to insert the same wave function in momentum space
\eref{mom-loc} into the general superposition \eref{wave-rad} of
physical states. Doing so, we get
\begin{equation}
  \label{loc-no-go}
   \psi(\tau,r,\ph) = \sum_s \int k \, \dd k  \, 
          \expo{-\ii\mm(k)\tau} \, \expo{\ii s(\ph-\ph_0)}
          \,   \zeta(k,s;r_0) \, \zeta(k,s;r). 
\end{equation}
Now, what about the functions $\zeta(k,s;r)$ for fixed $s$? Do they
define an orthonormal basis of the space of all radial wave function?
If this was the case, then we would immediately recover the wave
function \eref{rpp-loc} for $\tau=0$, thus we have a localized pair of
particles at that moment of time. Unfortunately, however, the
functions $\zeta(k,s;r)$ do not provide an orthonormal basis. This is
not obvious from the representation \eref{zeta-r}, but it can be seen
by going back to the $X$-representation \eref{zeta-x}. The Bessel
functions provide an orthonormal basis if and only if the index is
fixed. For the free particles this is the case, since in
\eref{rpp-zeta} the index is $|s|$.

In \eref{zeta-r}, however, the index is $|s|/(1-4\lpl\mm(k))$, and
thus not only the argument but also the index of the Bessel function
depends on $k$. The functions $\zeta(k;s,r)$ for fixed $s$ in
\eref{loc-no-go} are not orthonormal, and therefore the resulting wave
function for $\tau=0$ is not of the form \eref{rpp-loc}, describing a
localized state. Now, one could argue that there might be another
superposition of physical states describing a localized pair of
particles. However, if there were such localized states, then we could
use them as a basis of the physical Hilbert space, and this would be
exactly the basis that we defined in the beginning of this section,
when we tried to set up a Schr\"odinger quantization. We saw, however,
that due to the global structure of the classical phase space such a
basis does not exist. 

So, what is the conclusion? Obviously, the spacetime in which the
quantized particles are moving has some kind of \emph{foamy}
structure. Maybe it is possible to analyze this structure more
explicitly, by having a closer look at the wave functions
$\zeta(k,s;r)$. For a single particle system, such an analysis has
been carried out in some detail in \cite{matwel}, finding a
semi-discrete structure of spacetime and an explicit uncertainty
relation, which forbids the particle to be localized. Here things are
more involved and therefore it is more difficult to make the idea of a
foamy spacetime more precise. But at least we can see that it is
impossible to localize the particles at a point in space, and this is
already a nice result. It gives us a hint to what kind of influence a
real theory of quantum gravity might have on the local structure of
spacetime in general.

\subsubsection*{The quantized cone}
Finally, let us consider yet another strange result, which has
actually not very much to do with the point particles. There is
quantized conical geometry of the spacetime at spatial infinity. Let
us once again look at the spinning cone \eref{spin-cone}. It defines
the asymptotic structure of the spacetime at infinity. We saw that it
has the following global geometric features. There is a deficit angle
of $8\pi GM$, and a time offset of $8\pi GS$. Now, consider the energy
momentum eigenstates spanning the physical Hilbert space. They are the
eigenstates of $M$ and $S$, with eigenvalues
\begin{equation}
   M  = \hbar \, \mm  ,\qquad
   S = \frac{ \hbar \, s}{1-4\lpl\mm}, \qquad
   0 < \mm < \frac1{4\lpl}, \qquad  s \in \ZZ .  
\end{equation}
For simplicity, we assumed that the particles are massless,
distinguishable, and that the statistics parameter is $\stat=0$. There
is nothing essentially different in a more general case, except that
the formulas below are slightly different. In the context of quantized
general relativity, the energy momentum eigenstates can also be
regarded as states with a \emph{sharp geometry} of the spacetime at
infinity, but with the locations of the particles in space smeared
out. However, the geometry of the spinning cone is thereby not
arbitrary. There is an obvious relation between the eigenvalues of $M$
and $S$, namely
\begin{equation}
  \label{quant-MS}
  (1-4GM) \,  S \in \ZZ \,  \hbar .
\end{equation}
As this results from the quantization of the canonical angular momentum
$J$, we expect that this is a general result, which does not depend,
for example, on the number of particles present. In fact, one can show
that for a general multi particle model there is the same relation
between the canonical angular momentum $J$ and the parameters of the
spinning cone $M$ and $S$, and the same is expected for any kind of
matter included \cite{matmul}.
 
Thus, we conclude that the geometry of the conical infinity itself is
quantized. Just like in a quantized atom, where not every classically
allowed combination of energy and angular momentum can be realized,
the conical geometry of the spacetime at infinity is restricted by
this quantum condition. It becomes a very simple relation when it is
written in geometric units, that is lengths and angles. We define the
total angle of the spinning cone to be $\alpha=2\pi-8\pi GM$. It is
the circumference of a unit circle around the central axis. And we
define the time offset to be $\tau=8\pi GS$. Then the quantization
condition becomes
\begin{equation}
  \label{quant-cone}
  \alpha \, \tau \in (4\pi)^2 \, \ZZ \, \lpl .  
\end{equation}
Only those conical spacetimes can be realized where the product of the
total angle and the time offset is an integer multiple of
$(4\pi)^2\lpl$.  Note that $\alpha$ is dimensionless and $\tau$ is a
length, or time, so that necessarily the Planck length shows up. If
$\alpha$ is close to $2\pi$, then $\tau$ is quantized is units of
$8\pi\lpl$, which is a kind of time quantization. If we assume that
this result is independent of the kind of matter being present inside
the universe, then what we have found is a real \emph{quantization of
geometry}.

\section*{Acknowledgments}
For helpful discussions and hospitality we would like to thank Ingemar
Bengtsson, S\"oren Holst, and Hermann Nicolai. 

\begin{appendix}

\section*{Appendix}
\setcounter{equation}{0} 

\def\thesection{A} 

Here we summarize some facts about the spinor representation of the
three dimensional Lorentz algebra $\algsl(2)$ of traceless $2\times2$
matrices, and the associated Lie group $\grpSL(2)$. A more
comprehensive collection of formulas, using the same notation, can be
found in \cite{matwel}. As a vector space, $\algsl(2)$ is isometric to
three dimensional Minkowski space. An orthonormal basis is given by
the gamma matrices
\begin{equation}
  \label{gamma}
  \gam_0 = \pmatrix{ 0 & 1 \cr -1 & 0 } , \qquad
  \gam_1 = \pmatrix{ 0 & 1 \cr 1 & 0 }, \qquad
  \gam_2 = \pmatrix{ 1 & 0 \cr 0 & -1 }.
\end{equation}
They satisfy the algebra 
\begin{equation}
  \label{gamma-alg}
  \gam_a \gam_b = \eta_{ab} \, \one - \eps_{abc} \, \gam^c, 
\end{equation}
where $a,b=0,1,2$, the metric $\eta_{ab}$ has signature $(-,+,+)$, and
for the Levi Civita symbol $\eps^{abc}$ we have $\eps^{012}=1$.
Expanding a generic matrix in terms of these gamma matrices, we obtain
an isomorphism of $\algsl(2)$ and Minkowski space, 
\begin{equation}
  \vv = v^a \, \gam_a  \equivalent
  v^a = \ft12\Trr{\vv\gam^a}.
\end{equation}
Some useful relations are that the scalar product of two vectors is
equal to the trace norm of the corresponding matrices, and the vector
product is essentially given by the matrix commutator,
\begin{equation}
  \label{vec-mat}
   \ft12\Trr{\vv\ww} = v_a w^a , \qquad
   \ft12 [ \vv , \ww ] = - \eps^{abc} \, v_a w_b \, \gam_c.
\end{equation}
Sometimes it is useful to introduce cylindrical coordinates in Minkowski
space, writing
\begin{equation}
  \label{vec-cyl}
   \vv = \tau \, \gam_0 + \rho \, \gam(\varphi),
\end{equation}
where $\tau$ and $\rho\ge0$ are real, and $\varphi$ is an angular
direction. The vector $\gam(\varphi)$ and its derivative
$\gam'(\varphi)$ form a rotated set of spacelike unit vectors, pointing
into the direction of $\varphi$ and the orthogonal direction,
\begin{equation}
  \label{gamma-rot}
  \gam(\varphi) = \cos\varphi \, \gam_1 + \sin\varphi \, \gam_2, \qquad
  \gam'(\varphi) = \cos\varphi \, \gam_2 - \sin\varphi \, \gam_1. 
\end{equation} 
Useful relations are 
\begin{equation}
  \gam_0 \gam(\varphi) = \gam'(\varphi), \qquad
  \gam_0 \gam'(\varphi) = - \gam(\varphi),
\end{equation}
and 
\begin{equation}
  \gam(\varphi_1) \, \gam(\varphi_2) = 
        \cos(\varphi_1-\varphi_2) \, \one + 
        \sin(\varphi_1-\varphi_2) \, \gam_0. 
\end{equation}
The Lie group $\grpSL(2)$ consists of matrices $\uu$ with unit
determinant. The group acts on the algebra in the adjoint
representation, so that 
\begin{equation}
  \vv \mapsto \uu^{-1} \vv \uu.
\end{equation}
This provides a proper Lorentz rotation of the vector $\vv$. The
conjugacy classes of the algebra are characterized by the invariant
length $\ft12\Trr{\vv^2}$. For timelike vectors $\vv$ with
$\ft12\Trr{\vv^2}<0$, we distinguish between positive timelike vectors
with $v^0=\ft12{\vv\gam^0}>0$, and negative timelike vectors with
$v^0=\ft12{\vv\gam^0}<0$. And similar for lightlike vectors. 

A special group element, which represents a clockwise rotations about
the $\gam_0$-axis by an angle $\alpha$, is
\begin{equation}
  \label{rot-mat}
  \uu = \expo{\alpha\gam_0/2} = 
         \cos(\alpha/2) \, \one + \sin(\alpha/2) \, \gam_0.
\end{equation}
This implies
\begin{equation}
  \gam_0 \mapsto \uu^{-1} \gam_0 \, \uu = \gam_0, \qquad
  \gam(\varphi) \mapsto  \uu^{-1} \gam(\varphi) \, \uu = 
                   \gam(\varphi-\alpha).  
\end{equation}
A boost is specified by a rapidity $\chi>0$ and a direction $\pdir$,
\begin{equation}
  \uu = \expo{\chi\gam(\pdir)} = 
         \cosh\chi \, \one + \sinh\chi \, \gam(\pdir).
\end{equation}
A generic element $\uu\in\grpSL(2)$ can be expanded in terms of the
unit and the gamma matrices, defining a scalar $u$ and a vector
$\pp\in\algsl(2)$,
\begin{equation}
  \label{proj}
  \uu = u \, \one + p^a \, \gam_a \follows
   \pp = p^a \, \gam_a.
\end{equation}
The scalar $u$ is basically the trace of $\uu$, and the vector $\pp$ is
called the \emph{projection} of $\uu$. The determinant condition implies 
that 
\begin{equation}
  u^2 = p^a p_a + 1. 
\end{equation}
According to the property of the vector $\pp$, we distinguish between
timelike, lightlike and spacelike group elements $\uu$. For a timelike
element we have $-1<u<1$, and it represents a rotation about some
timelike axis, which is then specified by the vector $\pp$. The angle
of rotation $\alpha$ is given by $u=\cos(\alpha/2)$, and the direction
is given by the sign of $p^0$. 

\end{appendix}

\end{document}